\documentclass[preprintnumbers,prd,showpacs,floatfix,superscriptaddress,nofootinbib,twocolumn, letterpaper]{revtex4-1}
\pdfoutput=1

\usepackage{longtable}
\usepackage{bm}
\usepackage{relsize}
\usepackage{amsfonts}
\usepackage{amsmath}
\usepackage{amssymb,epsf}
\usepackage{latexsym}
\usepackage{graphicx,epsfig}
\usepackage{amssymb}
\usepackage{float}
\usepackage{subfigure}
\usepackage[export]{adjustbox} %for valign in fig
\usepackage{epstopdf}
\usepackage[colorlinks=true,citecolor=blue,linkcolor=blue,urlcolor=black]{hyperref}
\usepackage{dcolumn}
\usepackage{psfrag}
\usepackage{wrapfig}
\usepackage{makeidx}
\usepackage{epsf}
\usepackage{color}
\usepackage{multirow}
\usepackage{mathtools}

\begin{document}
%%%%%%%%%%%%%%%%%%%%%%%%%%%%%%%%%%%%%%%%%%%%%%%%%%%%%%%%%%%%%%%%%%%%%%%%%%%%

\title{Observational imprints of gravastars from accretion disks and hot-spots}

\author{João Luís Rosa}
\email{joaoluis92@gmail.com}
\affiliation{Institute of Physics, University of Tartu, W. Ostwaldi 1, 50411 Tartu, Estonia}
\affiliation{Institute of Theoretical Physics and Astrophysics, University of Gda\'{n}sk, Jana Ba\.{z}y\'{n}skiego 8, 80-309 Gda\'{n}sk, Poland}

\author{Daniela S. J. Cordeiro}
\email{dani.j.cordeiro@gmail.com}
\affiliation{Instituto de Astrof\'{i}sica e Ci\^{e}ncias do Espa\c{c}o, Faculdade de Ci\^{e}ncias da Universidade de Lisboa, Edifício C8, Campo Grande, P-1749-016 Lisbon, Portugal}

\author{Caio F.  B. Macedo} 
\email{caiomacedo@ufpa.br}
\affiliation{Faculdade de F\'isica, Campus Salin\'opolis, Universidade Federal do Par\'a, 68721-000, Salin\'opolis, Par\'a, Brazil}

	\author{Francisco S. N. Lobo} \email{fslobo@ciencias.ulisboa.pt}
\affiliation{Instituto de Astrof\'{i}sica e Ci\^{e}ncias do Espa\c{c}o, Faculdade de Ci\^{e}ncias da Universidade de Lisboa, Edifício C8, Campo Grande, P-1749-016 Lisbon, Portugal}
\affiliation{Departamento de F\'{i}sica, Faculdade de Ci\^{e}ncias da Universidade de Lisboa, Edif\'{i}cio C8, Campo Grande, P-1749-016 Lisbon, Portugal}

\date{\today}

%%%%%%%%%%%%%%%%%%%%%%%%%%%%%%%%%%%%%%%%%%%%%%%%%%%%%%%%%%%%%%%%%%%%%%%%%%%%
\begin{abstract} 
In this work, we analyze the observational properties of thin-shell gravastars under two astrophysical frameworks, namely surrounded by optically-thin accretion disks and orbited by hot-spots. We consider the thin-shell gravastar model with two free parameters, the gravastar radius and ratio of mass allocated at the thin-shell, and produce the corresponding observables via the use of numerical backwards ray-tracing codes. Regarding the observations of accretion disks, our results indicate that, due to the absence of a strong gravitational redshift effect, smooth gravastar configurations cannot reproduce shadow observations when internal emission is assumed. We thus expect such models to be excluded as candidates for supermassive objects in galactic cores. Nevertheless, thin-shell gravastars with a large portion of their total mass allocated at the surface can produce such an effect and are thus adequate candidates for black-hole mimickers. In the context of hot-spot orbits, the astrometrical observational properties of ultra-compact gravastars resemble closely those of other ultra-compact objects e.g. fluid stars and bosonic stars. However, for low-compacticity configurations, the time-integrated fluxes feature additional contributions in the form of a high-intensity plunge through image. These qualitative differences in the observational properties of gravastars in comparison with black-hole spacetimes could potentially be discriminated by the next generation of interferometric experiments in gravitational physics.
\end{abstract}
%%%%%%%%%%%%%%%%%%%%%%%%%%%%%%%%%%%%%%%%%%%%%%%%%%%%%%%%%%%%%%%%%%%%%%%%%%%%

\maketitle

%%%%%%%%%%%%%%%%%%%%%%%%%%%%%%%%%%%%%%%%%%%%%%%%%%%%%%%%%%%%%%%%%%%%%%%%%%%%
\section{Introduction}\label{sec:intro}
%%%%%%%%%%%%%%%%%%%%%%%%%%%%%%%%%%%%%%%%%%%%%%%%%%%%%%%%%%%%%%%%%%%%%%%%%%%%

In recent years, the field of gravitational physics has witnessed a remarkable progress in its experimental endeavors, yielding compelling evidence for the existence of highly compact objects. Notably, the collaborative efforts of the LIGO/Virgo collaboration, the Event Horizon Telescope (EHT), and the GRAVITY collaboration have provided groundbreaking discoveries. These include the detection of gravitational wave signals stemming from binary mergers of compact objects \cite{LIGOScientific:2016aoc,LIGOScientific:2021djp}, the observation of a distinct shadow-like feature within the core of the M87 galaxy \cite{EventHorizonTelescope:2019dse,EventHorizonTelescope:2021bee} and close to Sgr A*\cite{EventHorizonTelescope:2022wkp}, as well as the detection of infrared flares in the vicinity of our own galactic center \cite{GRAVITY:2020lpa}. These experimental achievements align remarkably well with the theoretical predictions of black-hole spacetimes, reinforcing the validity of the Kerr hypothesis \cite{Will:2014kxa,Yagi:2016jml}, which describes the end state of a complete gravitational collapse within an appropriate astrophysical context as resulting in the formation of a rotating and electrically-neutral black hole \cite{Kerr:1963ud,Penrose:1964wq}.

Although black-hole spacetimes have been proven highly efficient in describing the aforementioned observations, these spacetimes are known to be inherently problematic from both a mathematical and physical point of view. Indeed, a direct consequence of complete gravitational collapse is the appearance of singularities in the spacetime \cite{Penrose:1964wq,Penrose:1969pc}, indicating a geodesic incompleteness of the spacetime and leading to a lack of understanding of the gravitational interaction. Furthermore, the existence of an event horizon and the information loss paradox that arises with it \cite{Hawking:1976ra} indicates a loss of predictability in gravitational collapse. To address these limitations of the black-hole spacetime, a wide variety of alternative models known as Exotic Compact Objects (ECOs) has been proposed \cite{Cardoso:2019rvt}, with the goal of replicating the observational properties of black-hole spacetimes while avoiding the previously mentioned problematic features\cite{Wang:2023nwd}.

Within the extensive array of proposed ECO models, a prominent category encompasses compact objects composed of one or more relativistic fluid components. Noteworthy examples in this category include relativistic fluid spheres \cite{Buchdahl:1959zz,Rosa:2020hex} and gravastars \cite{Mazur:2001fv,Mazur:2004fk,Lobo:2005uf,Visser:2003ge,Cattoen:2005he}. In fact, the gravastar picture is an alternative model to the concept of a black hole, where there is an effective phase transition at or near where the event horizon is expected to form, and the interior is replaced by a de Sitter condensate \cite{Mazur:2001fv,Mazur:2004fk}. Models belonging to this category have been rigorously demonstrated to satisfy several essential criteria for physical relevance, including the non-exoticity of matter (i.e., satisfying appropriate energy conditions) and linear stability \cite{Rosa:2020hex}, as well as exhibiting distinctive observational properties that deviate slightly from their black-hole counterparts \cite{Cardoso:2015zqa,Rosa:2023hfm,Chirenti:2007mk,Pani:2009ss,Harko:2009gc}, rendering them as viable alternatives to the black-hole paradigm, testable through the wealth of data gathered by the ongoing and the future generation of experiments in gravitational physics \cite{Cardoso:2016oxy,Cardoso:2017cqb,Postnikov:2010yn,Cardoso:2017cfl,Cardoso:2016rao}.

In this work, we focus on the study of observational properties of gravastar models in two main astrophysical settings, namely when such models are surrounded by optically-thin accretion disks, and orbited by relativistic hot-spots. The emission profiles of the accretion disks are modelled by the Johnson's-SU distribution \cite{Gralla:2020srx,Vincent:2022fwj} and the ray-tracing is performed with a Mathematica-based ray-tracing code previously used to produce the observed intensity profiles and images for other ECO models \cite{Rosa:2023hfm,Rosa:2022tfv,Rosa:2023qcv,Olmo:2023lil,Guerrero:2022msp,Guerrero:2022qkh,Olmo:2021piq,Guerrero:2021ues,Asukula:2023akj}. The orbits of hot-spots are simulated using the ray-tracing software GYOTO \cite{Vincent:2011wz,Grould:2016emo}, which was proven useful in the imaging of black-hole spacetimes in suitable astrophysical contexts \cite{Vincent:2020dij,Lamy:2018zvj,Vincent:2016sjq,Vincent:2012kn}, as well as ECO models \cite{Rosa:2023qcv,Rosa:2022toh,Tamm:2023wvn}.

This paper is organized as follows. In Sec. \ref{sec:theory}, we introduce the thin-shell gravastar model and analyze the geodesic structure and effective potential. In Sec. \ref{sec:disks}, we model the intensity profiles for two distinct accretion models and produce the observed intensity profiles and observed axial and inclined images. In Sec. \ref{sec:hotspot}, we simulate the orbits of relativistic hot-spots and produce the corresponding astrometric observables, and finally in Sec. \ref{sec:concl}, we summarize and trace out our conclusions.

%%%%%%%%%%%%%%%%%%%%%%%%%%%%%%%%%%%%%%%%%%%%%%%%%%%%%%%%%%%%%%%%%%%%%%%%%%%%
\section{Theory and setup}\label{sec:theory}
%%%%%%%%%%%%%%%%%%%%%%%%%%%%%%%%%%%%%%%%%%%%%%%%%%%%%%%%%%%%%%%%%%%%%%%%%%%%

\subsection{Thin-shell gravastar model}

In General Relativity (GR), the spacetime that describes a static and spherically symmetric thin-shell gravastar is composed of two distinct spacetime regions, $\mathcal{V}^\pm$, where we dub $\mathcal{V}^+$ as the exterior region and $\mathcal{V}^-$ as the interior region, separated by a spherical timelike hypersurface $\Sigma$. In the usual spherical coordinates $x^\mu=\{t,r,\theta,\phi\}$, the line elements $ds^2_\pm$ of the two spacetime regions can be generally written in the form
\begin{equation}\label{eq:genmetric}
ds^2_\pm=-f^\pm\left(r\right) dt^2+\frac{1}{h^\pm\left(r\right)}dr^2+r^2d\Omega^2,
\end{equation}
where $d\Omega^2=d\theta^2+\sin^2\theta d\phi^2$ is the surface element of the two-sphere. The metric functions $f\left(r\right)^\pm$ and $h\left(r\right)^\pm$ take the forms
\begin{equation}\label{eq:fhext}
f^+\left(r\right)=h^+\left(r\right)=1-\frac{2M}{r},\qquad r>R,
\end{equation}
\begin{equation}\label{eq:fhint}
f^-\left(r\right)=\alpha h^-\left(r\right) = \alpha\left(1-\frac{2m\left(r\right)}{r}\right),\qquad r<R,
\end{equation}
where $M$ is the total mass of the gravastar as measured by an observer in the exterior region $\mathcal{V}^+$, $\alpha$ is a constant free parameter whose effect on the model we clarify in what follows, $R$ is the radius of the gravastar, which coincides with the radius of the hypersurface $\Sigma$, and the mass function $m\left(r\right)$ is defined as
\begin{equation}
m\left(r\right)=\frac{4}{3}\pi \rho r^3,
\end{equation}
where $\rho$ is the constant energy density of the exotic fluid that populates the interior region $\mathcal{V}^-$, which satisfies an equation of state of the form $p=-\rho$, where $p$ is the isotropic pressure. 

For the whole spacetime $\mathcal V = \mathcal{V}^+\cup\mathcal{V}^-$ to be a solution of the Einstein's field equations, it is necessary that the interior and exterior metrics defined in Eqs. \eqref{eq:genmetric}--\eqref{eq:fhint} satisfy the so-called junction (or matching) conditions \cite{Israel:1966rt}. We define the induced metric $h_{ab}$ and the extrinsic curvature $K_{ab}$ at $\Sigma$ as
\begin{equation}
h_{ab}=e^\mu_a e^\nu_b g_{\mu\nu},\qquad K_{ab}=e^\mu_a e^\nu_b\nabla_\mu n_\nu,
\end{equation}
where $e^\mu_a\equiv \frac{\partial x^\mu}{\partial y^a}$ are the projection operators from the four-dimensional spacetime manifold $\mathcal{V}$ into the hypersurface $\Sigma$ described by a set of coordinates $y^a$ that excludes the direction orthogonal to $\Sigma$, $g_{\mu\nu}$ is the metric tensor, $\nabla_\mu$ is the covariant derivative operator, and $n_\mu$ is the normal vector to the hypersurface $\Sigma$. Thus, the two junction conditions in GR can be written in the form
\begin{equation}\label{eq:junctions}
\left[h_{ab}\right]=0,\qquad 8\pi S_{ab}=h_{ab}\left[K\right]-\left[K_{ab}\right],
\end{equation}
where we have introduced the jump notation $\left[X\right]=X^+|_\Sigma-X^-|_\Sigma$, $S_{ab}$ is the three-dimensional stress-energy tensor of the thin-shell at $\Sigma$, and $K=h^{ab}K_{ab}$ is the trace of the extrinsic curvature. 

Defining the surface energy density $\sigma$ and the surface tension $\Pi$ of the hypersurface as
\begin{equation}
S_a^b=\text{diag}\left(-\sigma,\Pi,\Pi\right),
\end{equation}
one finds that for the general spherically symmetric metric given in Eq. \eqref{eq:genmetric}, the first and second junction conditions in Eq. \eqref{eq:junctions} take the forms
\begin{equation}\label{eq:jc1}
\left[f\right]=0,
\end{equation}
\begin{equation}\label{eq:jc2}
\left[\sqrt{h}\right]=-4\pi R\sigma,\qquad \left[\frac{\sqrt{h}f'}{f}\right]=8\pi\left(\sigma + 2\Pi\right),
\end{equation}
respectively, where the second junction condition contributes with two restrictions, arising from the radial and angular components of $K_{ab}$. Note that the angular component of $h_{ab}$ does not contribute with an additional constraint since the angular parts of the metrics in Eq. \eqref{eq:genmetric} coincide. Introducing the following definitions for the volumetric mass $M_\rho$ and surface mass $M_\sigma$ as
\begin{equation}
M_\rho=\frac{4}{3}\pi R^3 \rho,\qquad M_\sigma = 4\pi R^2\sigma,
\end{equation}
the first junction condition in Eq. \eqref{eq:jc1} sets the value of $\alpha$ for a given gravastar model, whereas the second junction condition in Eq. \eqref{eq:jc2} sets the values of the exterior mass $M$ and the surface tension $\Pi$, which take the forms
\begin{equation}\label{eq:defalpha}
\alpha=\frac{1-2\bar M}{1-2\bar M_\rho},
\end{equation}
\begin{equation}\label{eq:defM}
\bar M=\bar M_\rho+\bar M_\sigma\sqrt{1-2\bar M_\rho}-\frac{\bar M_\sigma^2}{2},
\end{equation}
\begin{equation}\label{eq:defPi}
\Pi=\frac{1}{8\pi R}\left(\frac{1-\bar M}{\sqrt{1-2\bar M}}-\frac{1-4\bar M_\rho}{\sqrt{1-2\bar M_\rho}}\right),
\end{equation}
where we have introduced the dimensionless quantities $\bar M = M/R$, $\bar M_\rho = M_\rho/R$, and $\bar M_\sigma = M_\sigma / R$. Note that the expression for $\Pi$ in Eq. \eqref{eq:defPi} is provided just for completeness, but it is not relevant in the analysis that follows.

\begin{figure*}
    \centering
    \includegraphics[scale=0.95]{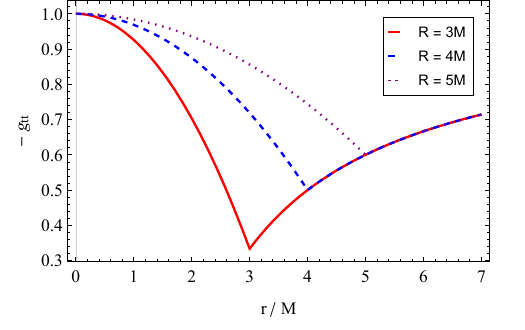}\qquad
    \includegraphics[scale=0.95]{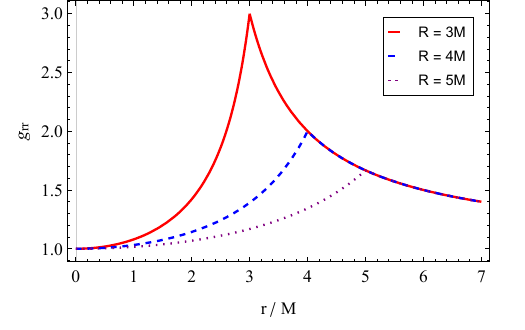}\\
    \includegraphics[scale=0.95]{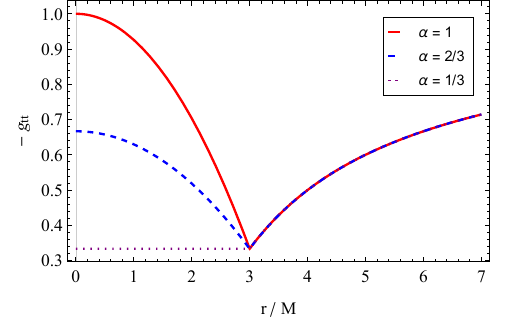}\qquad
    \includegraphics[scale=0.95]{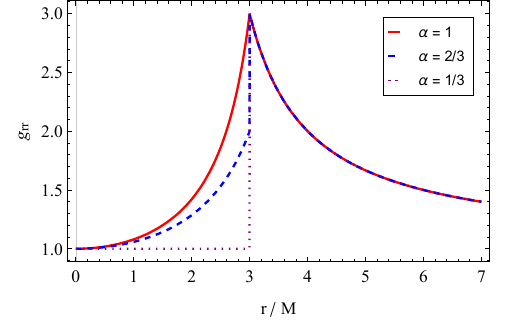}
    \caption{Components $g_{tt}$ (left column) and $g_{rr}$ (right column) of the metric in Eq.\eqref{eq:genmetric} for $\alpha=1$ with varying $R$ (top row) and for $R=3M$ with varying $\alpha$ (bottom row). Changing the value of $R$ alters the radius at which the matching is performed while maintaining the boundary conditions, while changing the value of $\alpha$ alters the central boundary value of $g_{tt}$ and induces a discontinuity in $g_{rr}$. }
    \label{fig:metric}
\end{figure*}

Summarizing, the thin-shell gravastar model can thus be described by two free parameters, the radius $R$ and the constant $\alpha$, the latter controlling the ratio of the total mass $M$ of the gravastar that is distributed in its volume and surface. In Figure \ref{fig:metric}, it is shown how the $g_{tt}$ and $g_{rr}$ metric components change with variations of $R$ and $\alpha$. It can be seen that changing the value of $R$ alters the radius at which the matching is performed while maintaining the boundary conditions at the origin $r=0$, whereas a change in the value of $\alpha$ alters the central boundary value of $g_{tt}$ and induces a discontinuity in $g_{rr}$. For a set radius $R$, the choice $\alpha=1$ corresponds to a model for which the thin-shell has no mass, i.e., the entire mass of the gravastar is distributed in its volume. This can be seen from Eq. \eqref{eq:defalpha}, which sets $M=M_\rho$, and consequently Eq. \eqref{eq:defM} sets $M_\sigma=0$. The same conclusion could be reached from the first of Eq. \eqref{eq:jc2}, since for $\alpha=1$ the function $h$ is continuous and thus $\sigma=0$. For any $\alpha\in\left]1-2\bar M,1\right]$, the total mass $M$ of the gravastar is distributed both in its volume and its surface, whereas in the limit $\alpha=1-2\bar M$ the volumetric mass vanishes, i.e., $M_\rho=0$. Note that values of $\alpha<1-2\bar M$ are allowed from a mathematical point of view, but these configurations feature negative energy densities $\rho<0$ which violate the weak energy condition, and thus are of limited physical relevance. 

\subsection{Geodesic structure and effective potential}\label{sec:geodesics}

To obtain the equations of motion for test particles in a given background spacetime, it is usual to recur to the Lagrangian formalism, in which the Lagrangian density reads $\mathcal L = g_{\mu\nu}\dot x^\mu\dot x^\nu=-\delta$, where overdots denote derivatives with respect to the affine parameter along the geodesics and $\delta$ is a constant parameter that takes the values $\delta=0$ or $1$ for timelike (massive) and null (massless) test particles, respectively. The Lagrangian analysis can be simplified via the use of the spacetime symmetries. In particular, given that the spacetime is spherically symmetric, one can restrict their analysis to the equatorial plane $\theta=\pi/2$ and $\dot \theta=0$ without loss of generality. Following this assumption, two conserved quantities can be defined, namely the energy per unit mass $E=-g_{tt}\dot t$ and angular momentum per unit mass $L=r^2\dot\phi$. The radial component of the equation of motion for a test particle then reads
\begin{equation}\label{eq:eom_geodesic}
{-g_{tt}g_{rr}}\dot r^2+V\left(r\right)=E^2,
\end{equation}
\begin{equation}\label{eq:def_potential}
V\left(r\right)=-g_{tt}\left(\frac{L^2}{r^2}+\delta\right),
\end{equation}
i.e., it reduces to the form of a (mass dependent) classical test particle moving along a one-dimensional potential $V\left(r\right)$. The description above is very useful for finding the path of null geodesics from a specific source and understanding the geometry of accretion disks through the orbital stability analysis of time-like circular geodesics. We shall delve further into this below.

\begin{figure*}
    \centering
    \includegraphics[scale=1]{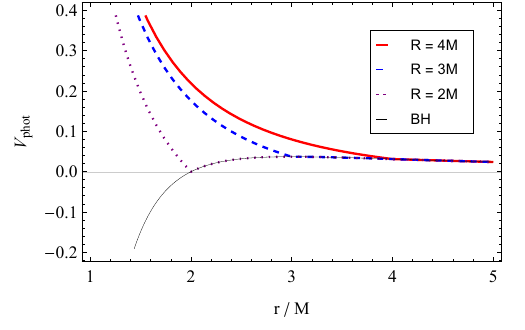}
    \includegraphics[scale=1]{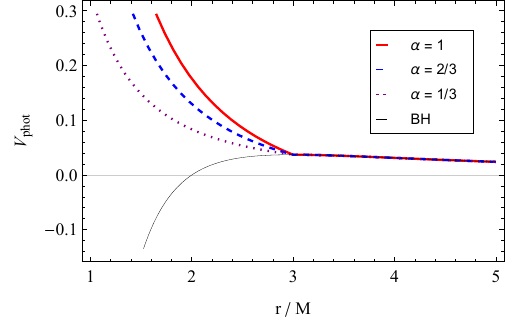}
    \caption{Photon effective potential $V_{\rm phot}$ from Eq. \eqref{eq:Vphot} as a function of the normalized radial coordinate $r/M$ for different values of $R$ and fixed $\alpha=1$ (left panel)\footnote{We note that the value $R=2M$ is not physically allowed due to the vanishing of the $g_{tt}$ component of the metric at that radius. Nevertheless, and given that any value of $R\gtrsim 2M$ is allowed, the case $R=2M$ is to be considered as a limit $R\to 2M$.}, and for different values of $\alpha$ and fixed $R=3M$ (right panel). The thin black line represents the photon-effective potential for the Schwarzschild black-hole. The number of stationary points is directly affected by the parameter $R$ but not by the parameter $\alpha$.}
    \label{fig:Vphot}
\end{figure*}

First, let us focus on null geodesics. We are particularly interested in circular orbits, which are defined by $\dot r=\ddot r=0$. From Eq. \eqref{eq:eom_geodesic}, this implies that $V\left(r\right)=E^2$, and $V'\left(r\right)=0$. Taking $\delta=0$ allows one to find circular null geodesics, also known as light-rings (LRs). From the effective potential, we can define a more convenient form
\begin{equation}\label{eq:Vphot}
V_{\rm phot}\left(r\right)=\frac{V\left(r\right)}{L^2}=-\frac{g_{tt}}{r^2},
\end{equation}
If the potential $V_{\rm phot}$ features any stationary points, these correspond to the radii of the LRs. In Fig. \ref{fig:Vphot} we plot the potential $V_{\rm phot}$ for different values of the gravastar radius $R$ while keeping $\alpha$ constant, and also for different values of $\alpha$ for $R$ constant. One observes that if $R>3M$, the potential $V_{\rm phot}$ does not present any stationary point, and thus no LRs are present in the spacetime; if $R=3M$ then the potential $V_{\rm phot}$ features a saddle point at $R=3M$ corresponding to a degenerate pair of LRs, and if $R<3M$ the potential $V_{\rm phot}$ has two extremum points, a maximum at $r^+_{\rm LR}=3M$ corresponding to an unstable LR, and a minimum at $r^-_{\rm LR}=R$ corresponding to a stable LR.\footnote{Although the derivative of the potential is discontinuous at the surface, this is due to the assumption that a thin-shell is present. For thick shells, which would be more realistic, we expect that our analysis would be equally valid.} The number of LRs in the spacetime thus depends directly on the radius of the gravastar. On the other hand, one observes that a variation in $\alpha$, although it alters the shape of the potential quantitatively, does not affect the number of LRs present for a given radius $R$.

For time-like circular geodesics, we have that the effective potential can be written as
\begin{equation}
V_{\rm part}=-g_{tt}\left(\frac{L^2}{r^2}+1\right).
\end{equation}
To study circular time-like geodesics, we can follow a procedure similar to the photon case. Assuming that there is a geodesic such that $r=r_{\rm p}={\rm constant}$, we can use Eq.~\eqref{eq:def_potential} to find the angular momentum and energy for these possible circular orbits. We have
\begin{align}
L_{\rm p}=\left.\sqrt{\frac{r^3g_{tt}'}{2g_{tt}-rg_{tt}'}}\ \right|_{r=r_{\rm p}},\\
E_{\rm p}=\left.\sqrt{\frac{2g_{tt}}{2g_{tt}-rg_{tt}'}}\ \right|_{r=r_{\rm p}}.
\end{align}
Notice that both $L_{\rm p}$ and $E_{\rm p}$ diverge when $2g_{tt}-rg_{tt}'\to 0$, which corresponds precisely to the position of light-like circular geodesics. Circular geodesics exist as long as the energy and angular momentum of the particle are positive and real-valuated. 

In horizonless compact objects, like boson stars, it is natural to investigate whether time-like circular geodesics inside the stellar matter exist~\cite{Rosa:2023qcv}. In such a situation, accretion disks could extend towards the center of the star or have additional inner edges depending on the stellar structure~\cite{Olivares:2018abq,Herdeiro:2021lwl}. Therefore, it is instructive to analyze if time-like circular orbits are possible inside the gravastars. This can be directly inferred from the computation of the angular momentum for a given $r_{\rm p}<R$. We thus have
\begin{equation}\label{eq:ang}
    L_{\rm p}=\frac{r_{\rm p}^2}{M}\sqrt{-\frac{2M+R(\alpha-1)}{R^3\alpha}}.
\end{equation}
As $1-2\bar M<\alpha\leq 1$, the angular momentum described above is purely imaginary and, therefore, circular orbits are not possible inside the star. From the physical point of view, this is expected: due to the gravitational repulsion generated by the interior de-Sitter core of gravastars, there is no support for circular geodesics. Note however that if the particles are allowed to pass through the shell without directly interacting with it, closed ``eccentric'' orbits that cross the interior of the star are possible (which can be seen through the effective potential analysis).\footnote{The quotes in eccentric here are justified as these orbits are not elliptical, and thus the word eccentricity does not have the usual meaning.} However, as accretion disk structures usually are constructed from continuous circular orbital motion up to the innermost stable orbit, their structures in gravastars are, in principle, similar to that of a BH, as long as $R<6M$.

\begin{figure*}
    \includegraphics[width=1\columnwidth]{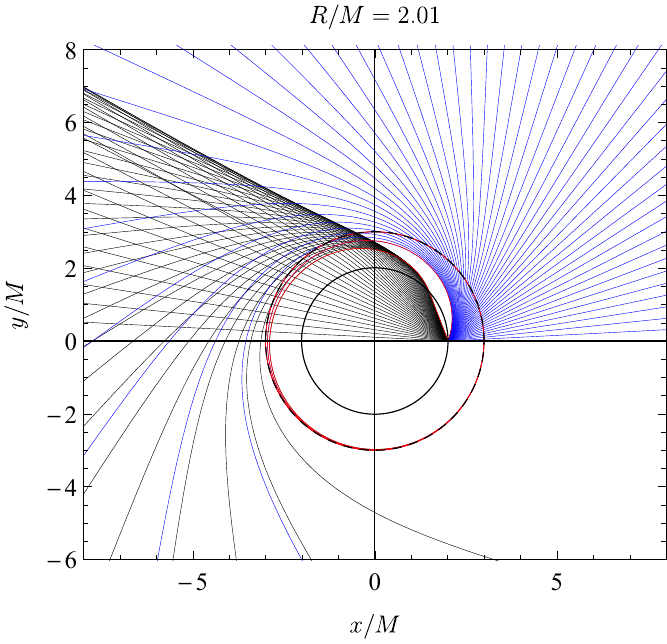}\includegraphics[width=1\columnwidth]{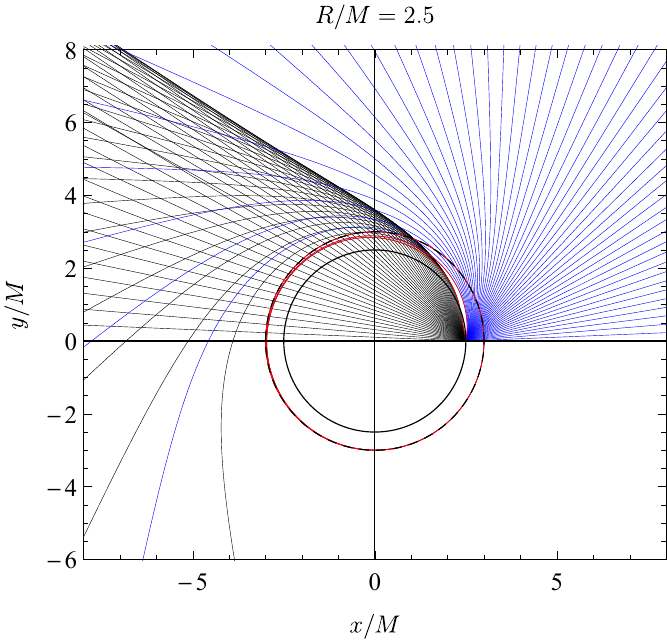}
    \caption{Light-rays coming from a point at the stellar surface for $R=2.01M$ (left panel) and $R=2.5M$ (right panel) with $\alpha=1$. In these two cases, there are light-rays that propagate inward (black lines) and outward (blue lines) that asymptotically approach the light-ring at $r_{\rm LR}=3M$, represented by the red lines. Notice that light rays with certain impact parameters are trapped in a nontrivial orbital motion around the compact object.}\label{fig:ray_spot}
\end{figure*}

Although gravastars might have similar disk structures, there could be signatures associated with the existence of a hard surface. Because matter can fall into the stellar surface and there is no horizon -- with instead a de Sitter repulsion inside -- it could accumulate at the surface of the star. This accumulation can lead to possible emissions at the stellar surface, generating distinctive bright spots. Note however that if the star is very compact, these bright spots might be overshadowed by the accretion disk emission, due to the high-redshift factor at the surface of the gravastar. In Sec.~\ref{sec:hotspot} we study in more detail the existence and observational consequences of bright spots. Here we can get a glimpse by analyzing the propagation of light from a source near the stellar surface of the gravastar. The result can be seen in Fig.~\ref{fig:ray_spot}, where we plot light rays coming from a point at the stellar surface for $R=2.01M$ and $R=2.5M$ for $\alpha=1$. Because of the structure of the star -- having a de-Sitter core -- light rays inside the star follow a straight line regardless of the value of $\alpha$. There is a range for the impact parameter, $b$, in which light rays are trapped orbiting the star. The boundary of this range is denoted by light rays emitted inward and outward with precisely $b=3\sqrt{3}M$, which asymptotically orbit at $r_{ LR}=3M$ (dotted line in the plot). In the limit $R\to 3M$ this range converges to a point given by this critical impact parameter, and for $R>3M$ the light-ring is non-existent.

\begin{figure}
    \centering
    \includegraphics[width=\columnwidth]{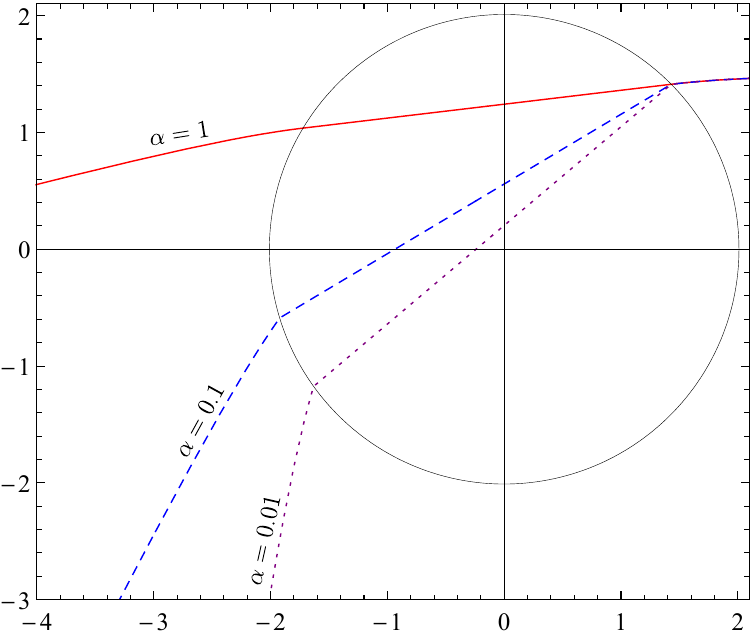}
    \caption{Light ray deflection considering a gravastar with $R/M=2.01$ and different values of $\alpha$. We see that the deflection angle at the stellar surface depends on the value of $\alpha$.}
    \label{fig:deflect}
\end{figure}

We note here that, although the light rays always follow straight lines, we have refraction-type phenomena between the vacuum exterior and the de Sitter core. This is caused by a discontinuity in $\dot{r}$. From Eq.~\eqref{eq:eom_geodesic} we have
\begin{equation}
    [\dot{r}]=\mp \sqrt{\frac{E^2-V(R)}{g_{tt}(R)}}\left[\frac{1}{\sqrt{g_{rr}}}\right],
\end{equation}
where the $\mp$ sign depends whether the light-ray is ingoing (-) or outgoing (+). As the above condition depends on the value of $\alpha$, it implies that light deflects differently for different values of $\alpha$ as it enters and leaves the star. To illustrate this effect, in Fig.~\ref{fig:deflect} we consider a light ray deflection with a given impact parameter for stars with the same radius, but different values of $\alpha$. We see that the deflection angle is larger for smaller values of $\alpha$.

The analysis of the effects of the free parameters $R$ and $\alpha$ in the effective photon potential $V_{\rm phot}$ indicate that the geodesic structure of the gravastar spacetime is highly dependent on the radius $R$, and thus one expects that the observational properties of these spacetimes should vary significantly for different values of $R$. On the other hand, the parameter $\alpha$ appears to be subdominant in affecting the geodesic structure, and thus one expects that variations in $\alpha$ should not induce any significant changes in the observational properties except for gravitational redshift effects. These statements are supported by the results that follow. 

%%%%%%%%%%%%%%%%%%%%%%%%%%%%%%%%%%%%%%%%%%%%%%%%%%%%%%%%%%%%%%%%%%%%%%%%%%%%
\section{Accretion disks}\label{sec:disks}
%%%%%%%%%%%%%%%%%%%%%%%%%%%%%%%%%%%%%%%%%%%%%%%%%%%%%%%%%%%%%%%%%%%%%%%%%%%%

\subsection{Intensity profiles}

\begin{figure}
    \centering
    \includegraphics[scale=0.9]{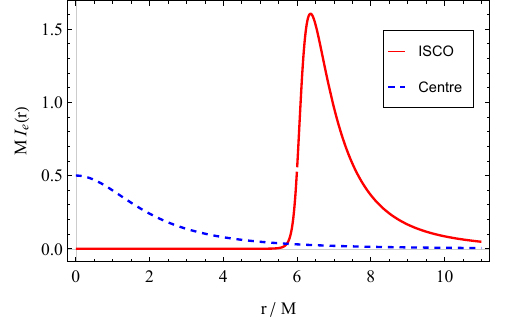}
    \caption{Intensity profiles in the reference frame of the emitter given in Eq.\eqref{eq:GLM} for the two accretion disk models considered. The ISCO disk model is characterized by $\gamma=-2$, $\mu=6M$, and $\sigma=M/4$, whereas the Center disk model is characterized by $\gamma=\mu=0$ and $\sigma=2M$.}
    \label{fig:disks}
\end{figure}

Let us consider the observational properties of gravastars surrounded by optically thin accretion disks. To perform ray-tracing simulations of this system, we recur to a Mathematica-based code previously used in several publications to study observational properties of ECOs \cite{Rosa:2023hfm,Rosa:2022tfv,Rosa:2023qcv,Olmo:2023lil,Guerrero:2022msp,Guerrero:2022qkh,Olmo:2021piq,Guerrero:2021ues}. In this code, the accretion disk is infinitesimally thin and it lies at the equatorial plane $\theta=\pi/2$. The emission profile of the accretion disk as a function of the radial coordinate $r$ is modelled through the recently introduced Gralla-Lupsasca-Marrone (GLM) model based on a Johnson's-SU distribution \cite{Gralla:2020srx}. The main advantage of the GLM model is that it has been shown to closely reproduce the intensity profiles obtained by a full general-relativistic magneto-hydrodynamic simulations of accretion disks in adequate astrophysical settings \cite{Vincent:2022fwj}. The intensity profile in the reference frame of the emitter $I_e$ for the GLM model is given by
\begin{equation}\label{eq:GLM}
I_e\left(r,\gamma,\mu,\sigma\right)=\frac{\exp\left\{-\frac{1}{2}\left[\gamma+\text{arcsinh}\left(\frac{r-\mu}{\sigma}\right)\right]^2\right\}}{\sqrt{\left(r-\mu\right)^2+\sigma^2}},
\end{equation}
where the free parameters $\gamma$, $\mu$, and $\sigma$ control the shape of the profile, namely the rate of increase, the position of the central peak, and the dilation of the profile, respectively. These parameters can then be manipulated in order to obtain intensity profiles that are suitable to test different astrophysical properties of the spacetime. In this work, we select two different intensity profiles (see Fig. \ref{fig:disks}) motivated by the structure of the gravastar spacetimes, namely
\begin{enumerate}
    \item ISCO model: given that the gravastar spacetimes considered in this work feature an ISCO at a radius $r_{ISCO}=6M$, below which circular orbits become unstable, it is natural to assume that the accretion disk should not extend to a radius smaller than the radius of the ISCO. Furthermore, given that the orbital velocity of massive test particles undergoing circular orbits $\Omega_c$ increases with a decrease in the radius in this regime, it is also intuitive to assume that the intensity of the emission should increase monotonically from infinity down to $r_{ISCO}$ and then decrease abruptly. We call this the ISCO model and it is characterized by the parameter values $\gamma=-2$, $\mu=6M$, and $\sigma=M/4$.
    \item Center model: given that the gravastar spacetimes considered in this work do not feature an event horizon, one cannot exclude the possibility of the accretion of particles into the gravastar to produce an accumulation of matter in the interior of the star, which in a general situation is expected to radiate and to be detectable by an exterior observer. In such a case, one would expect the density of the agglomerated matter to peak at the center of the gravastar and decrease monotonically with the radius. We thus select an emission profile that follows the same shape, as one would expect the emission intensity to be proportional to the density of matter. We denote this as the Center model and it is characterized by the parameters $\gamma=\mu=0$ and $\sigma=2M$.
\end{enumerate}
We note that, in a more realistic astrophysical situation, we expect both of the phenomena described by each of the disk models separately, i.e., the truncation of the accretion disk in the region where circular orbits are stable, plus the accumulation of matter in the inner regions of the spacetime, to happen simultaneously. Nevertheless, and given the additive properties of the intensity profiles, we chose to analyze the two models and their implications separately.

\subsection{Axial observations}

\begin{figure*}
    \centering
    \includegraphics[scale=0.62]{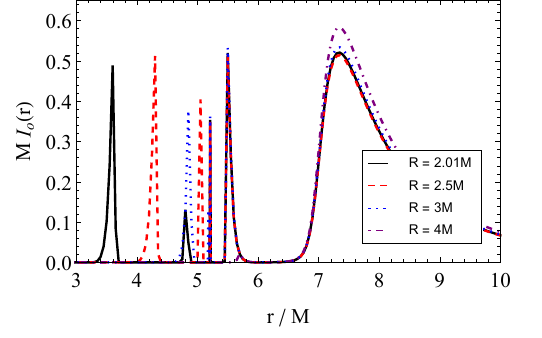}\quad
    \includegraphics[scale=0.62]{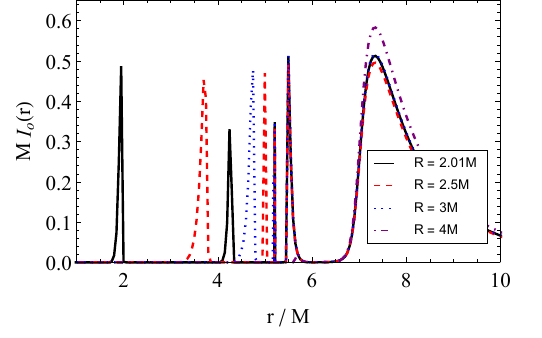}\quad
    \includegraphics[scale=0.62]{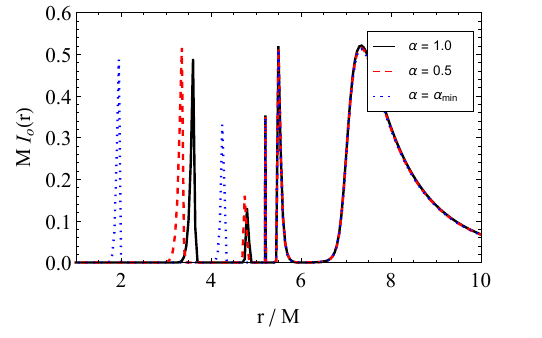}\\
    \includegraphics[scale=0.62]{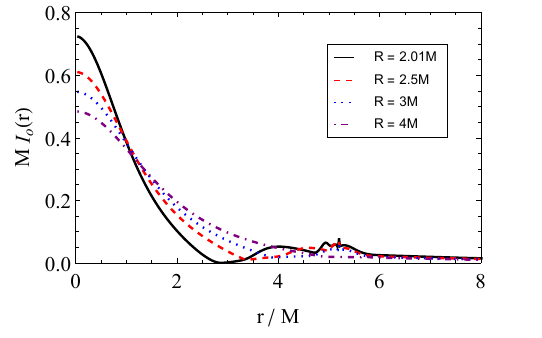}\quad
    \includegraphics[scale=0.62]{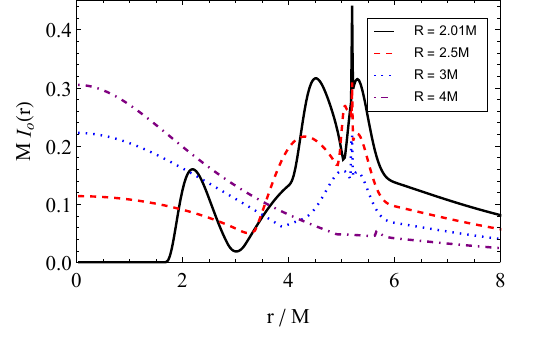}\quad
    \includegraphics[scale=0.62]{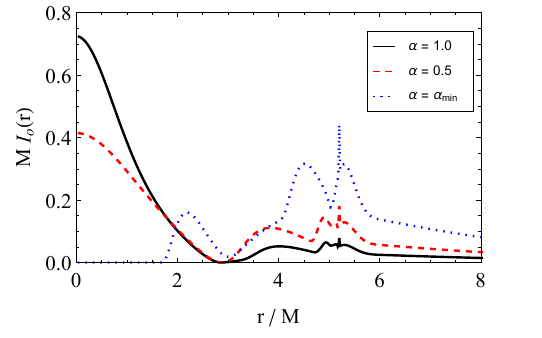}
    \caption{Observed intensity profiles $I_o\left(r\right)$ for the ISCO disk model (top row) and for the Centre disk model (bottom row) for $\alpha=1$ (left column) and $\alpha=\alpha_{\rm min}$ (middle column) with varying $R$, and for $R=2.01M$ (right column) with varying $\alpha$.}
    \label{fig:intobs}
\end{figure*}

\begin{figure*}
    \centering
    \includegraphics[scale=0.37]{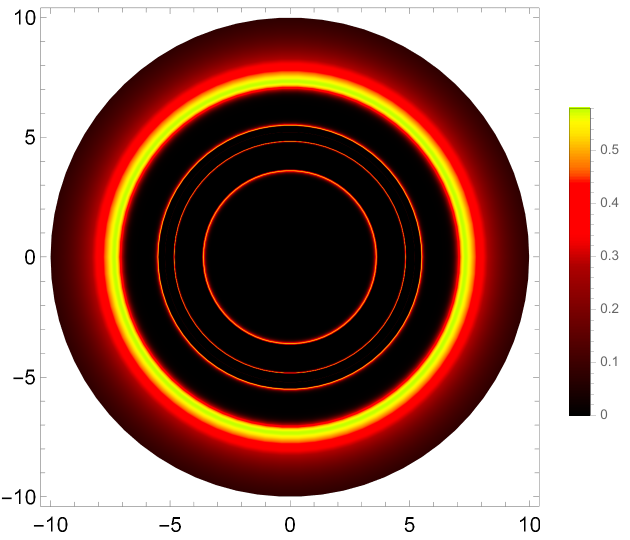}\qquad
    \includegraphics[scale=0.37]{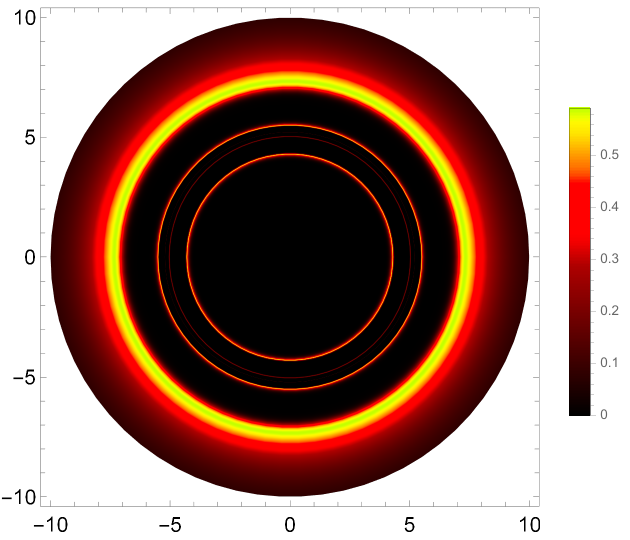}\qquad
    \includegraphics[scale=0.37]{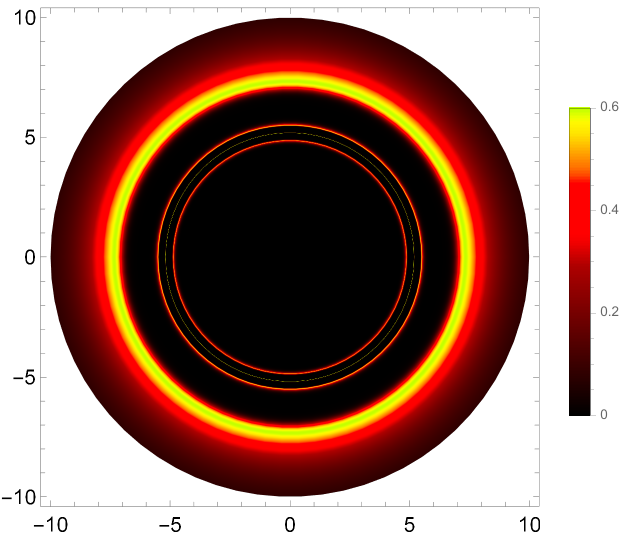}\qquad
    \includegraphics[scale=0.37]{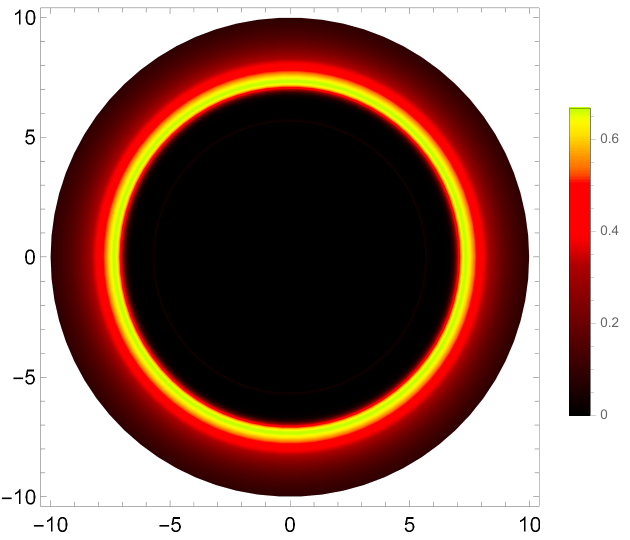}\\
    \includegraphics[scale=0.37]{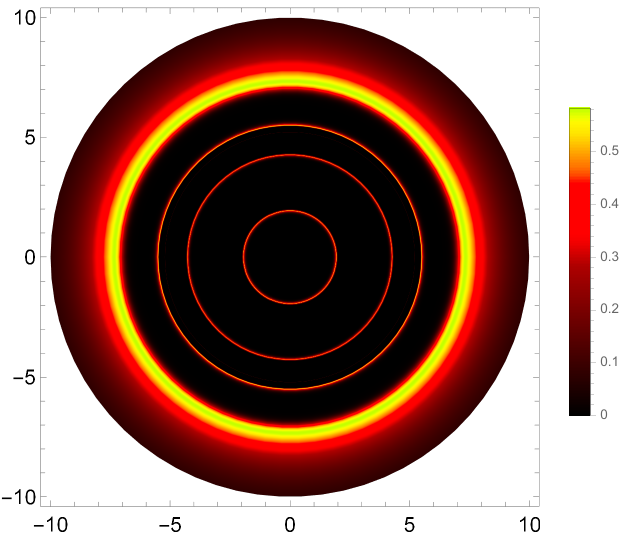}\qquad
    \includegraphics[scale=0.37]{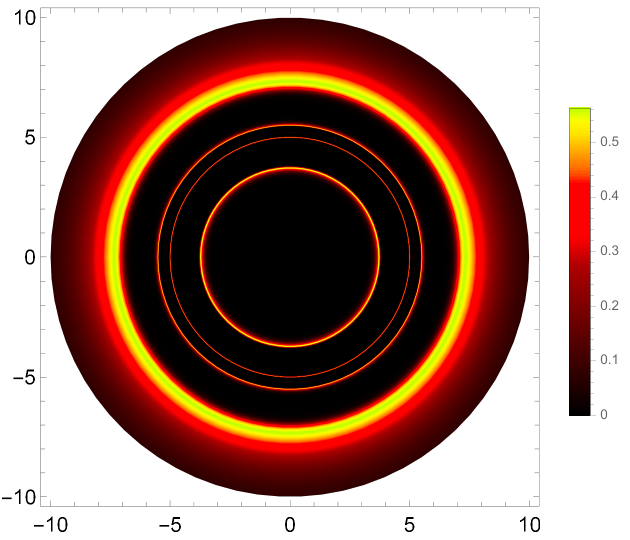}\qquad
    \includegraphics[scale=0.37]{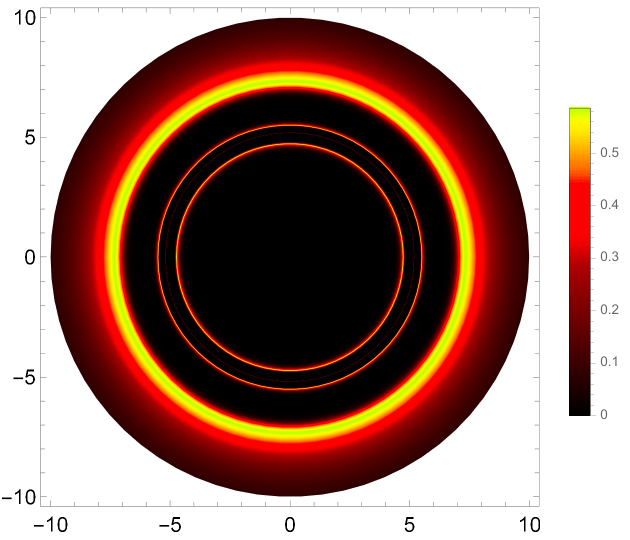}\qquad
    \includegraphics[scale=0.37]{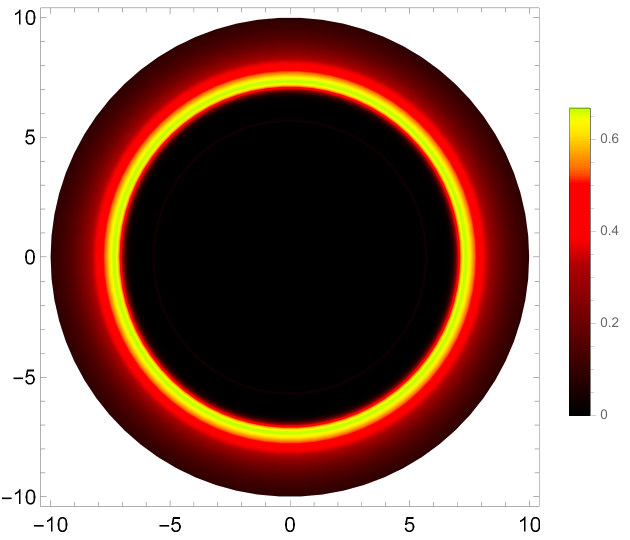}\\
    \includegraphics[scale=0.37]{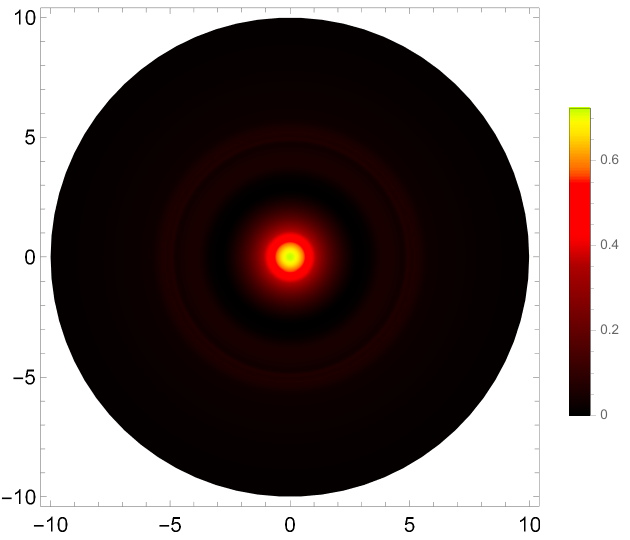}\qquad
    \includegraphics[scale=0.37]{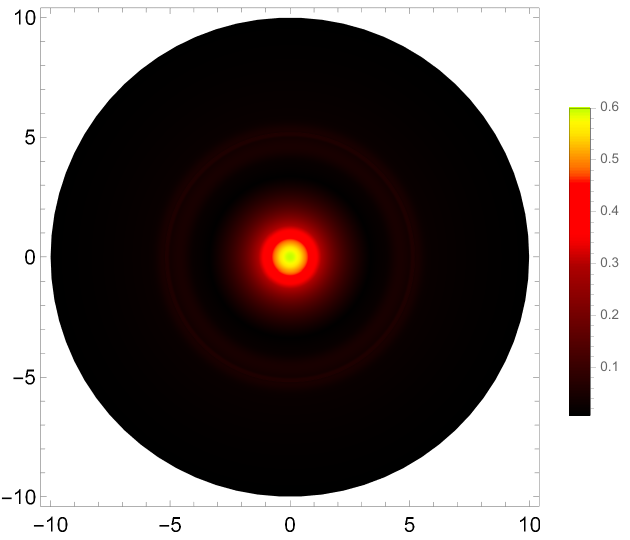}\qquad
    \includegraphics[scale=0.37]{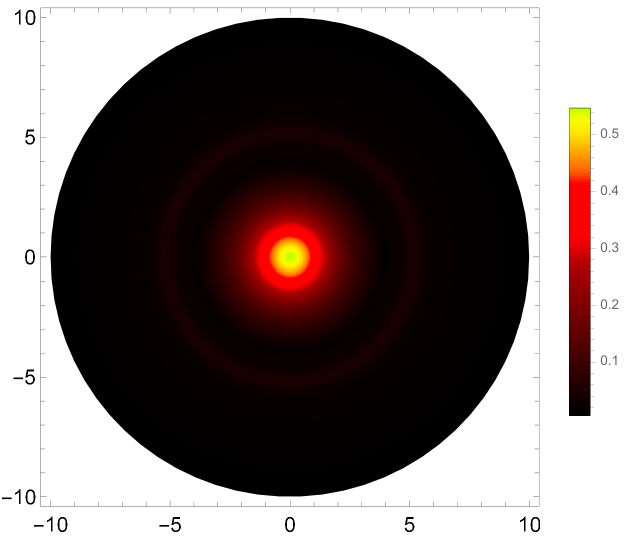}\qquad
    \includegraphics[scale=0.37]{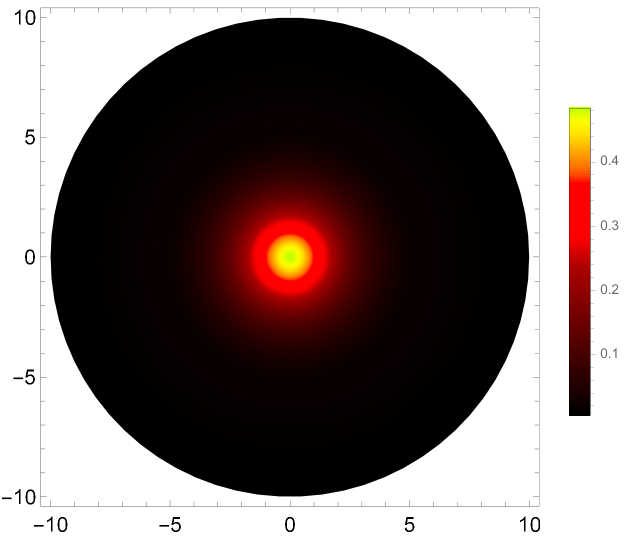}\\
    \includegraphics[scale=0.37]{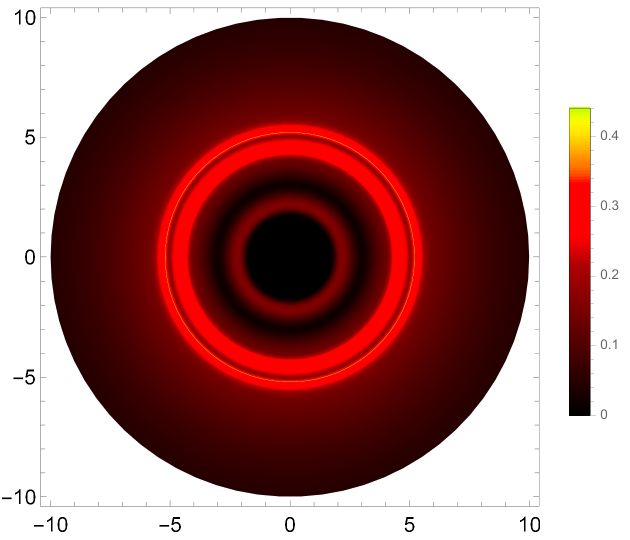}\qquad
    \includegraphics[scale=0.37]{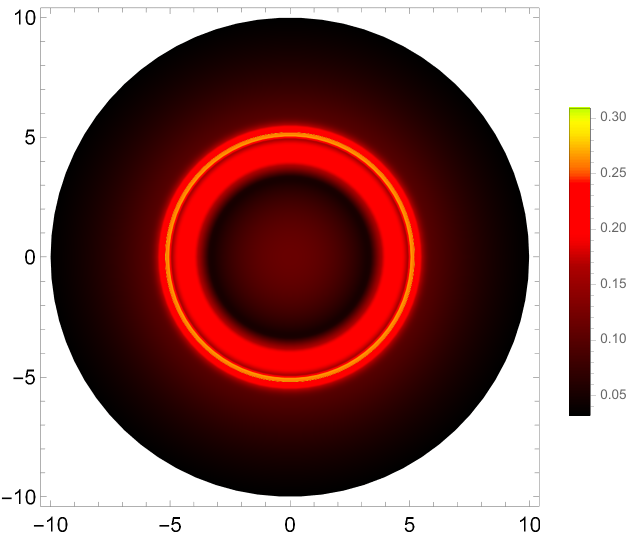}\qquad
    \includegraphics[scale=0.37]{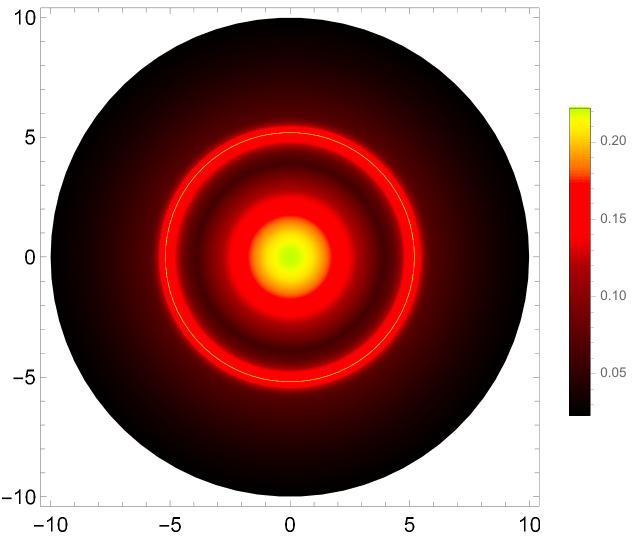}\qquad
    \includegraphics[scale=0.37]{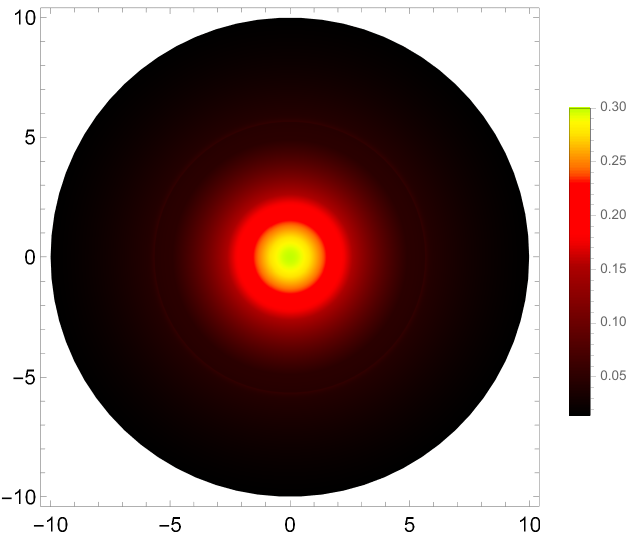}\\
    \caption{Observed axial images with $\theta=0$ for the ISCO accretion disk model (top two rows) and the Center accretion disk model (bottom two rows) for $R=2.01M$ (left column), $R=2.5M$ (middle left), $R=3M$ (middle right) and $R=4M$ (right column), and $\alpha=1$ (first and third rows) and $\alpha=\alpha_{\rm min}$ (second and fourth rows).}
    \label{fig:shadowaxial}
\end{figure*}

The intensity profiles plotted in Fig. \ref{fig:disks} correspond to the intensity profiles on the reference frame of the emitter, i.e., the accretion disk. Let $\nu_e$ be the frequency of the emitted photons in the reference frame of the emitter. Due to the effects of gravitational redshift, the observed frequency in the reference frame of the observer $\nu_o$ is shifted with respect to the emitted one as $\nu_o=\sqrt{-g_{tt}}\nu_e$, which induces a modification in the shape of the observed intensity profile $I_o$ as
\begin{equation}
I_o\left(r\right)=g_{tt}^2\left(r\right)I_e\left(r\right).
\end{equation}
In Fig. \ref{fig:intobs} we plot the observed intensity profiles for different accretion disk models and different combinations of the gravastar radius $R=\{2.01M, 2.5M, 3M, 4M\}$, as well as the values of $\alpha=\{1,\alpha_{\rm min}\}$, where $\alpha_{\rm min}$ corresponds to the value of $\alpha$ for which the entire mass of the gravastar is distributed at the thin-shell, see Eq. \eqref{eq:defalpha}. Furthermore, the observed images for the same combinations of models and parameters are given in Fig.\ref{fig:shadowaxial}.

An analysis of the observed intensity profiles indicates that the observational properties of gravastars for dilute configurations, say $R\gtrsim 4M$, are qualitatively and quantitatively similar, i.e., for models with low compacticity, the radius of the star is not a dominant parameter. Indeed, for those models, only the redshifted primary image, corresponding to a single peak of intensity in the observed profiles, is present. However, as one approaches more compact configurations, one observes the rise of additional peaks of intensity, corresponding to the secondary and light-ring contributions. This feature is visible for both the ISCO and the Center disk models, for which the intensity profiles corresponding to the configurations with $R\lesssim3M$ feature several additional contributions in comparison to the dilute configurations. Furthermore, a decrease in the parameter $\alpha$, i.e., a decrease in the volumetric distribution of mass and increase in the surface density of the thin-shell, is shown to decrease the relative intensity of the primary image and increase the contribution of the secondary peaks, while moving them closer to the center of the observation. Again, this feature is visible for both the ISCO and Center disk models, as seen in the right panels of Fig. \ref{fig:intobs}. In particular, for the Center disk model, one observes that the effects of gravitational redshift that decrease the intensity of the primary peak of intensity are stronger for lower values of $\alpha$.

The observed shadow images (see Fig. \ref{fig:shadowaxial}) confirm the previous analysis of the intensity profiles, i.e., that the gravastar radius is a subdominant parameter for dilute configurations, and that a decrease in $\alpha$ induces an increase in the secondary contributions to the image. An important drawback of the smooth gravastar solution in comparison with other ECO models, e.g., fluid stars \cite{Rosa:2023hfm} and bosonic stars \cite{Rosa:2022tfv,Rosa:2023qcv}, is that a shadow-like feature only arises if one considers that the accretion disk is truncated at a certain finite radius, e.g., for the ISCO disk model. Indeed, if one considers that the accretion disk might feature a central peak of intensity caused by some accumulation of matter via accretion in the center of the gravastar, e.g., for the Center disk model, the effects of the gravitational redshift are not strong enough to sufficiently decrease the observed intensity of the central peak. This is an expected result for the configurations with $\alpha=1$, as the time component of the gravastar metric $g_{tt}$ satisfies the boundary condition $g_{tt}\left(0\right)=-1$, see Fig. \ref{fig:metric}. Nevertheless, for the models with $\alpha=\alpha_{\rm min}$, for which the entire mass of the gravastar is located at the thin-shell, and thus the central value of $g_{tt}$ achieves the largest possible deviation from the previously mentioned boundary condition while maintaining positive energy densities, one observes the rise of a central shadow-like feature for sufficiently compact models, with $r\lesssim 2.5M$. 

\subsection{Inclined observations}

Let us now turn our attention to non-axial observations and, in particular, focus on observations close to the equatorial plane with an observation angle of $\theta=80^\circ$ with respect to the axial axis. Due to the increase in computational times to produce non-axial observations, instead of producing such observations for the entire plethora of configurations explored in the previous section, we use those results to filter those models who are more relevant from an observational point of view.

\begin{figure*}
    \centering
    \includegraphics[scale=0.37]{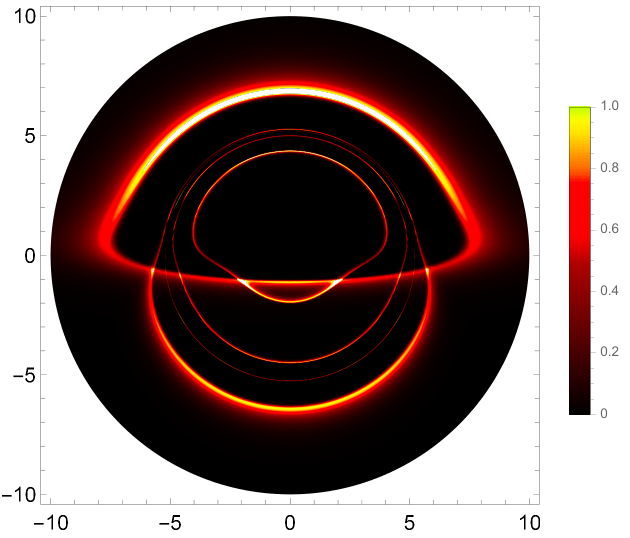}\qquad
    \includegraphics[scale=0.37]{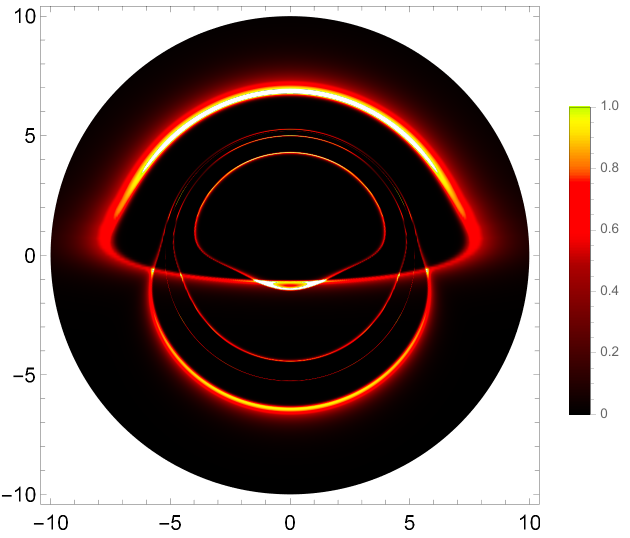}\qquad
    \includegraphics[scale=0.37]{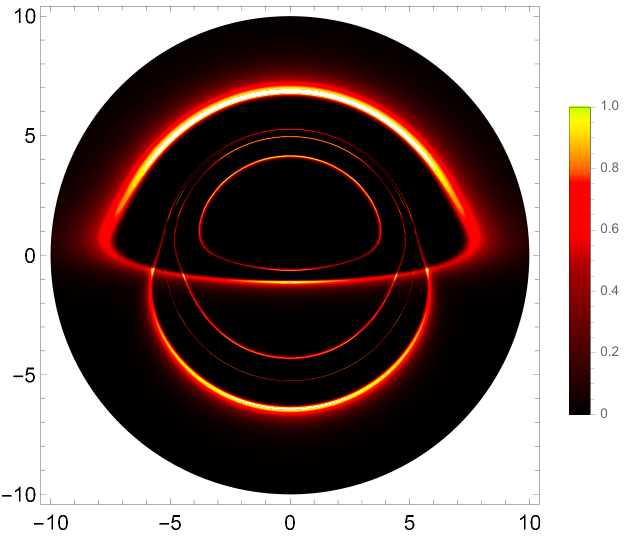}\qquad
    \includegraphics[scale=0.37]{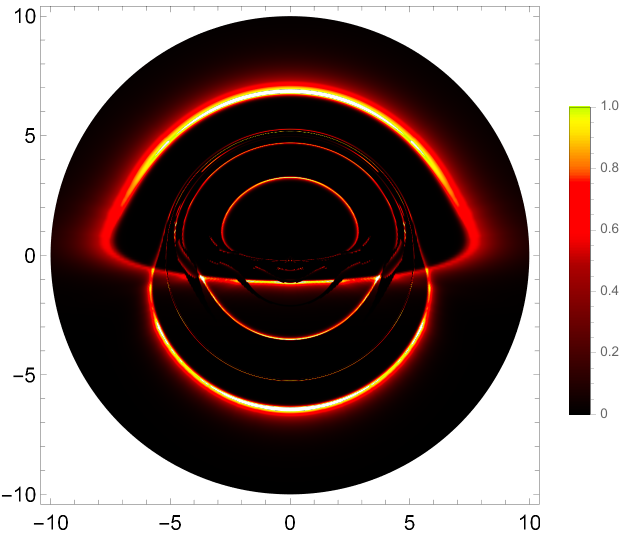}\\
    \includegraphics[scale=0.37]{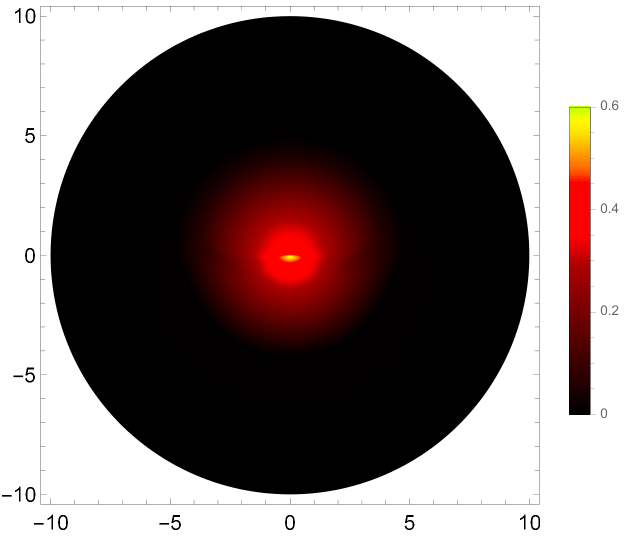}\qquad
    \includegraphics[scale=0.37]{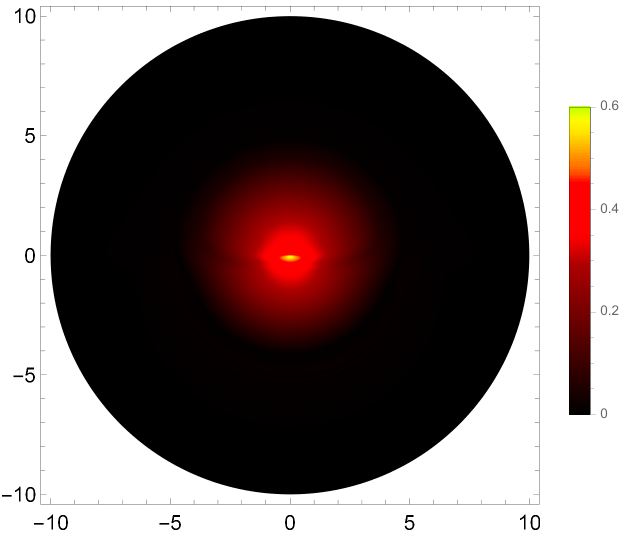}\qquad
    \includegraphics[scale=0.37]{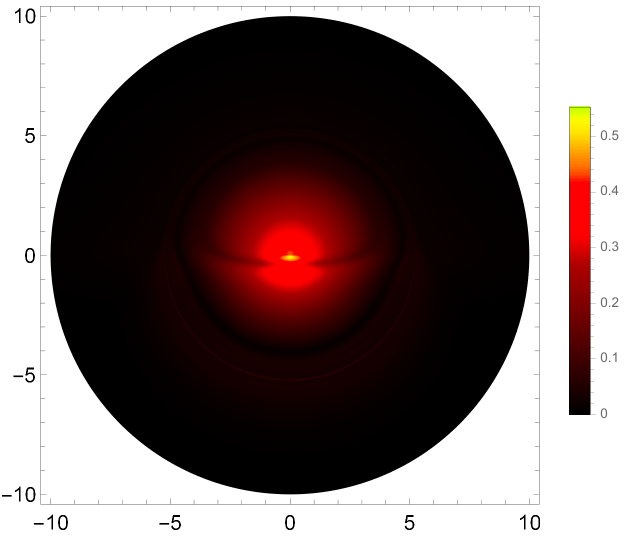}\qquad
    \includegraphics[scale=0.37]{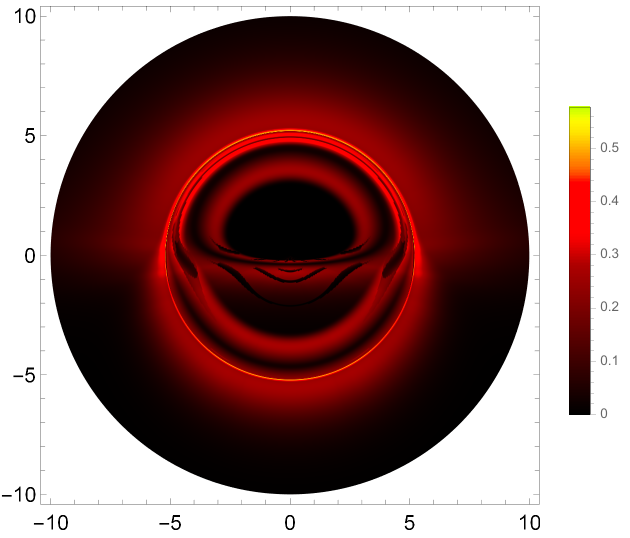}
    \caption{Observed inclined images with $\theta=80^\circ$ for the ISCO accretion disk model (top row) and the Center accretion disk model (bottom row) for $R=2.01M$ and with $\alpha=1$ (left column), $\alpha=\frac{2}{3}$ (middle left), $\alpha=\frac{1}{3}$ (middle right) and $\alpha=\alpha_{\rm min}$ (right column).}
    \label{fig:shadowinclined}
\end{figure*}

The axial images produced in the previous section indicate that, in order for the gravastar model to reproduce observational properties consistent with the appearance of a shadow-like feature, two requirements are necessary: the gravastar must have a large compacticity, a feature achievable through the selection or a gravastar radius close to the Schwarzschild radius $R\gtrsim 2M$; and a large portion of the gravastar mass must be allocated at the thin-shell, a result attainable by reducing the value of $\alpha\gtrsim \alpha_{\rm min}$. We thus consider in this section the gravastar model with $R=2.01M$ introduced in the previous section, with different values of $\alpha=\{\alpha_{\rm min}, \frac{1}{3}, \frac{2}{3},1\}$, where $\alpha_{\rm min}\simeq 0.00498$. The observed images by inclined observers for these configurations are given in Fig. \ref{fig:shadowinclined}.

\begin{figure*}
    \centering
    \includegraphics[scale=0.37]{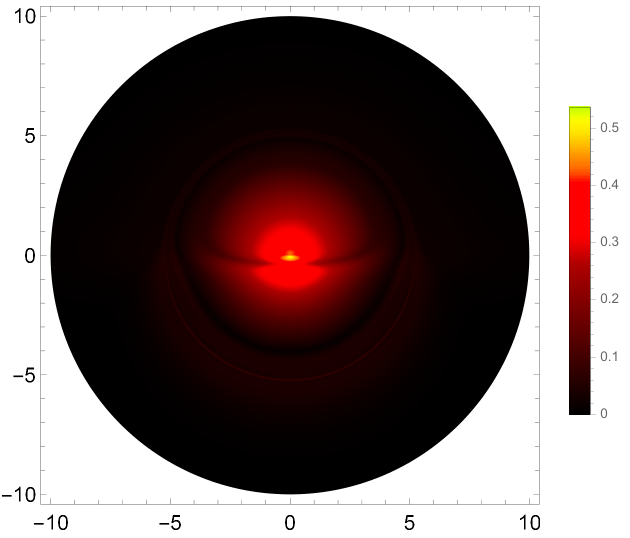}\qquad
    \includegraphics[scale=0.37]{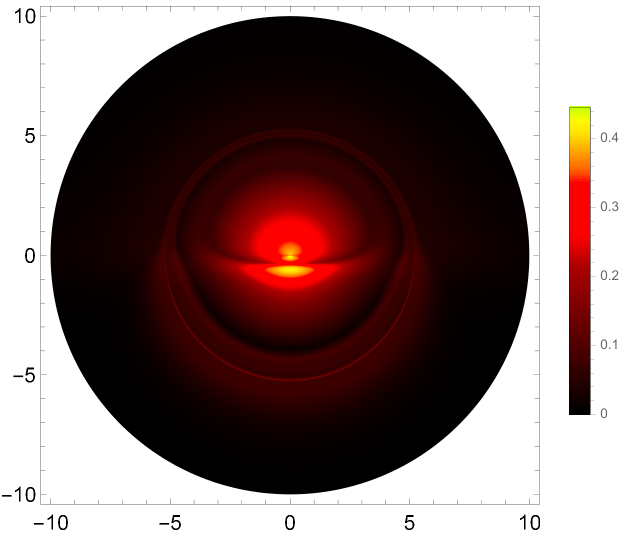}\qquad
    \includegraphics[scale=0.37]{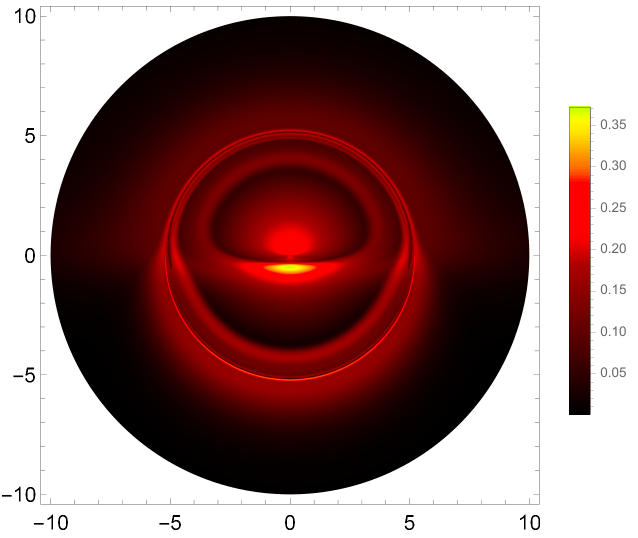}\qquad
    \includegraphics[scale=0.37]{R2_a0_80deg_C.pdf}\\
    \caption{Observed inclined images with $\theta=80^\circ$ for the Center accretion disk model for $R=2.01M$ and with $\alpha=0.3$ (left column), $\alpha=0.2$ (middle left), $\alpha=0.1$ (middle right) and $\alpha=\alpha_{\rm min}$ (right column).}
    \label{fig:shadow}
\end{figure*}

Our results indicate that for values of $\alpha$ far from the minimum value $\alpha_{\rm min}$, variations in this parameter do not alter significantly the observational properties of the gravastar configurations. However, as one approaches $\alpha_{\rm min}$, the effects of the gravitational redshift induce significant changes in the observed image, increasing the number of visible secondary images and decreasing the size of the innermost one, as can be seen particularly clearly for the ISCO disk model, and also producing a shadow-like feature in the case of the Center disk model. The quick transition that gives rise to the central shadow-like feature is emphasized in Fig. \ref{fig:shadow}, where the observations for the values of $\alpha=\left\{0.3, 0.2, 0.1, \alpha_{\rm min}\right\}$ are given. These results seem to indicate that not only smooth gravastar models are inadequate to reproduce the astrophysical observations of shadows and that a thin-shell is needed to fulfill that purpose, but also that a large portion of the mass of the gravastar must be allocated at the thin-shell, or equivalently, the volumetric mass of the gravastar must be much smaller than the surface mass.

%%%%%%%%%%%%%%%%%%%%%%%%%%%%%%%%%%%%%%%%%%%%%%%%%%%%%%%%%%%%%%%%%%%%%%%%%%%%
\section{Orbits and hot spots}\label{sec:hotspot}
%%%%%%%%%%%%%%%%%%%%%%%%%%%%%%%%%%%%%%%%%%%%%%%%%%%%%%%%%%%%%%%%%%%%%%%%%%%%

Consider now the observational properties of hot spots orbiting a central gravastar. To perform simulations of this astrophysical system, we recur to the open source ray-tracing software GYOTO \cite{Vincent:2011wz,Grould:2016emo,Vincent:2020dij,Lamy:2018zvj,Vincent:2016sjq,Vincent:2012kn,Rosa:2023qcv,Rosa:2022toh,Tamm:2023wvn}, where the hot-spot is modelled as a spherical isotropically emitting source orbiting the central gravastar along a circular geodesic of a given orbital radius $r_o$ in the equatorial plane $\theta=\pi/2$. The radius of the hot spot is taken to be $r_H=M/2$. The observer is taken to be at a distance of $r=1000M$ from the gravastar and stands above the equatorial plane in different observation angles, namely $\theta=\{20^\circ,80^\circ\}$. Upon performing the ray-tracing, the software outputs a two-dimensional matrix of specific intensities $I^\nu_{lm}$ for each time instant $t_k\in\left[0,T\right[$, where $T$ is the orbital period of the hot spot. One thus obtains a cube of data $I_{klm}=\Delta\nu I^\nu_{lm}$, where $\Delta\nu$ is the spectral width. This cube of data can then be used to produce three observable quantities, namely the time integrated fluxes $\left<I\right>_{lm}$, the temporal fluxes $F_k$, and the temporal centroids $\vec{c}_k$, which are given explicitly by
\begin{equation}\label{eq:intflux}
\left<I\right>_{lm}=\sum_k I_{klm},
\end{equation}
\begin{equation}\label{eq:timeflux}
F_k=\sum_{l,m}\Delta\Omega I_{klm},
\end{equation}
\begin{equation}\label{eq:timecent}
\vec{c}_k=\sum_{l,m}\Delta\Omega I_{klm} \vec{r}_{lm},
\end{equation}
where $\Delta\Omega$ denotes the solid angle of a single pixel and the vector $\vec{r}_{lm}$ denotes the displacement of the pixel $\{l,m\}$ with respect to the center of the observed image. From the temporal fluxes $F_k$, one can define the more commonly used temporal magnitude $m_k$ as
\begin{equation}\label{eq:magnitude}
m_k=-2.5 \log\left(\frac{F_k}{\min F_k}\right).
\end{equation}

\begin{figure*}
    \centering
    \includegraphics[scale=0.47]{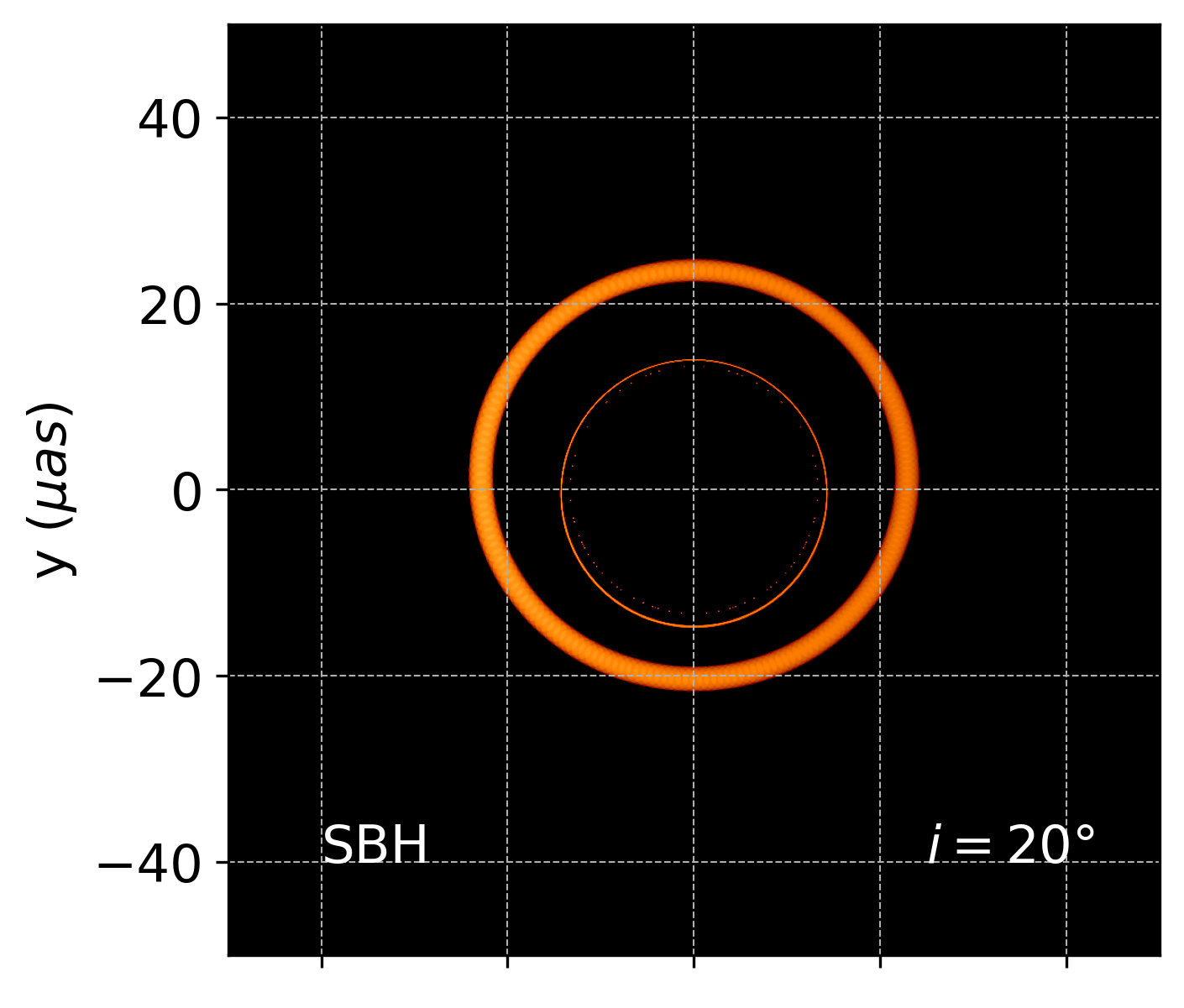} \hspace{-0.2cm}
    \includegraphics[scale=0.47]{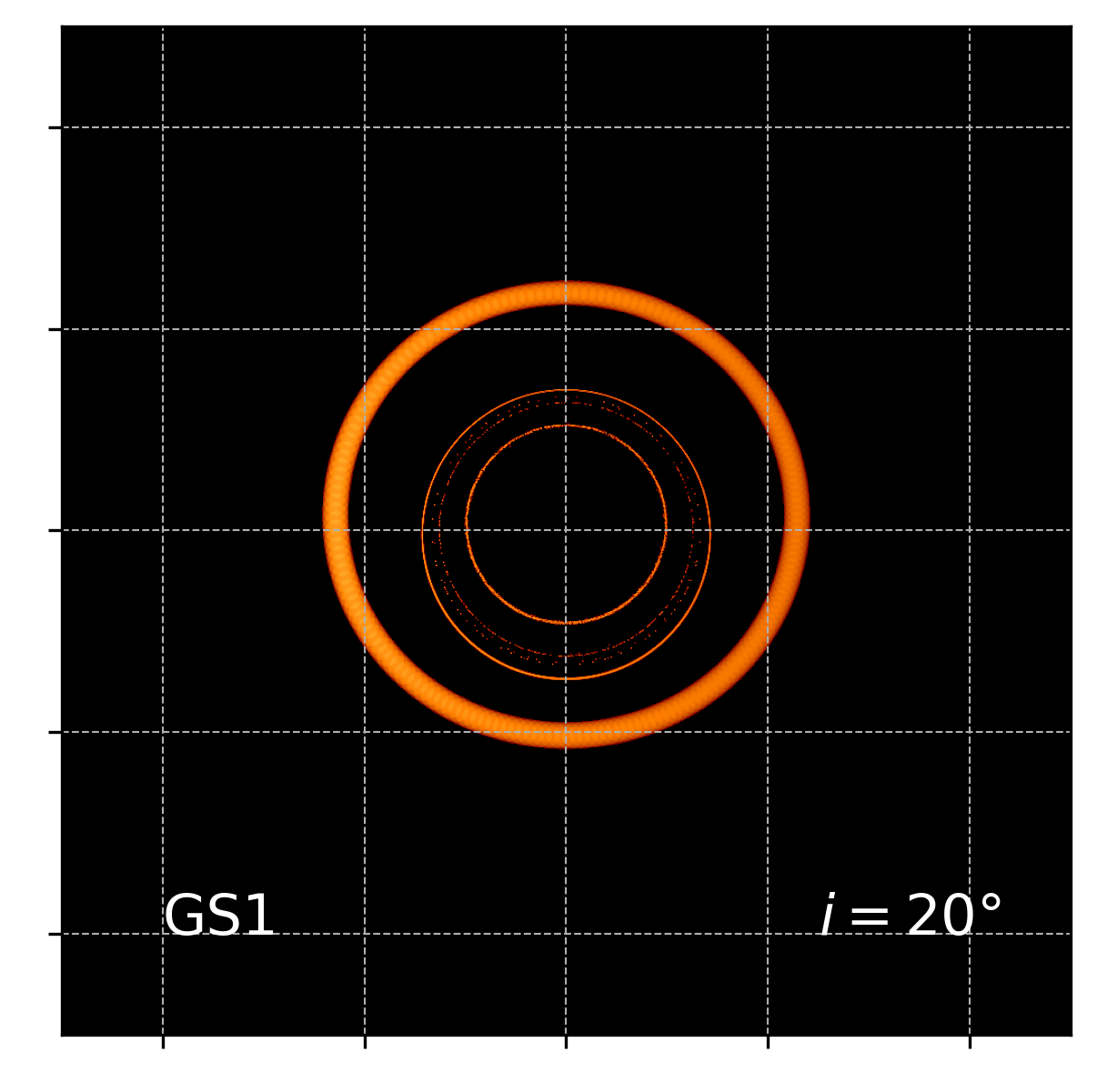} \hspace{-0.2cm}
    \includegraphics[scale=0.47]{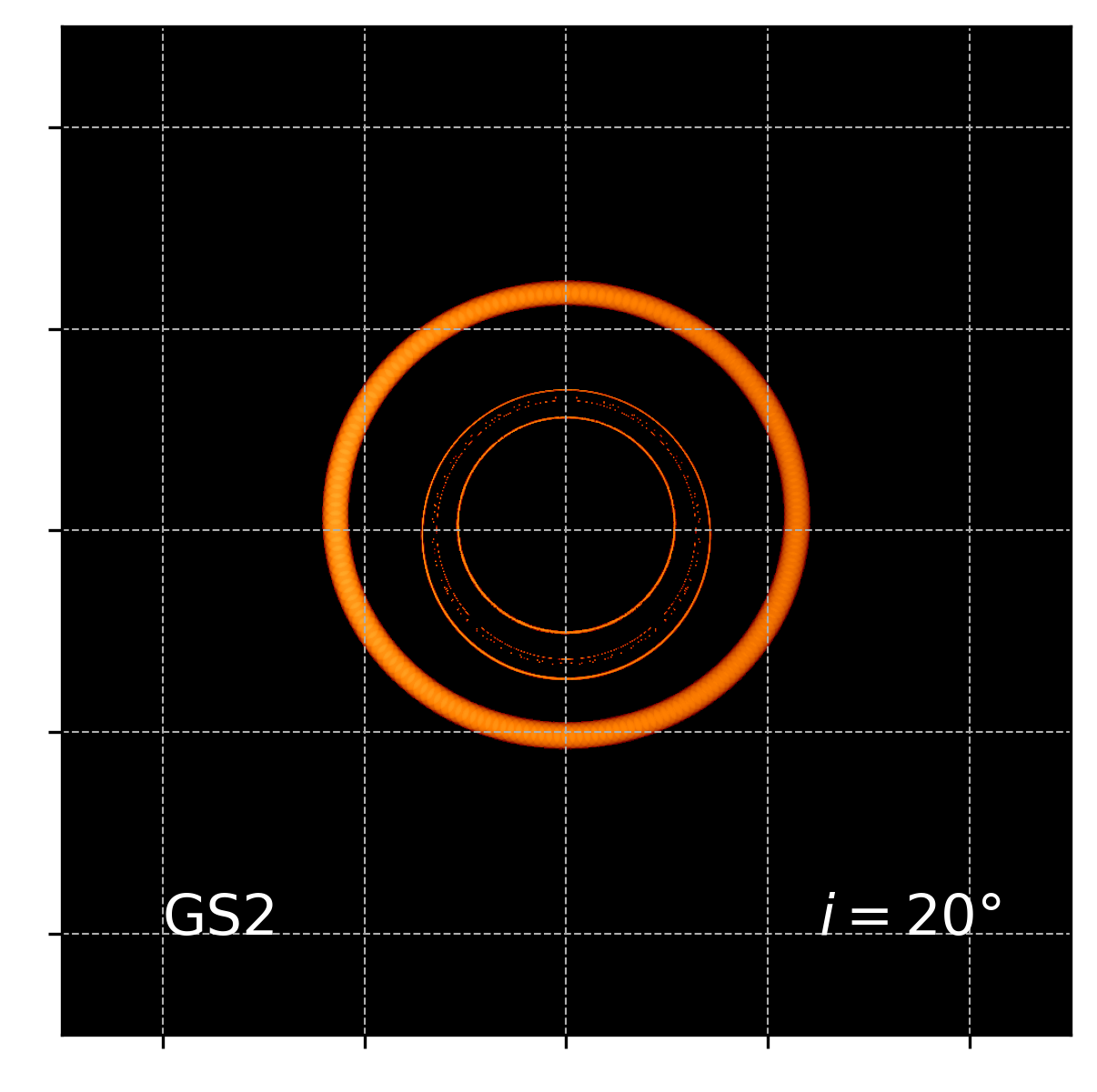}\\
    \includegraphics[scale=0.47]{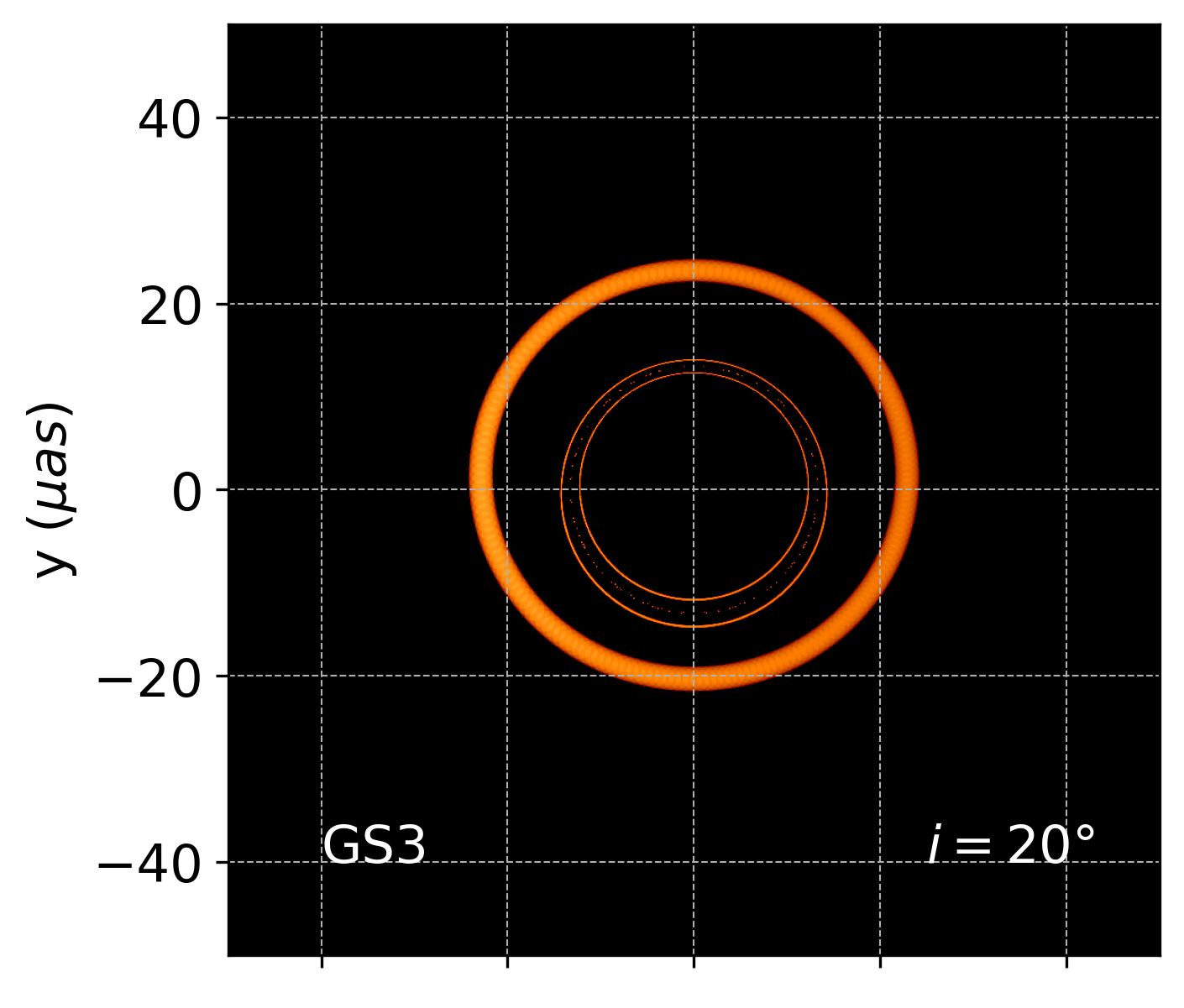} \hspace{-0.2cm}
    \includegraphics[scale=0.47]{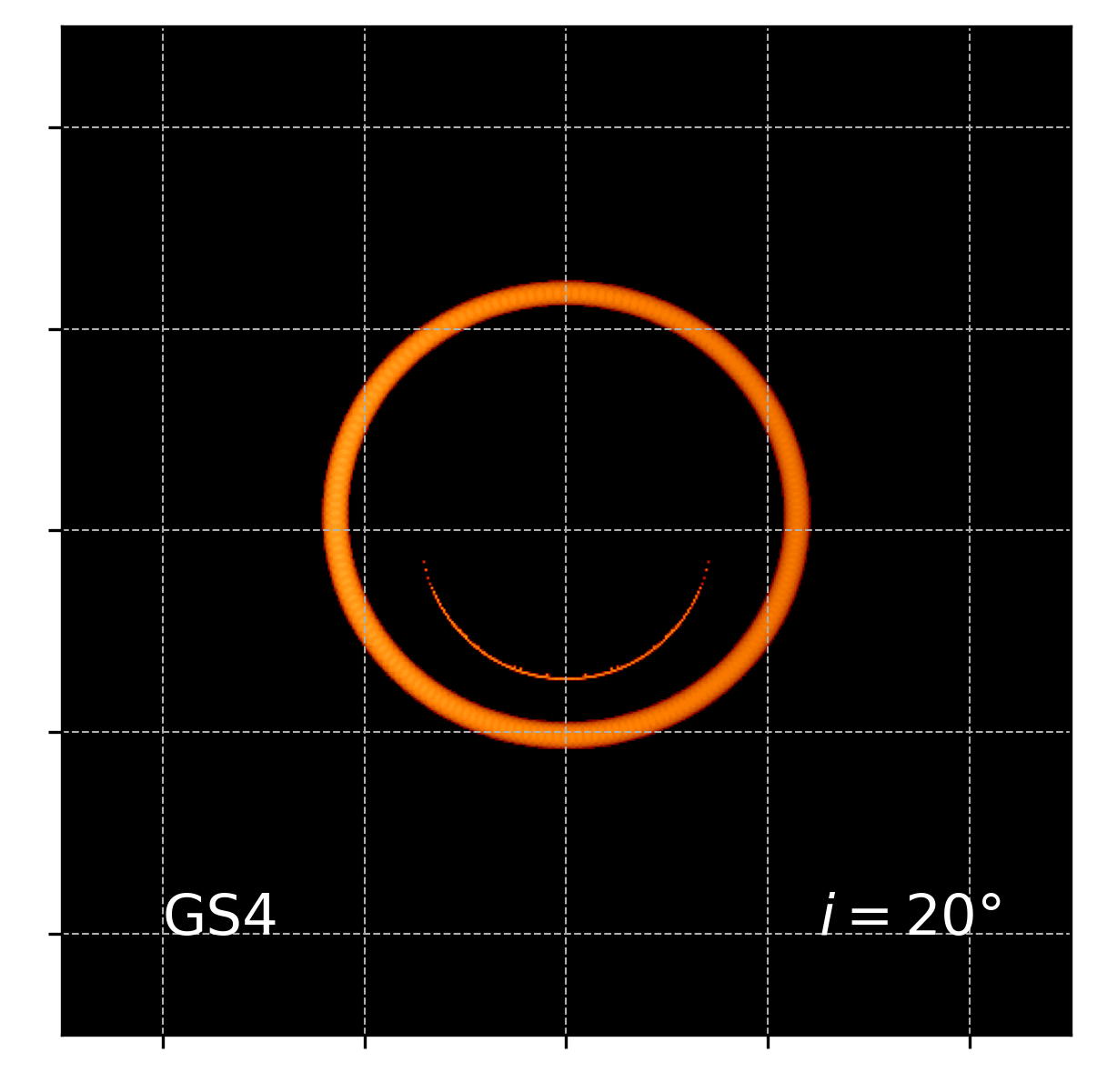} \hspace{-0.2cm}
    \includegraphics[scale=0.47]{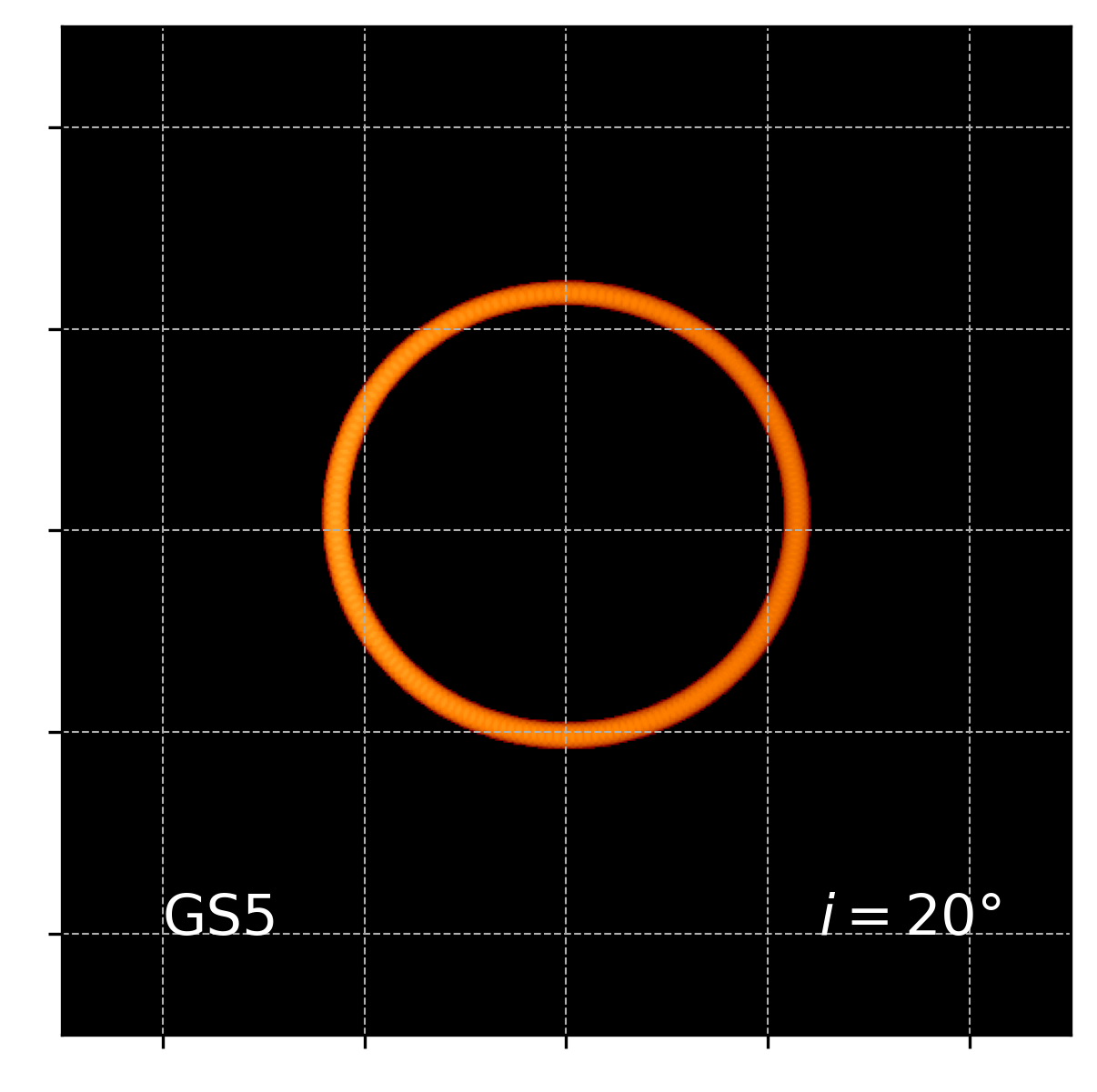}\\
    \includegraphics[scale=0.47]{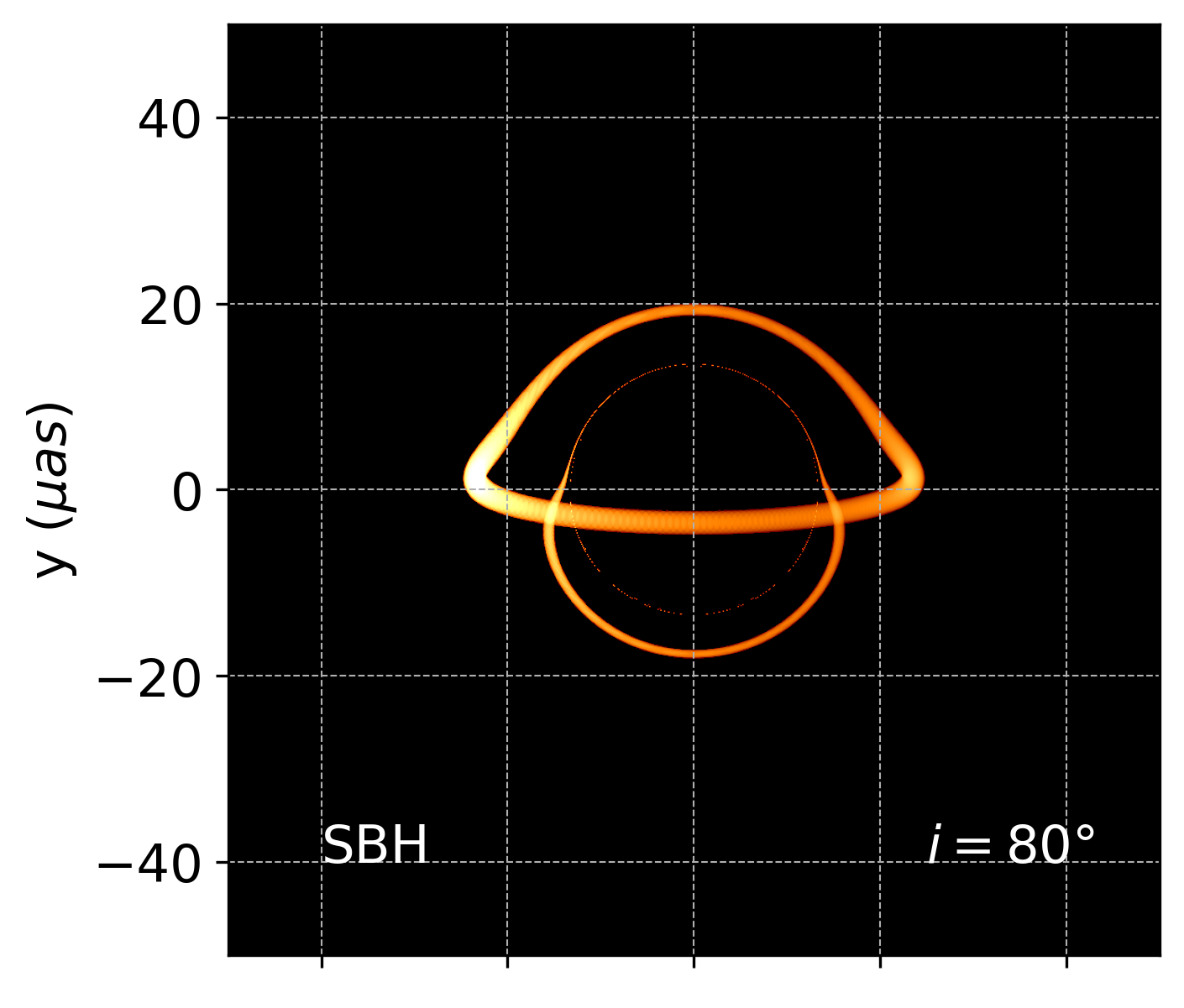} \hspace{-0.2cm}
    \includegraphics[scale=0.47]{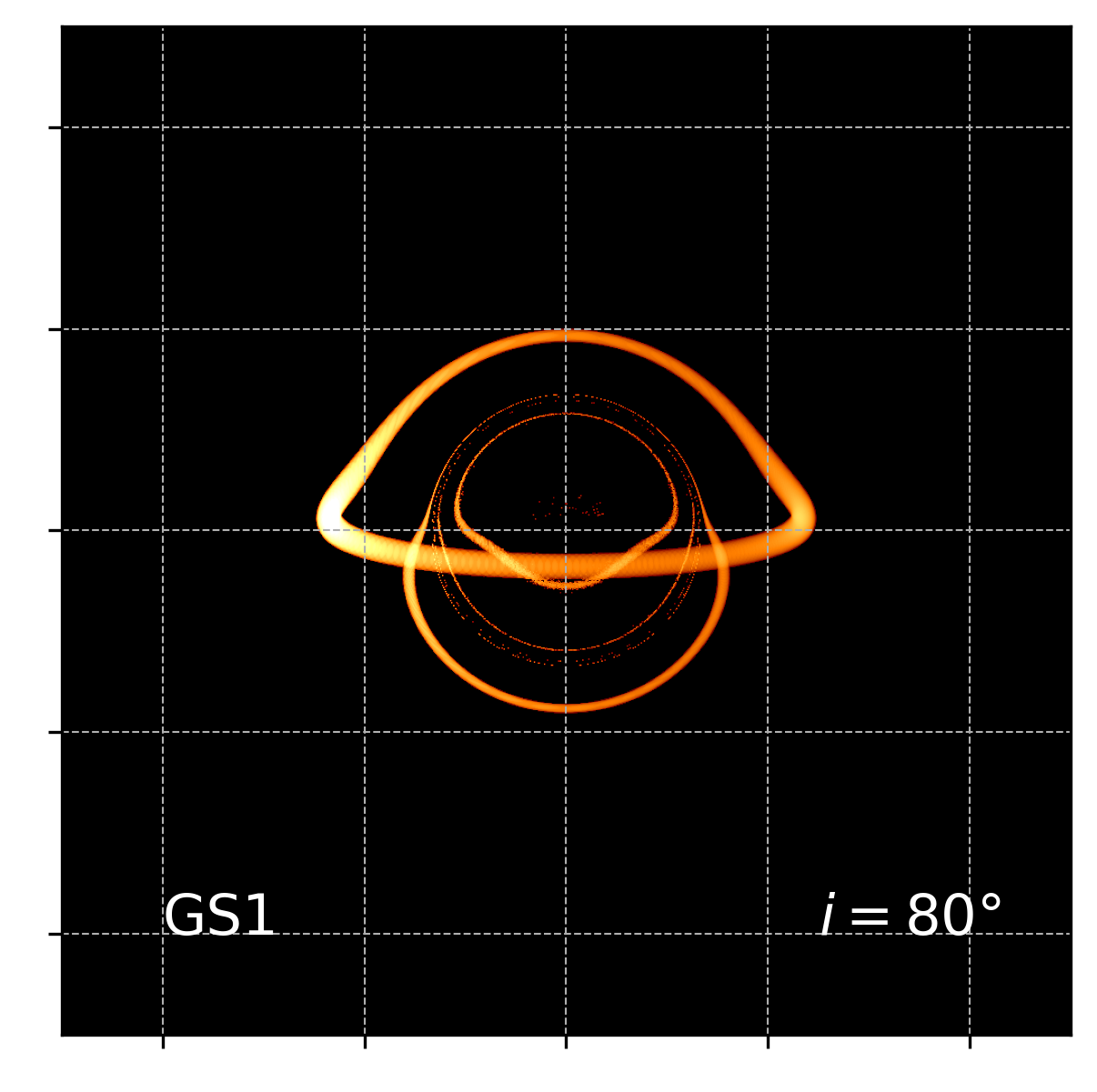} \hspace{-0.2cm}
    \includegraphics[scale=0.47]{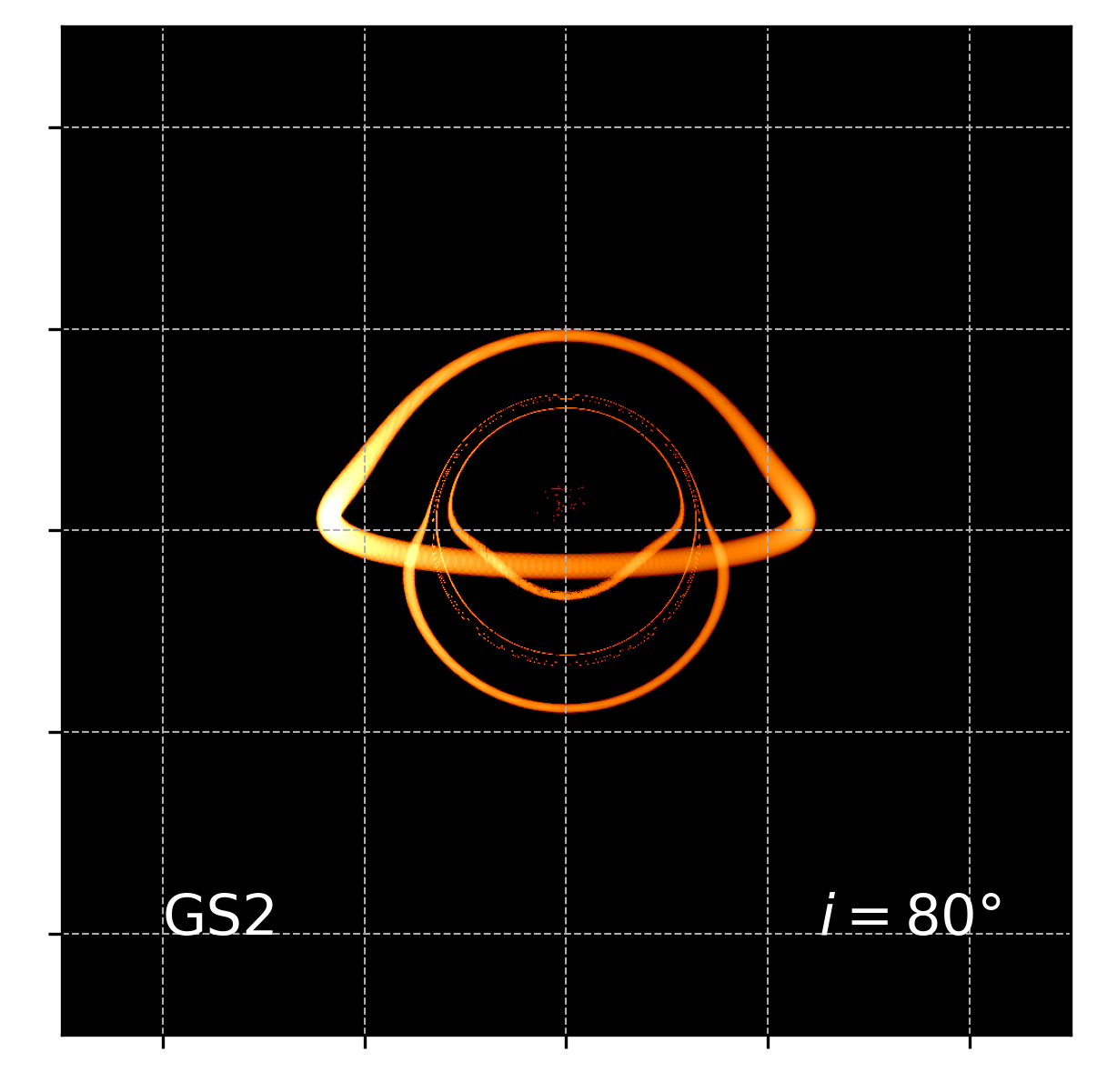}\\
    \includegraphics[scale=0.47]{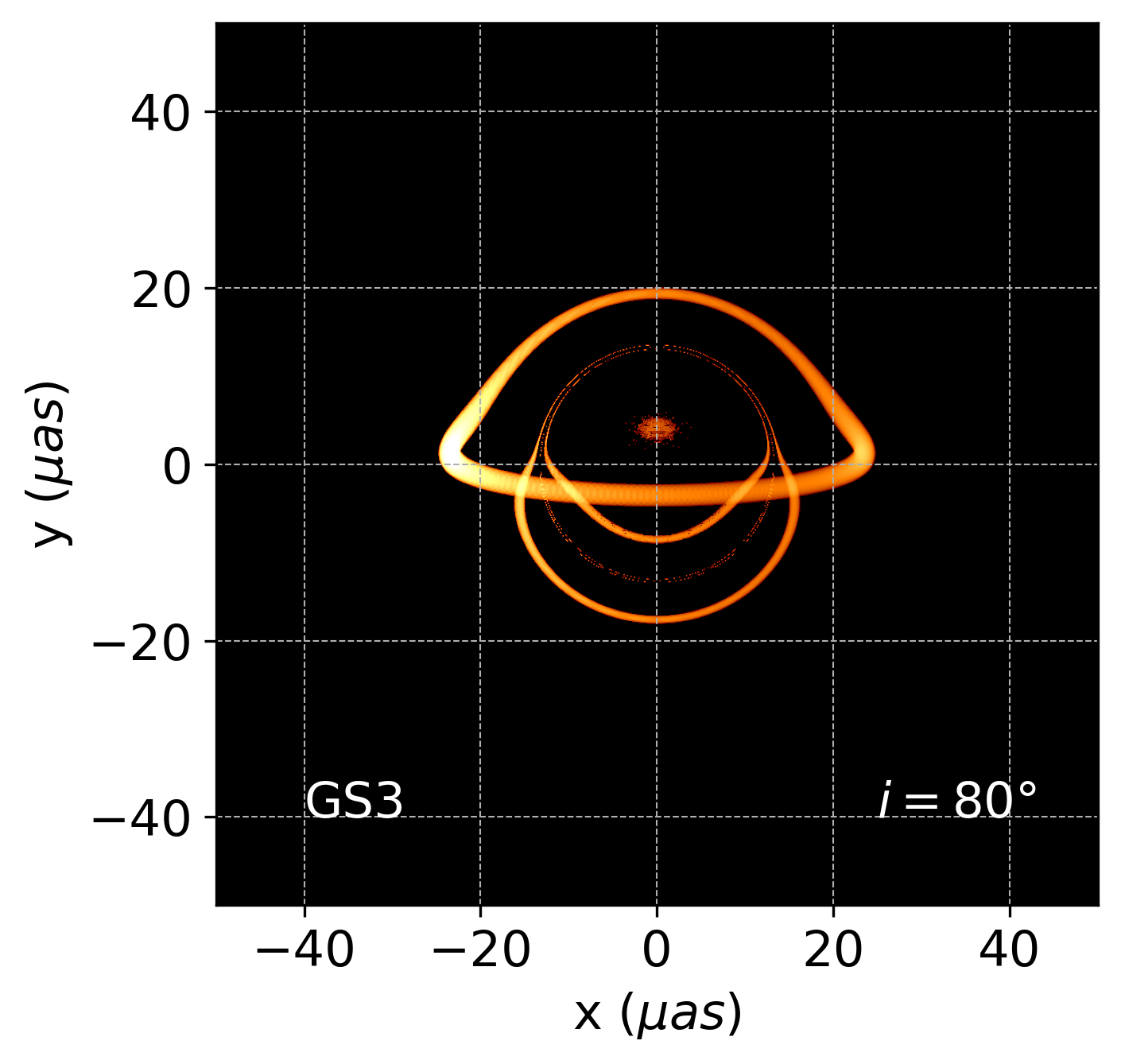} \hspace{-0.2cm}
    \includegraphics[scale=0.47]{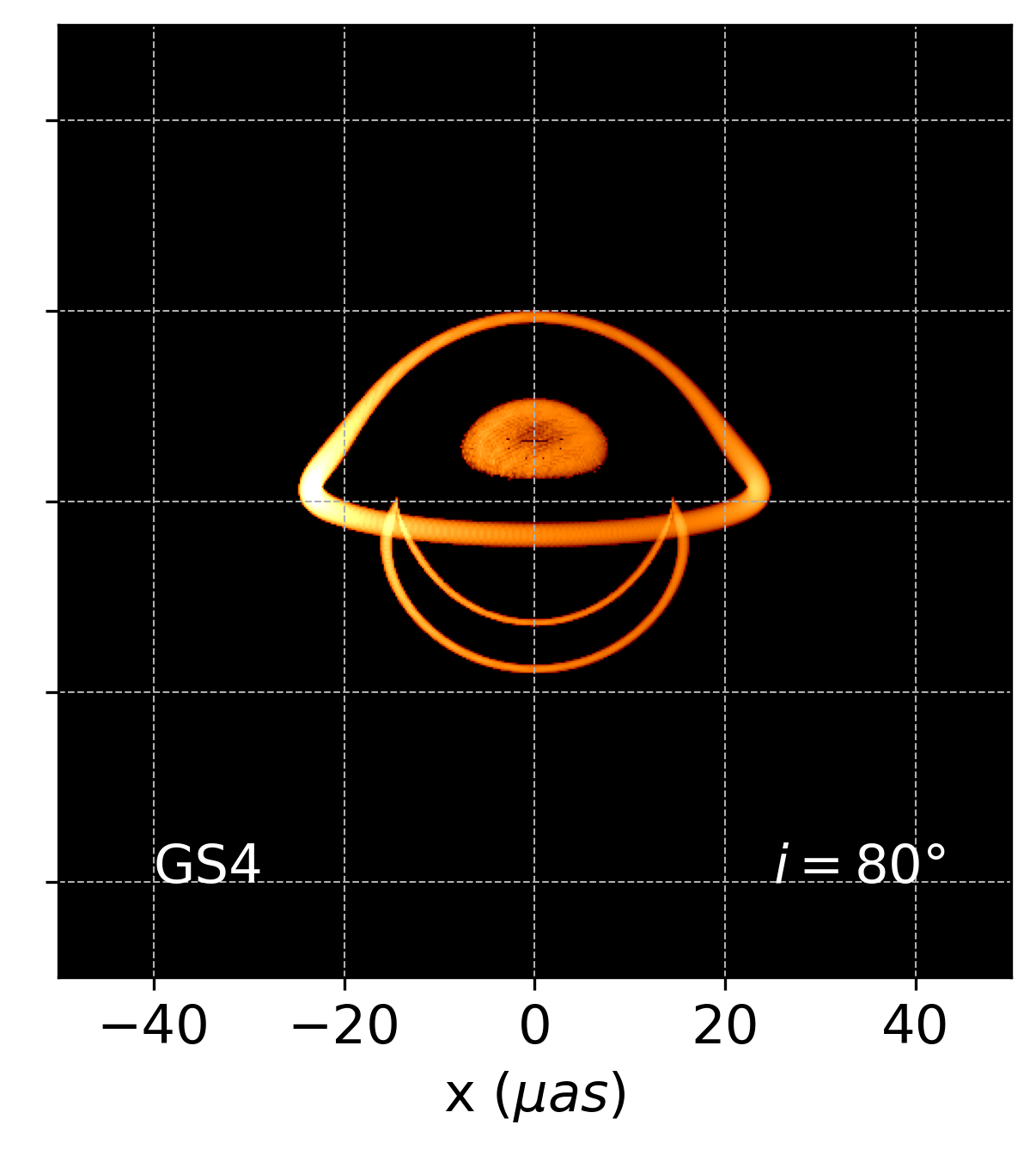} \hspace{-0.2cm}
    \includegraphics[scale=0.47]{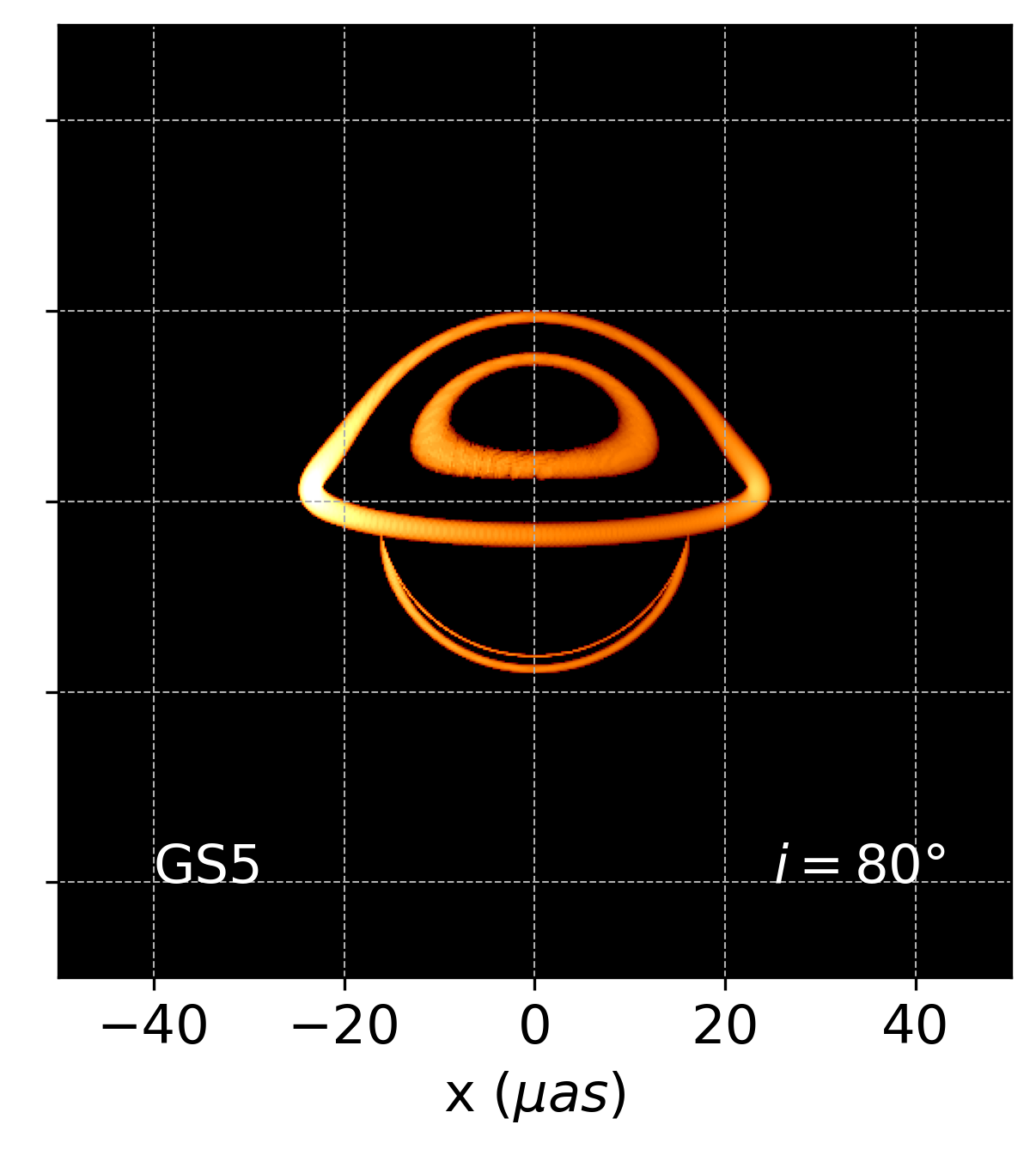} \\
\caption{Time-integrated fluxes of the orbital motion of hot-spots for the Schwarzschild BH and the models GS1 to GS5 for an observation inclination of $\theta=20^\circ$ (top two rows) and $\theta=80^\circ$ (bottom two rows), with $\alpha=1$ and $r_o=8M$. The ultra-compact configurations GS1 to GS3 present qualitative differences in the integrated fluxes in comparison with the non-ultracompact configurations GS4 and GS5.}
\label{fig:fluxincline}
\end{figure*}

\begin{figure*}
    \centering
    \includegraphics[scale=0.47]{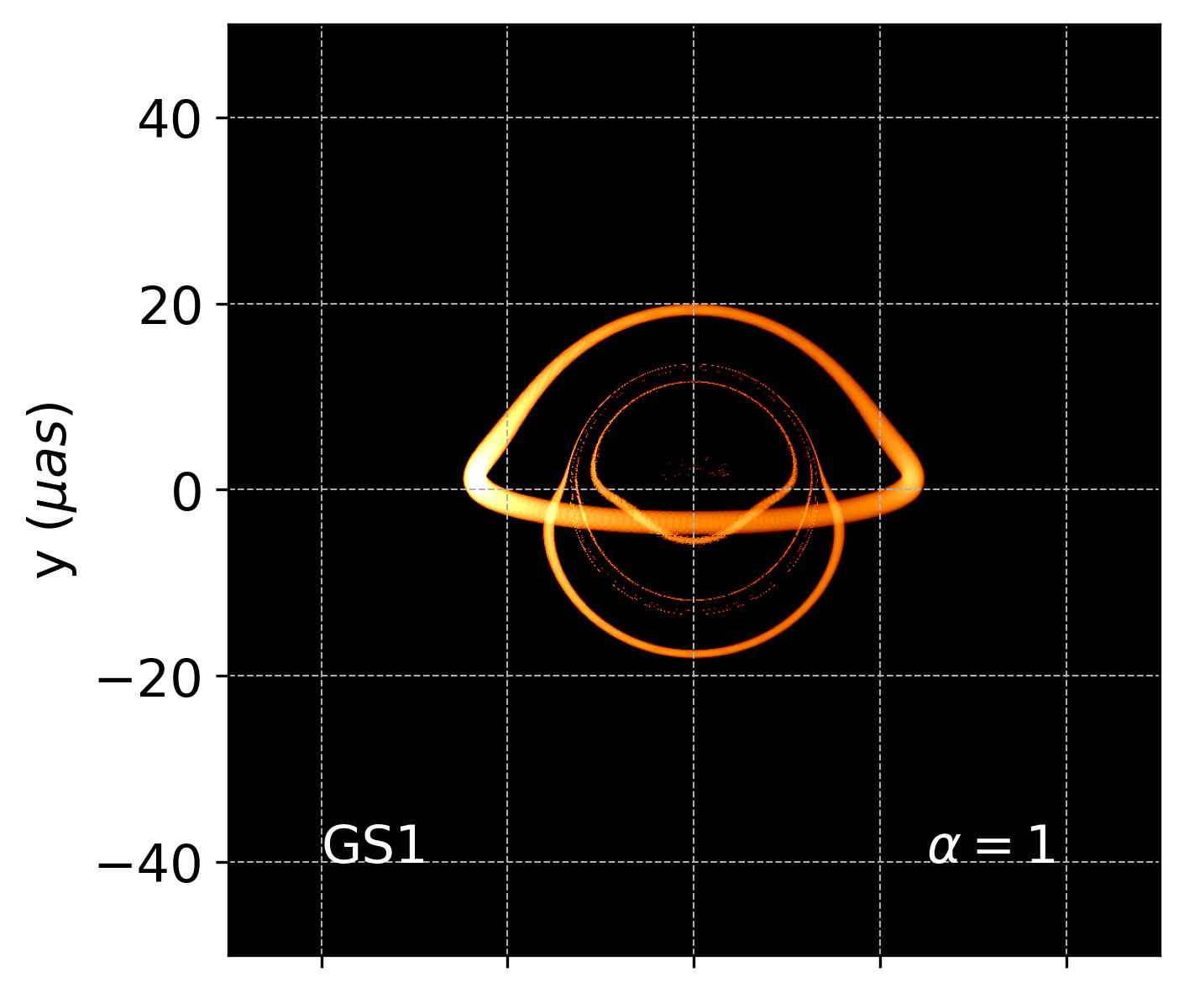} \hspace{-0.2cm}
    \includegraphics[scale=0.47]{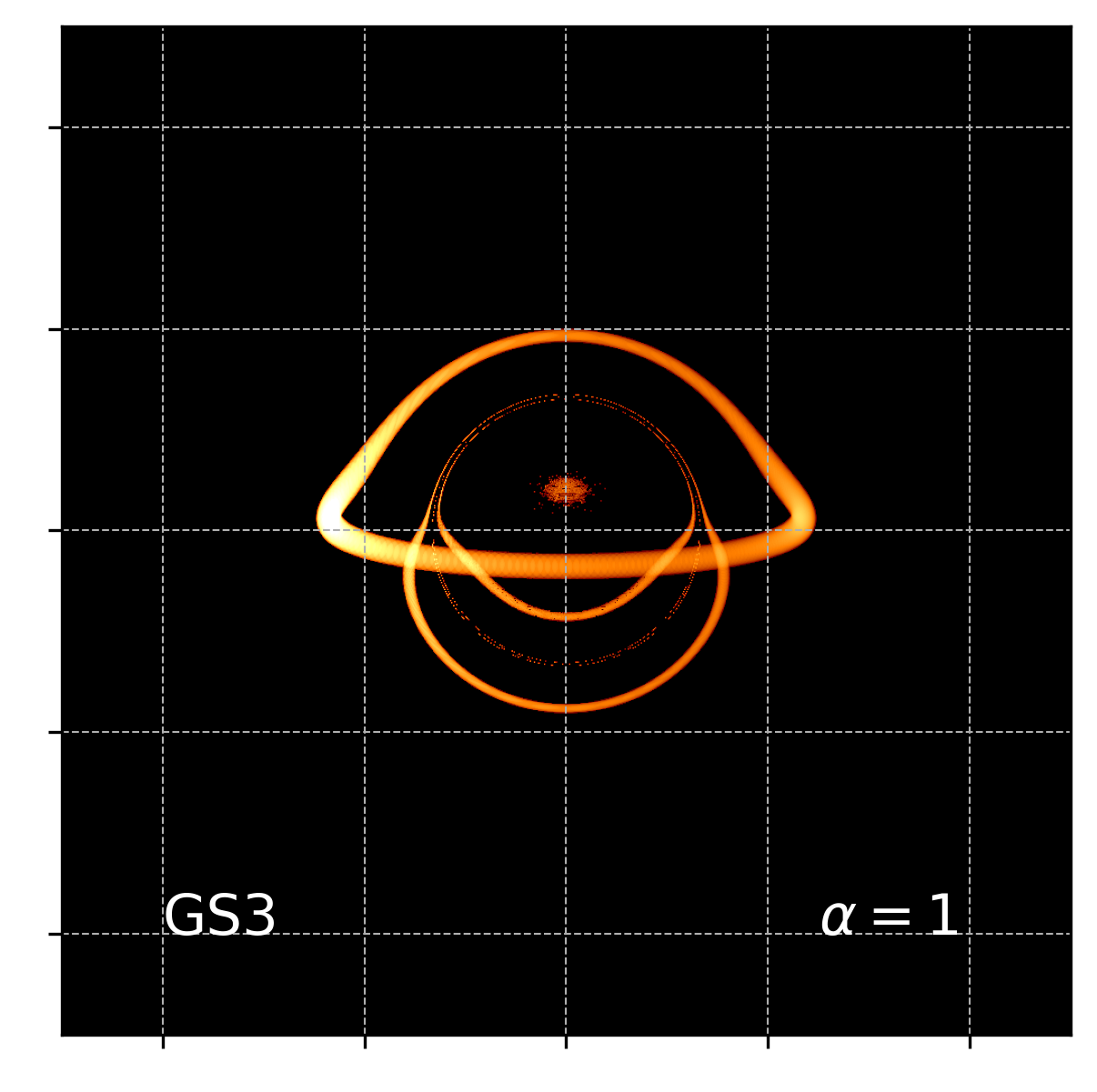} \hspace{-0.2cm}
    \includegraphics[scale=0.47]{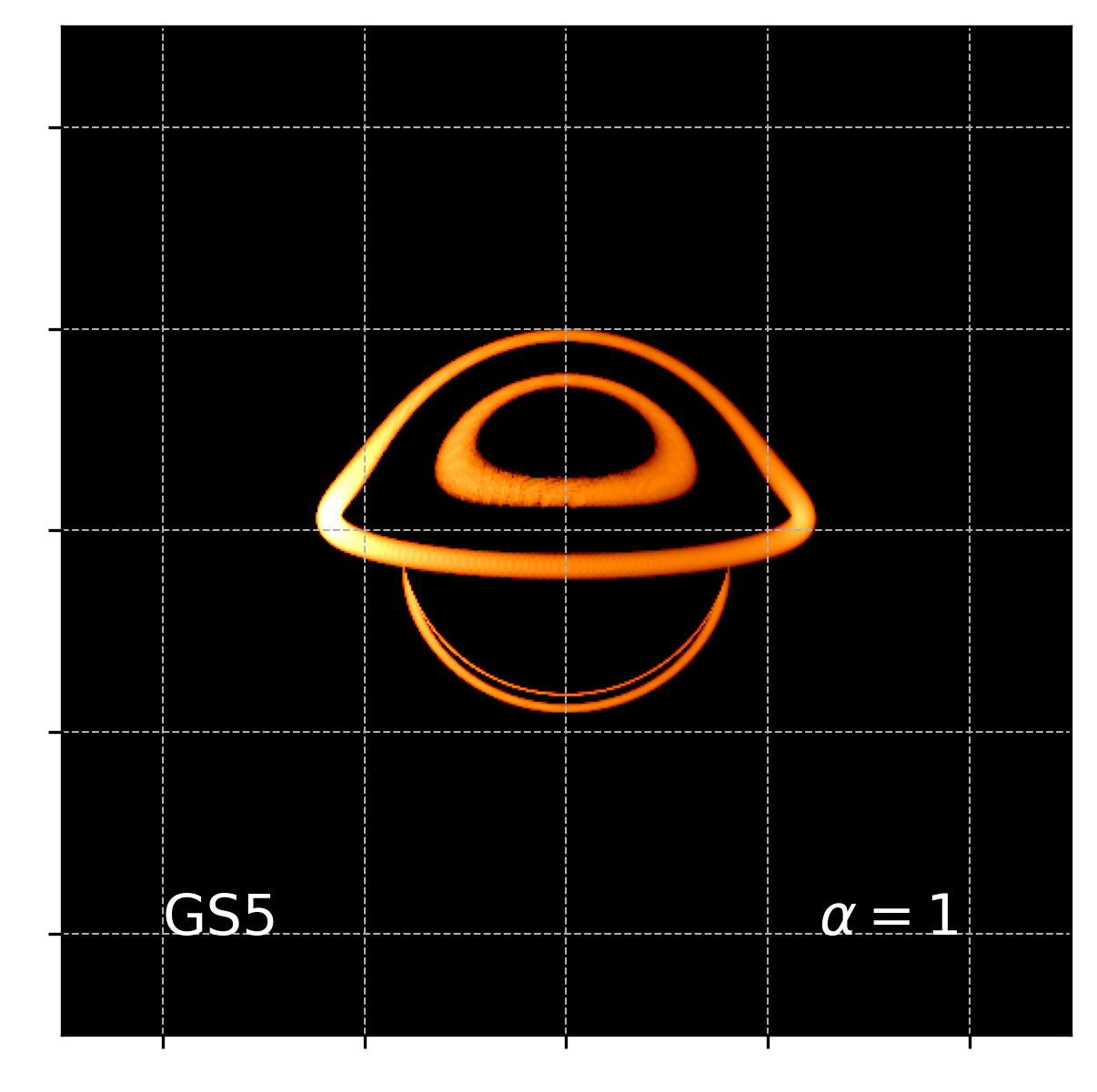}\\
    \includegraphics[scale=0.47]{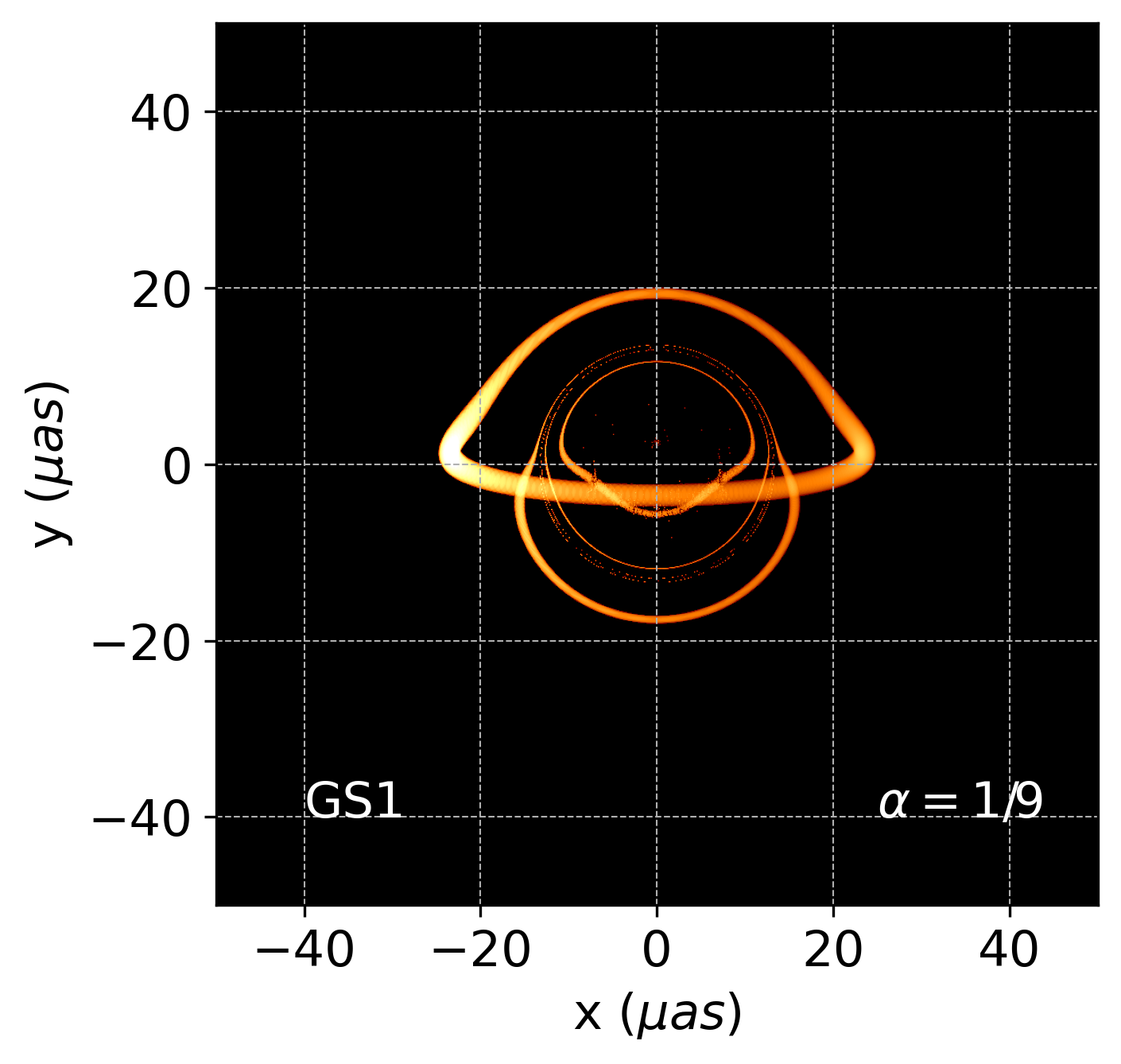} \hspace{-0.2cm}
    \includegraphics[scale=0.47]{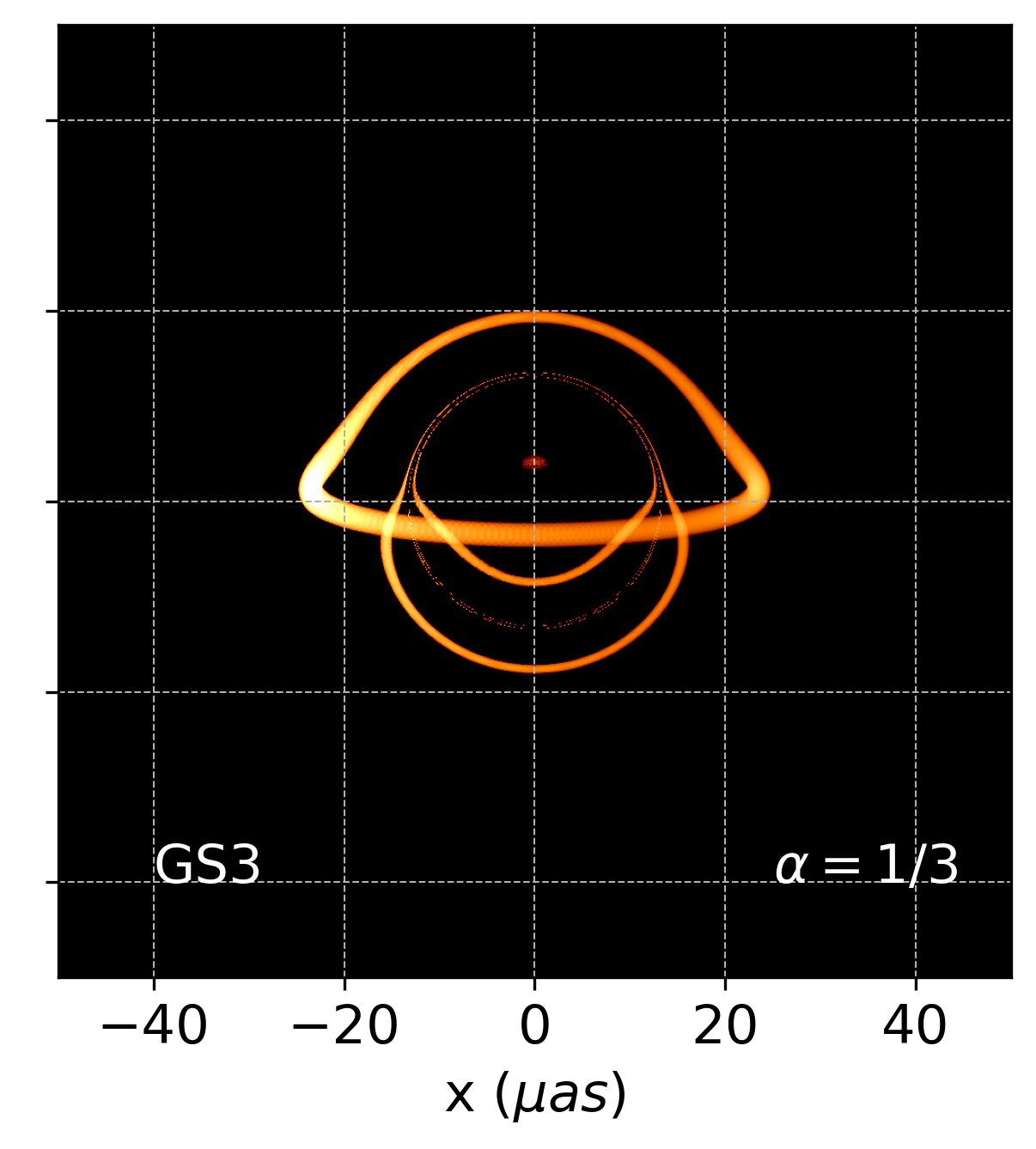} \hspace{-0.2cm}
    \includegraphics[scale=0.47]{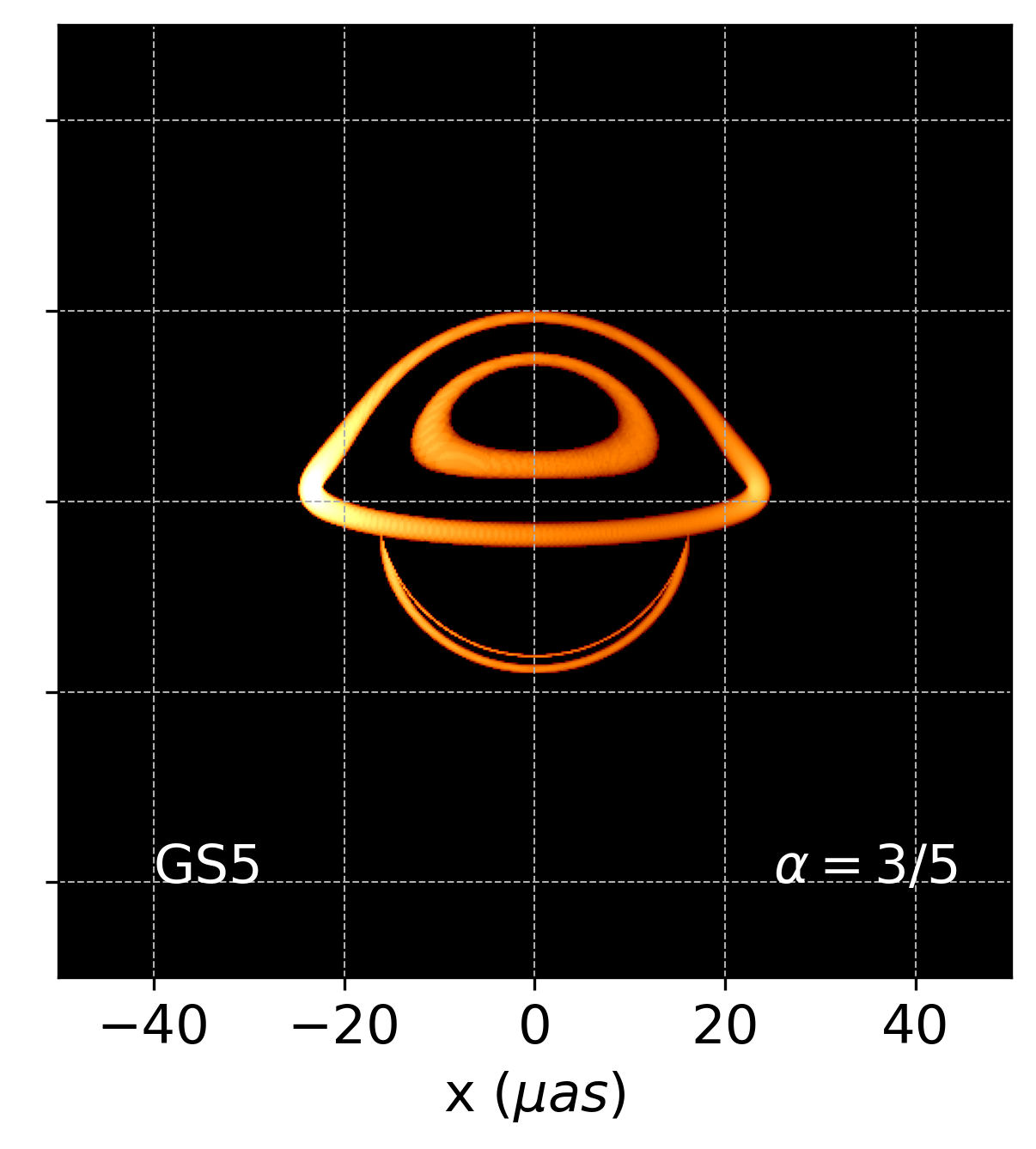}\\
    \caption{Time-integrated fluxes of the orbital motion of hot-spots for the models GS1, GS3 and GS5 for $\alpha=1$ (top row) and $\alpha=\alpha_{\rm min}$ (bottom row), with $\theta=80^\circ$ and $r_o=8M$. The parameter $\alpha$ is shown to affect negligibly the observational features of the integrated flux.}
    \label{fig:fluxalpha}
\end{figure*}

\begin{figure*}
    \centering
    \includegraphics[scale=0.47]{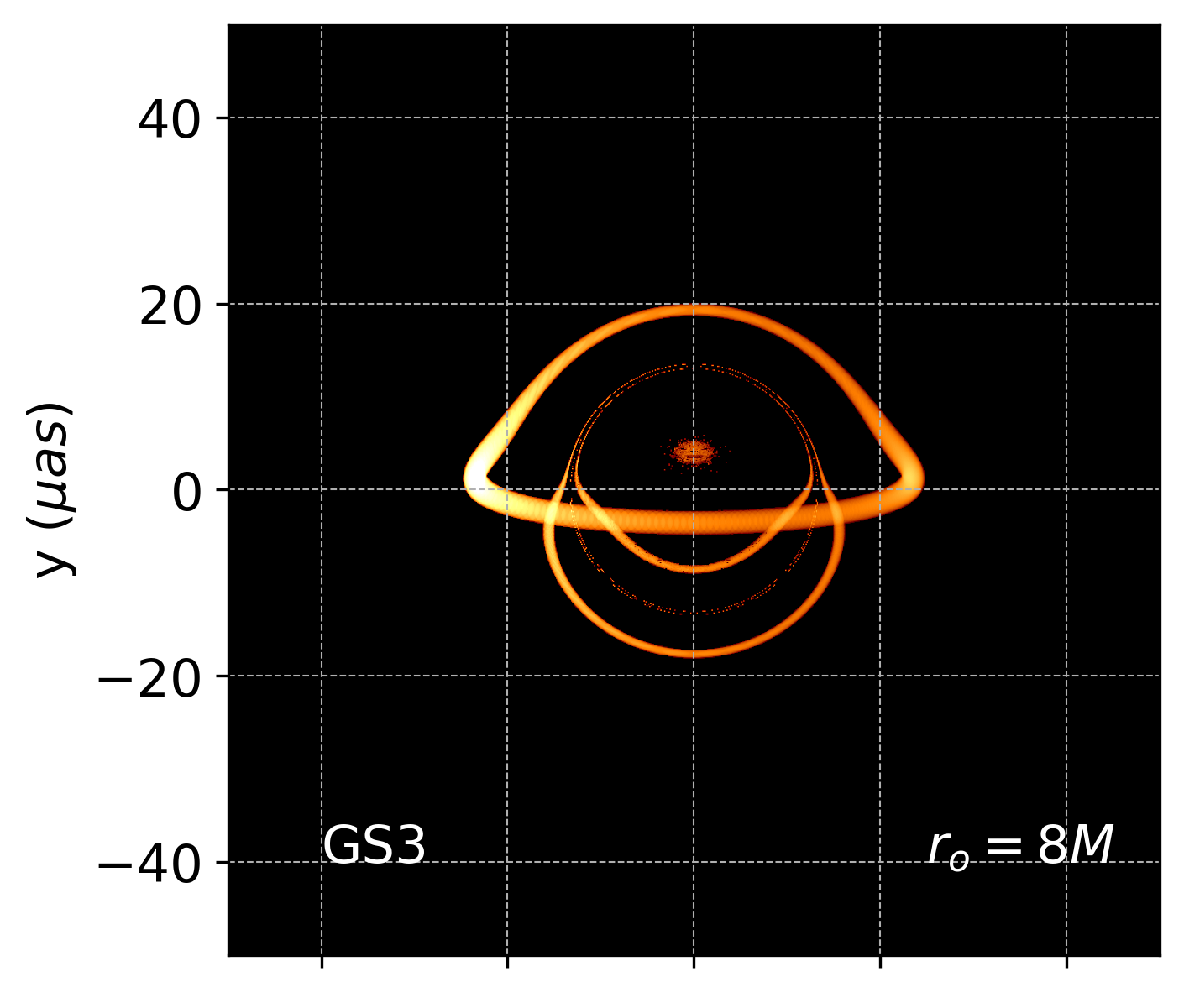} \hspace{-0.2cm}
    \includegraphics[scale=0.47]{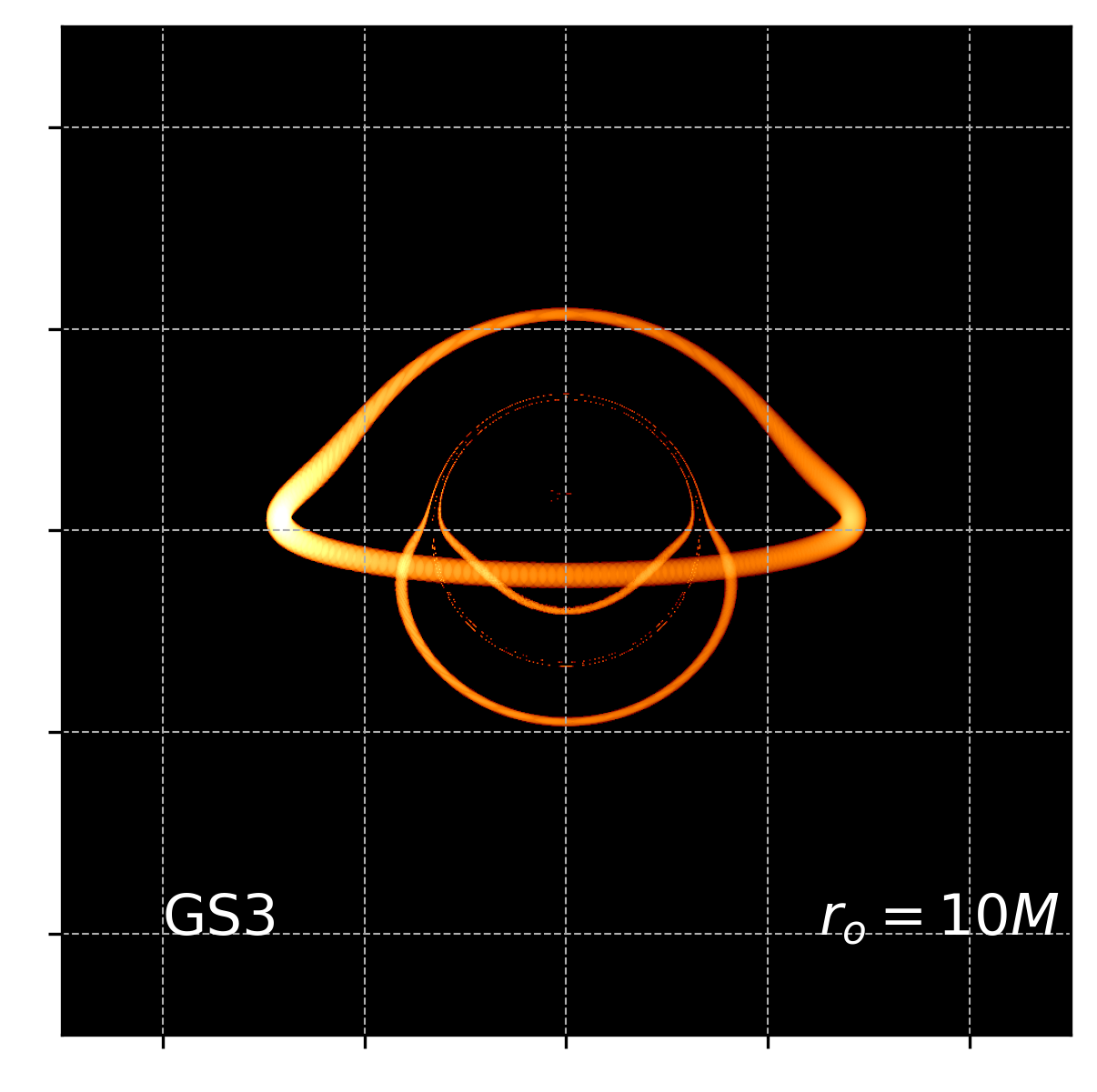} \hspace{-0.2cm}
    \includegraphics[scale=0.47]{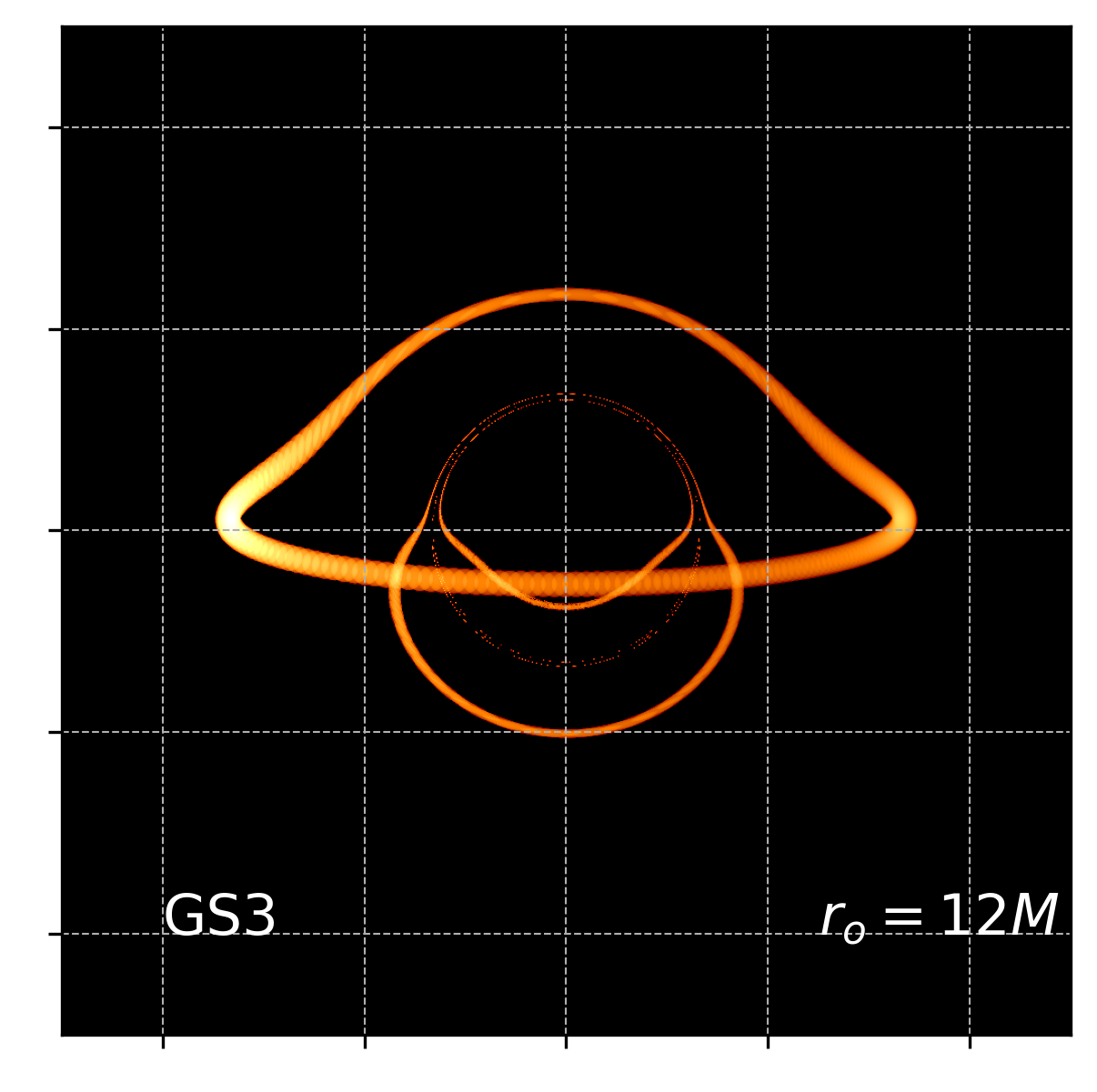}\\
    \includegraphics[scale=0.47]{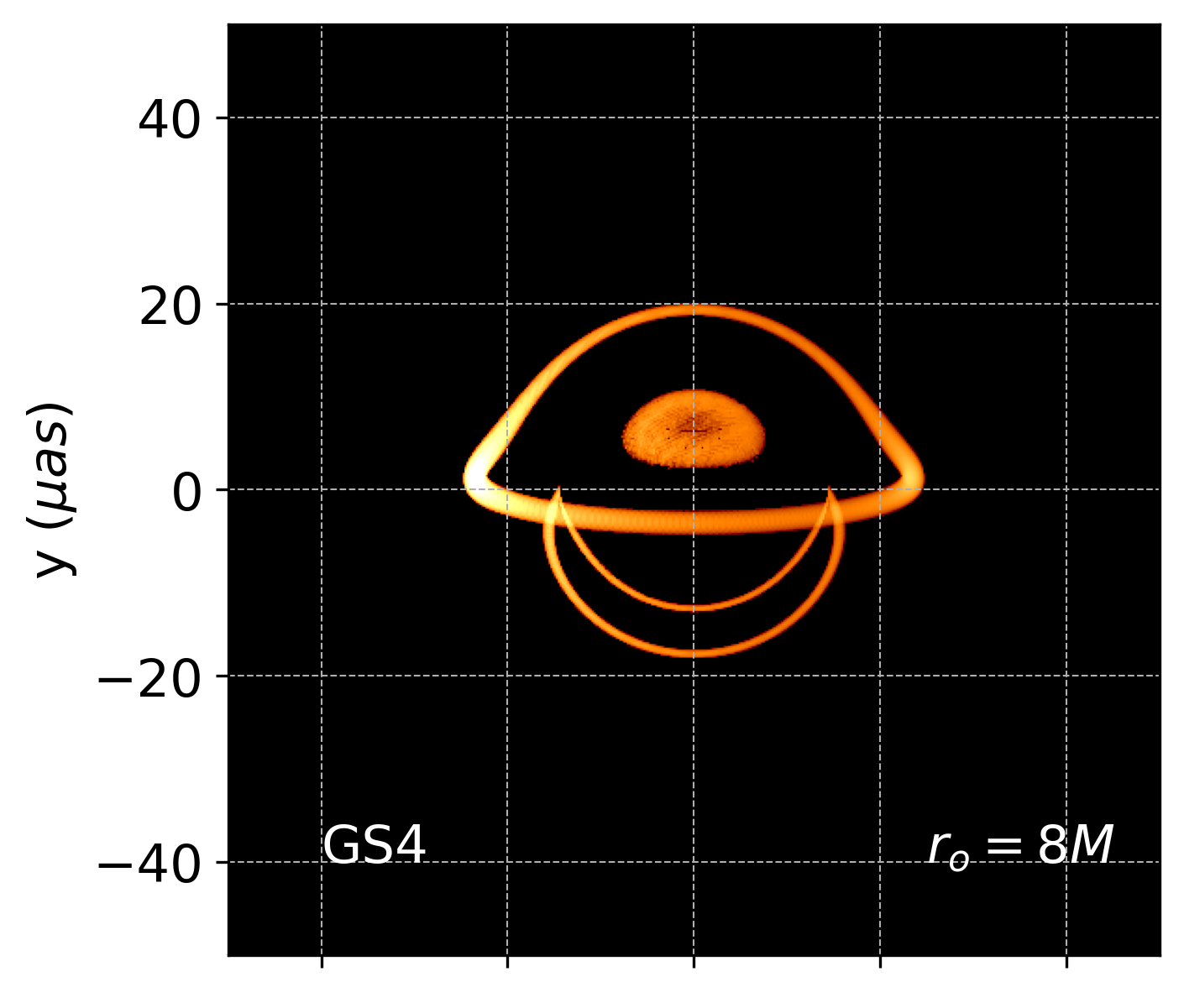} \hspace{-0.2cm}
    \includegraphics[scale=0.47]{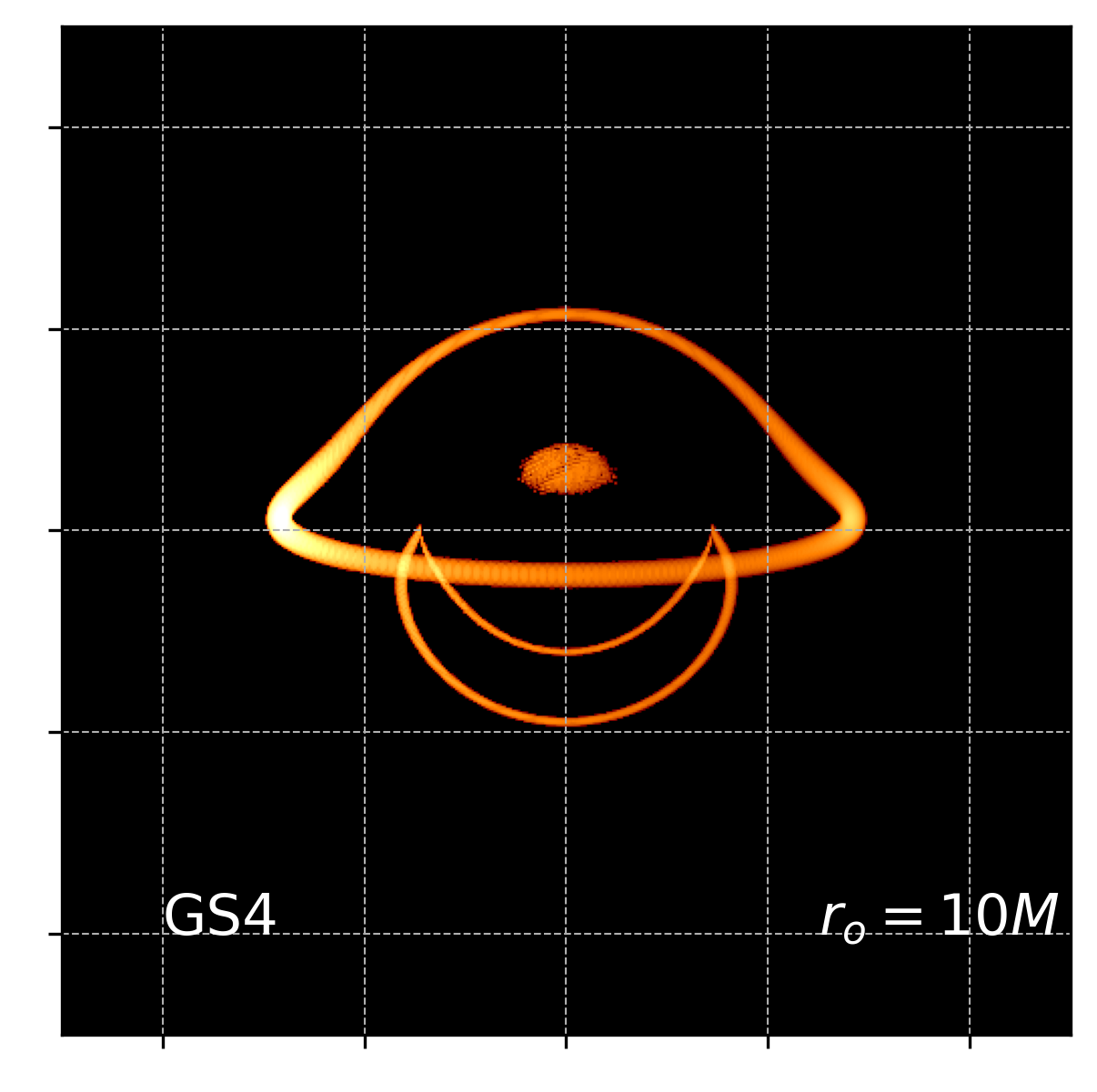} \hspace{-0.2cm}
    \includegraphics[scale=0.47]{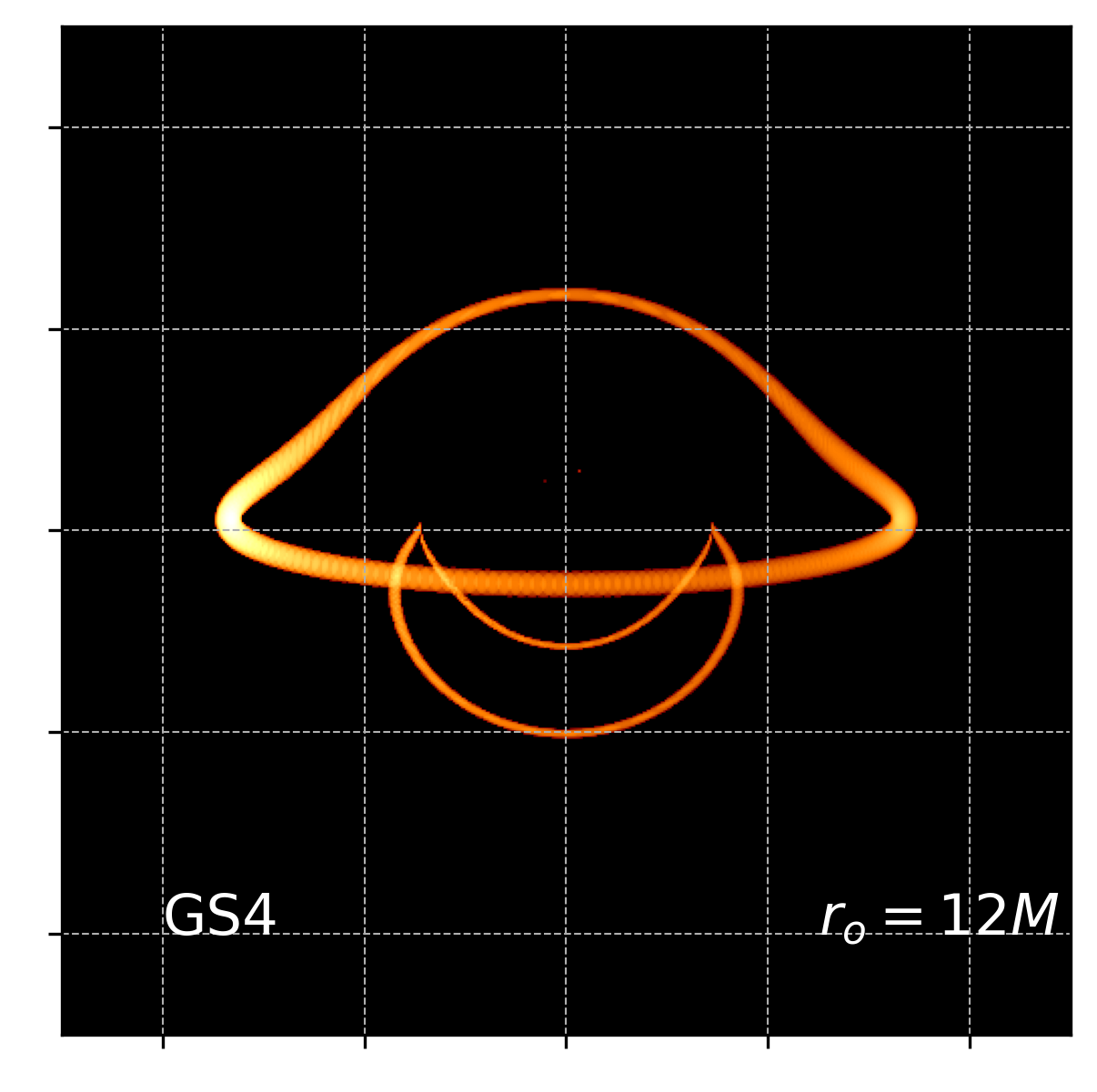}\\
    \includegraphics[scale=0.47]{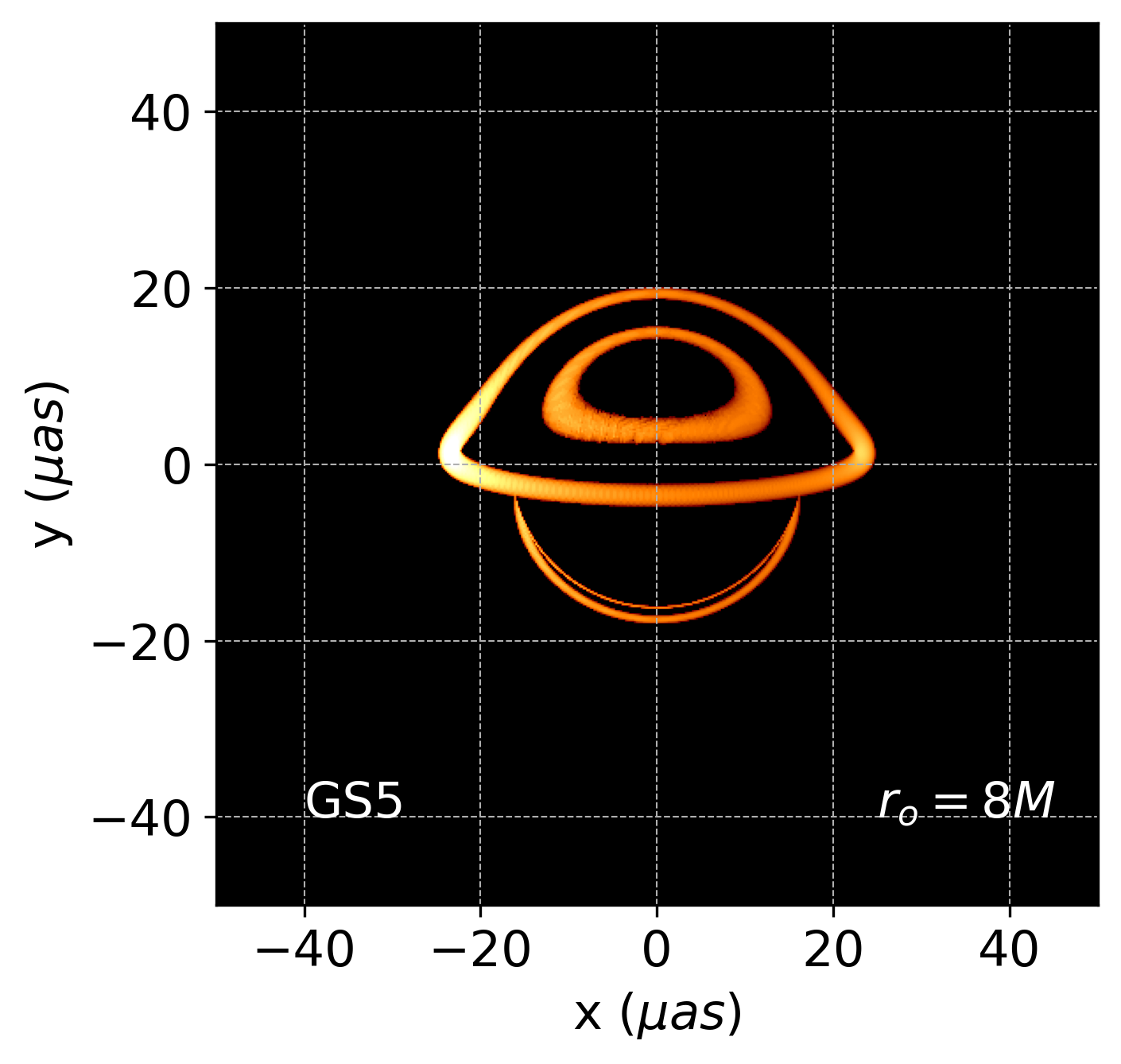} \hspace{-0.2cm}
    \includegraphics[scale=0.47]{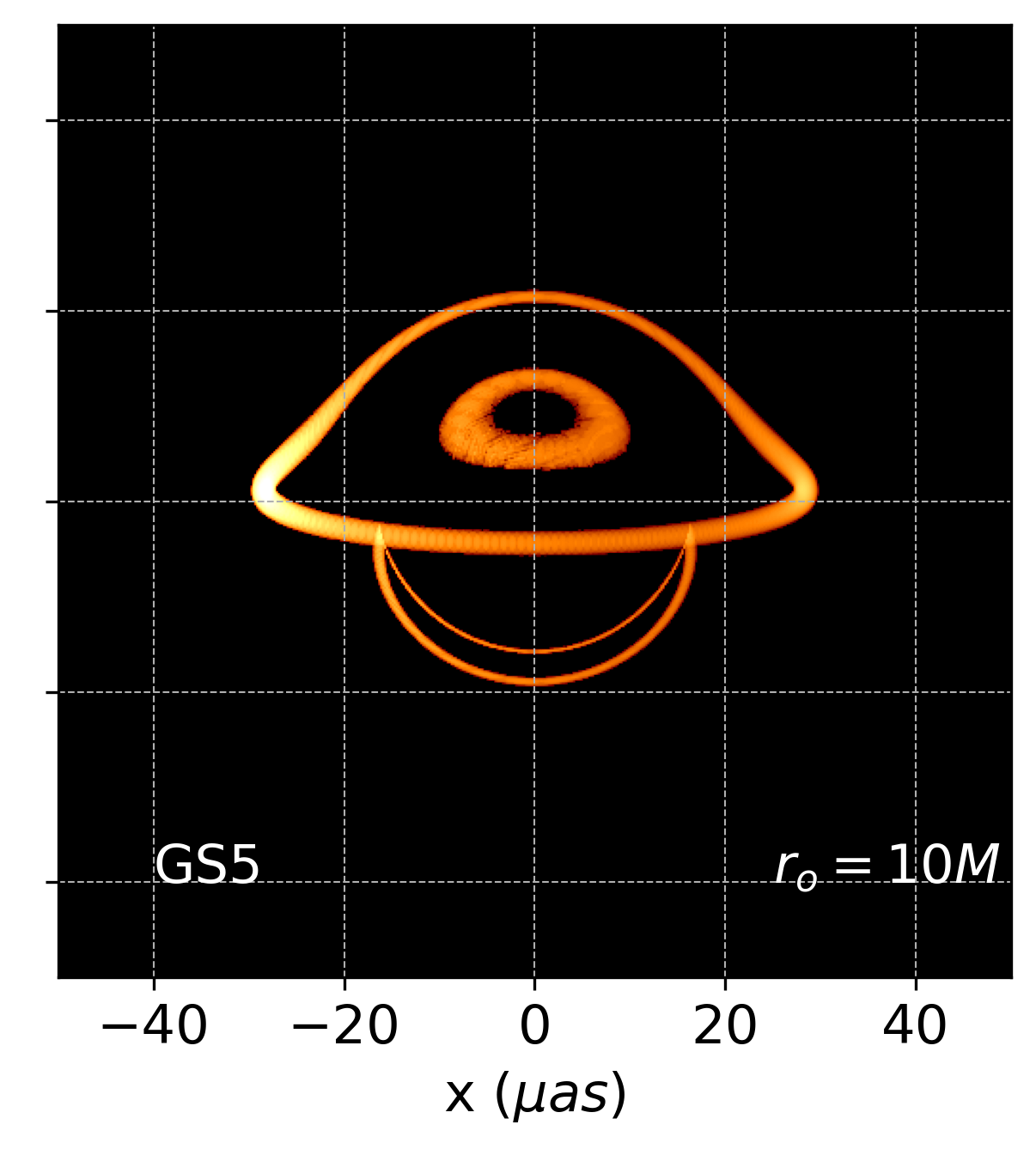} \hspace{-0.2cm}
    \includegraphics[scale=0.47]{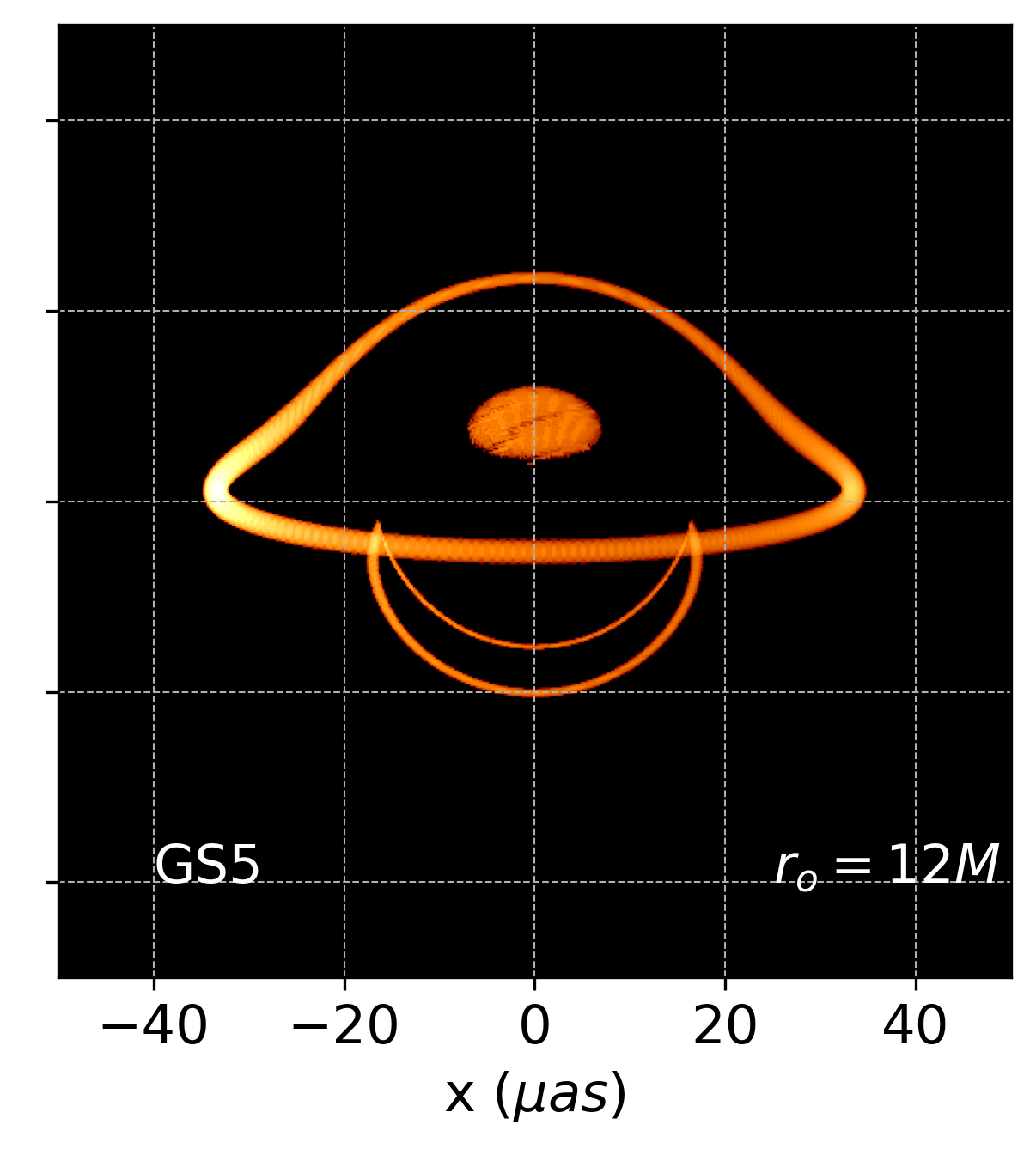}\\
    \caption{Time-integrated fluxes of the orbital motion of hot-spots for the models GS3 to GS5 (top to bottom) for an orbital radius of the hot-spot of $r_0=\{8M, 10M, 12M\}$ (left to right), with $\theta=80^\circ$ and $\alpha=1$. Models GS1 and GS2 were omitted since their observational properties are qualitatively equal to those of model GS3.}
    \label{fig:fluxradius}
\end{figure*}

\begin{figure*}
    \centering
    \includegraphics[scale=0.4]{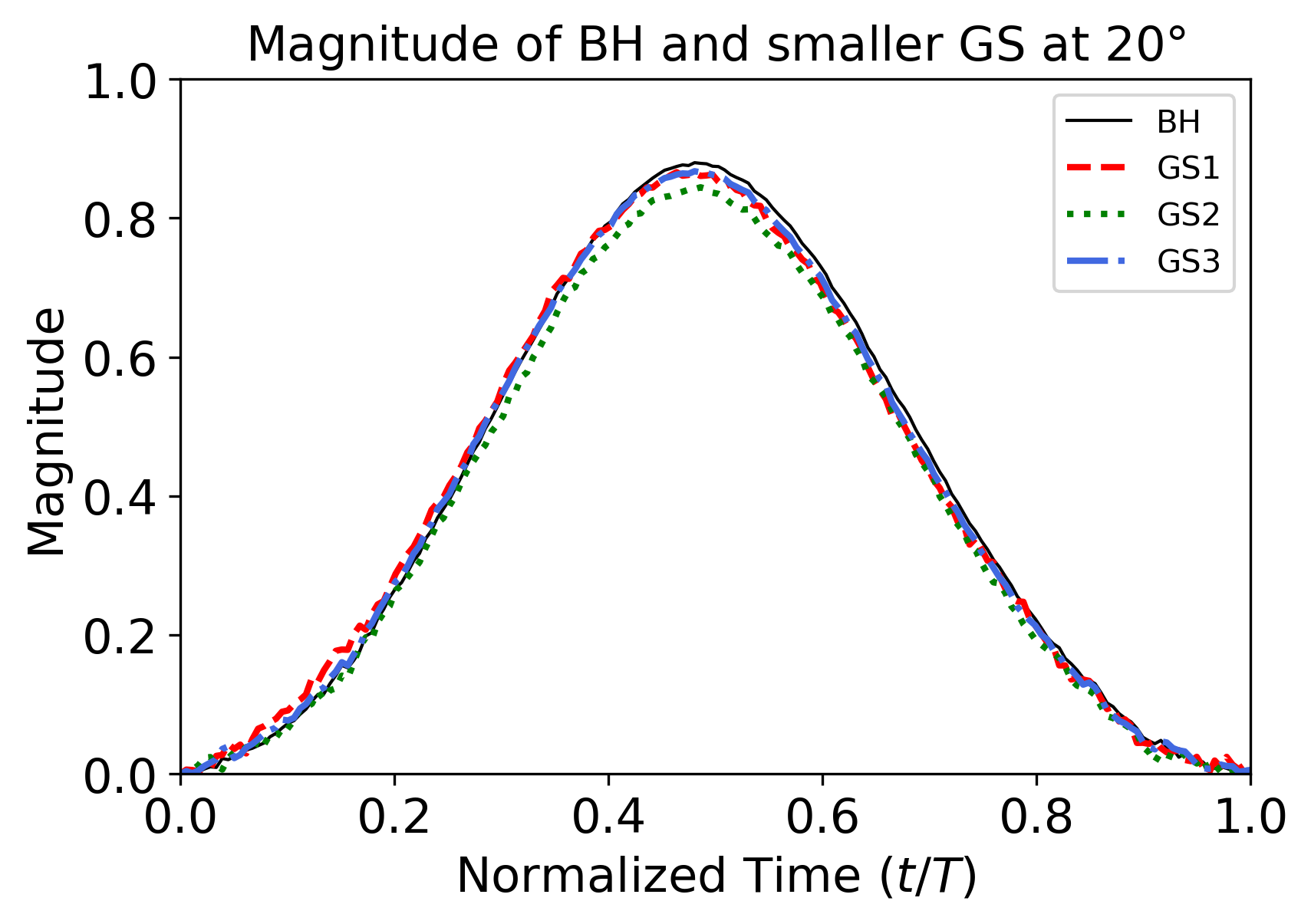} \qquad
    \includegraphics[scale=0.4]{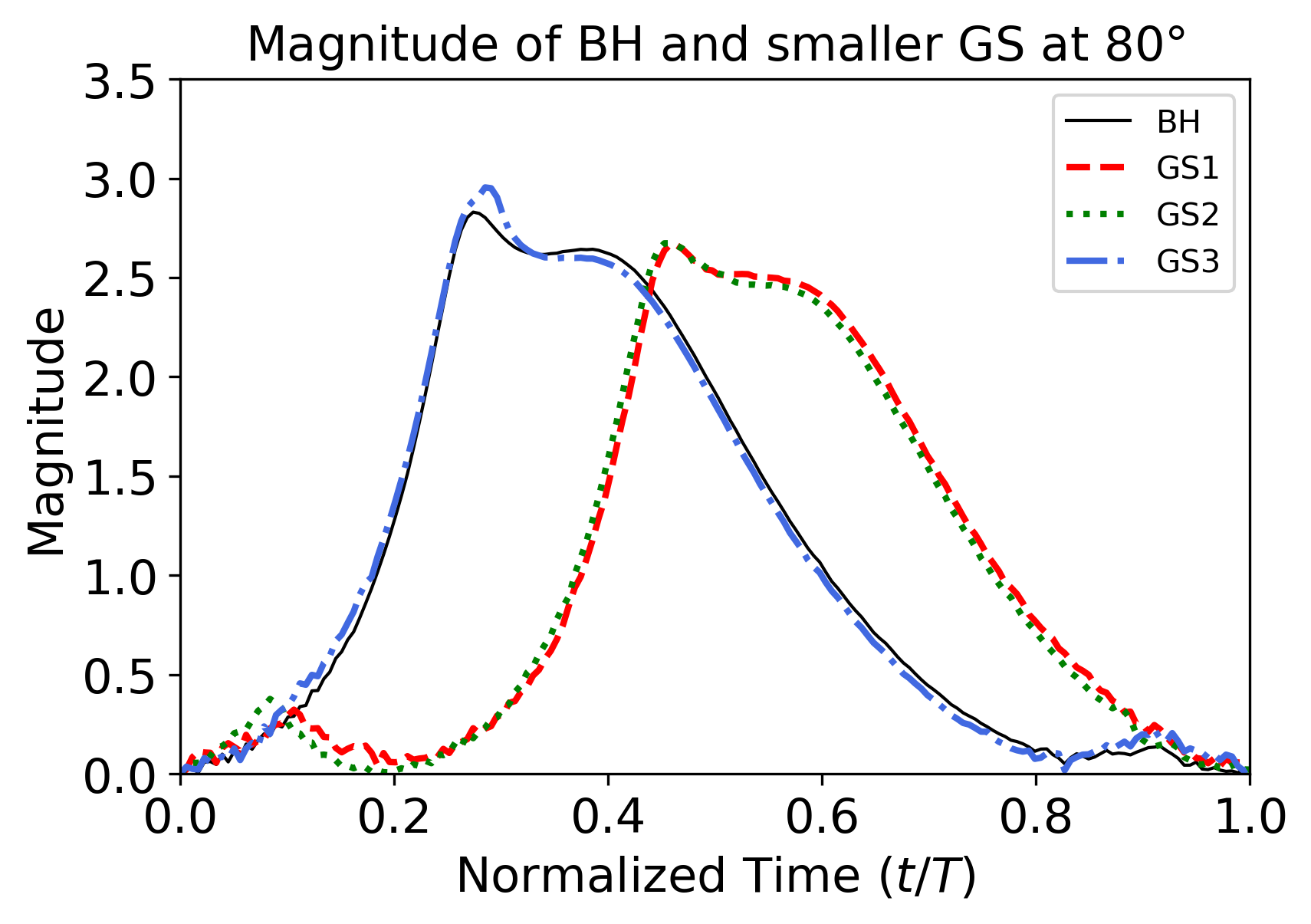}\\
    \includegraphics[scale=0.4]{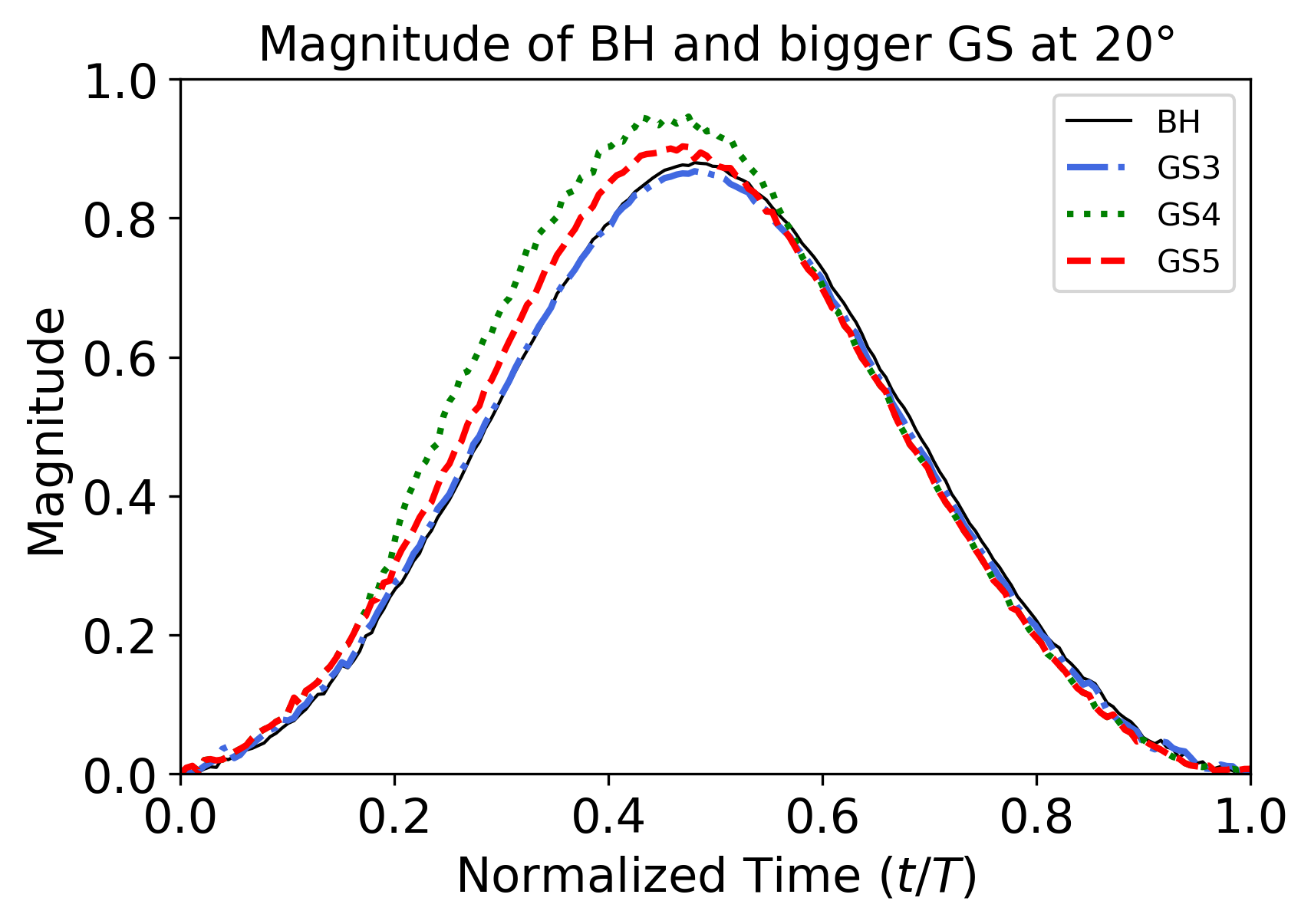} \qquad
    \includegraphics[scale=0.4]{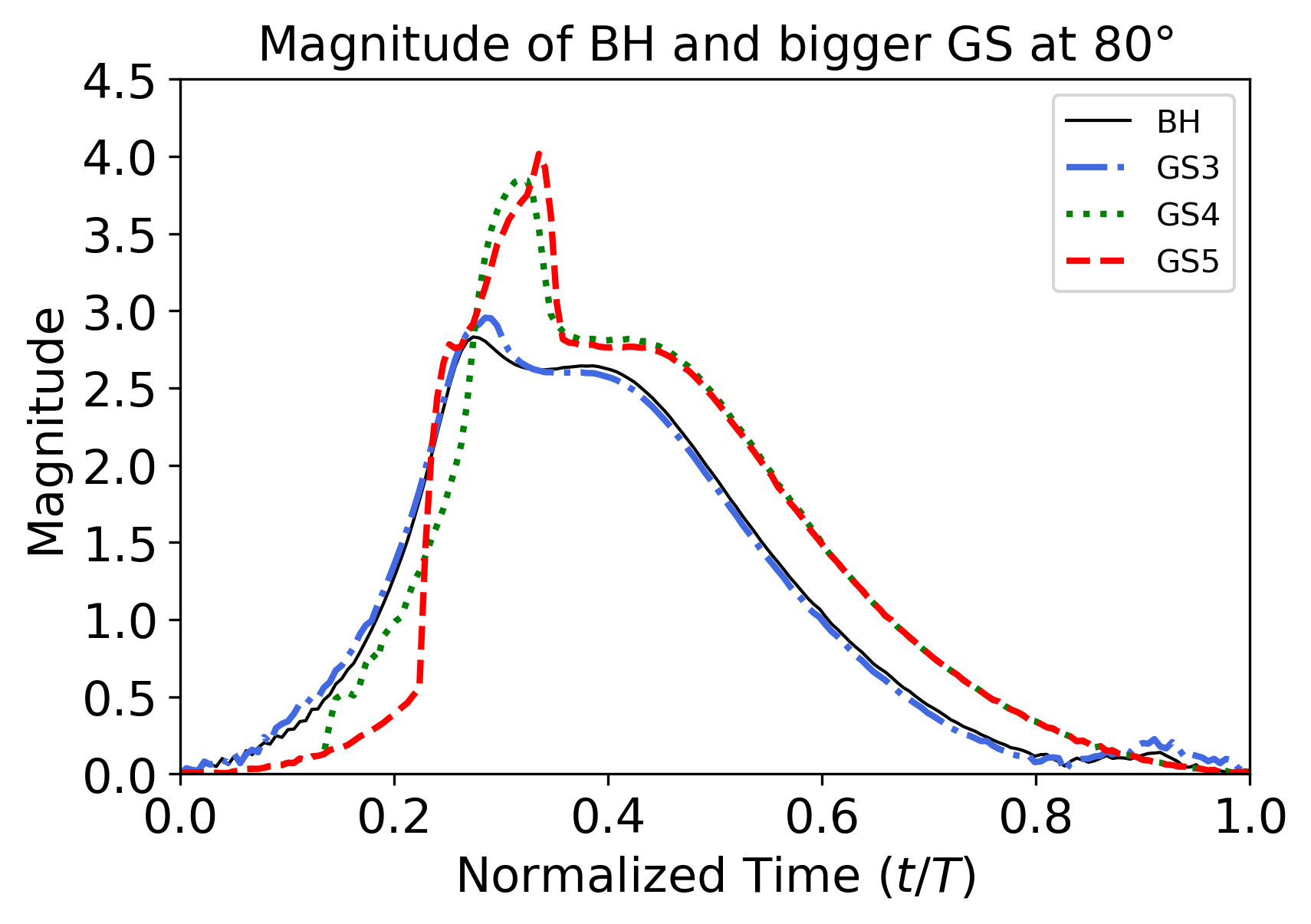}\\
    \caption{Temporal magnitudes of the orbital motion of hot-spots for the Schwarzschild BH and the models GS1 to GS5 for an observation inclination of $\theta=20^\circ$ (left column) and $\theta=80^\circ$ (right column), with $\alpha=1$ and $r_o=8M$. The non-ultra compact models GS4 and GS5 strongly deviate from their BH counterparts, while the ultra compact models GS1 to GS3 display comparable magnitudes.}
    \label{fig:magnitude}
\end{figure*}

\begin{figure*}
    \centering
    \includegraphics[scale=0.4]{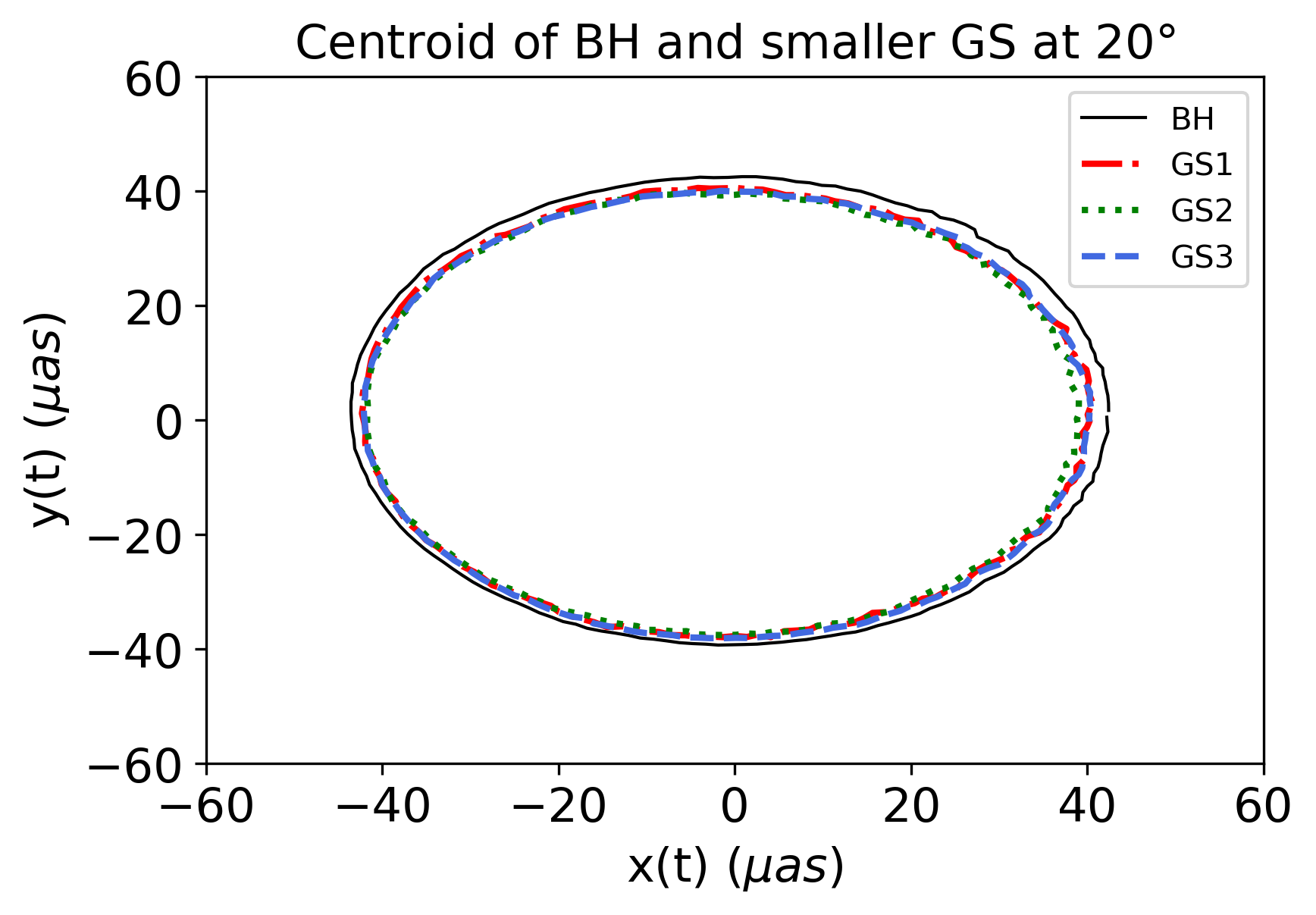} \qquad
    \includegraphics[scale=0.4]{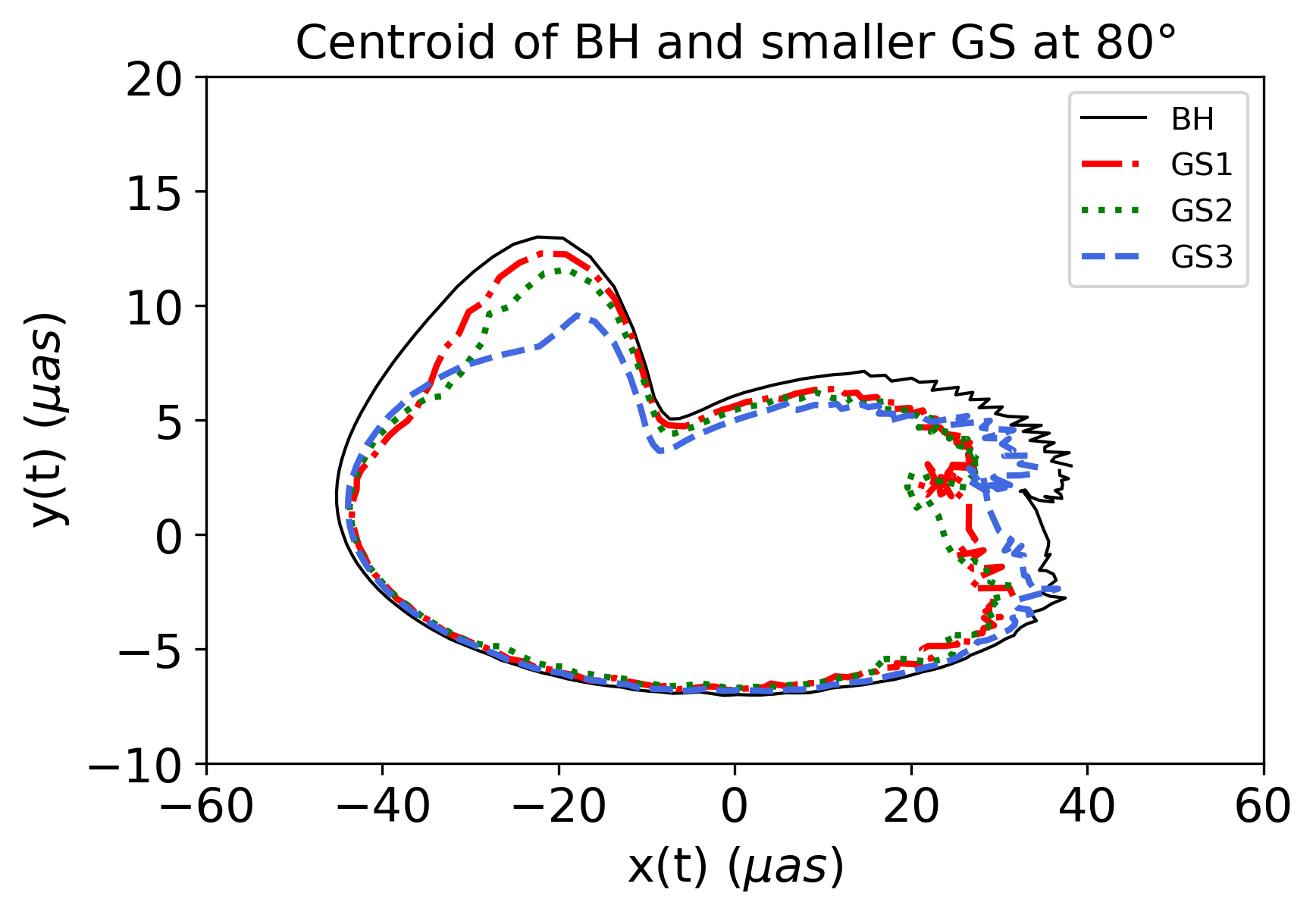}\\
    \includegraphics[scale=0.4]{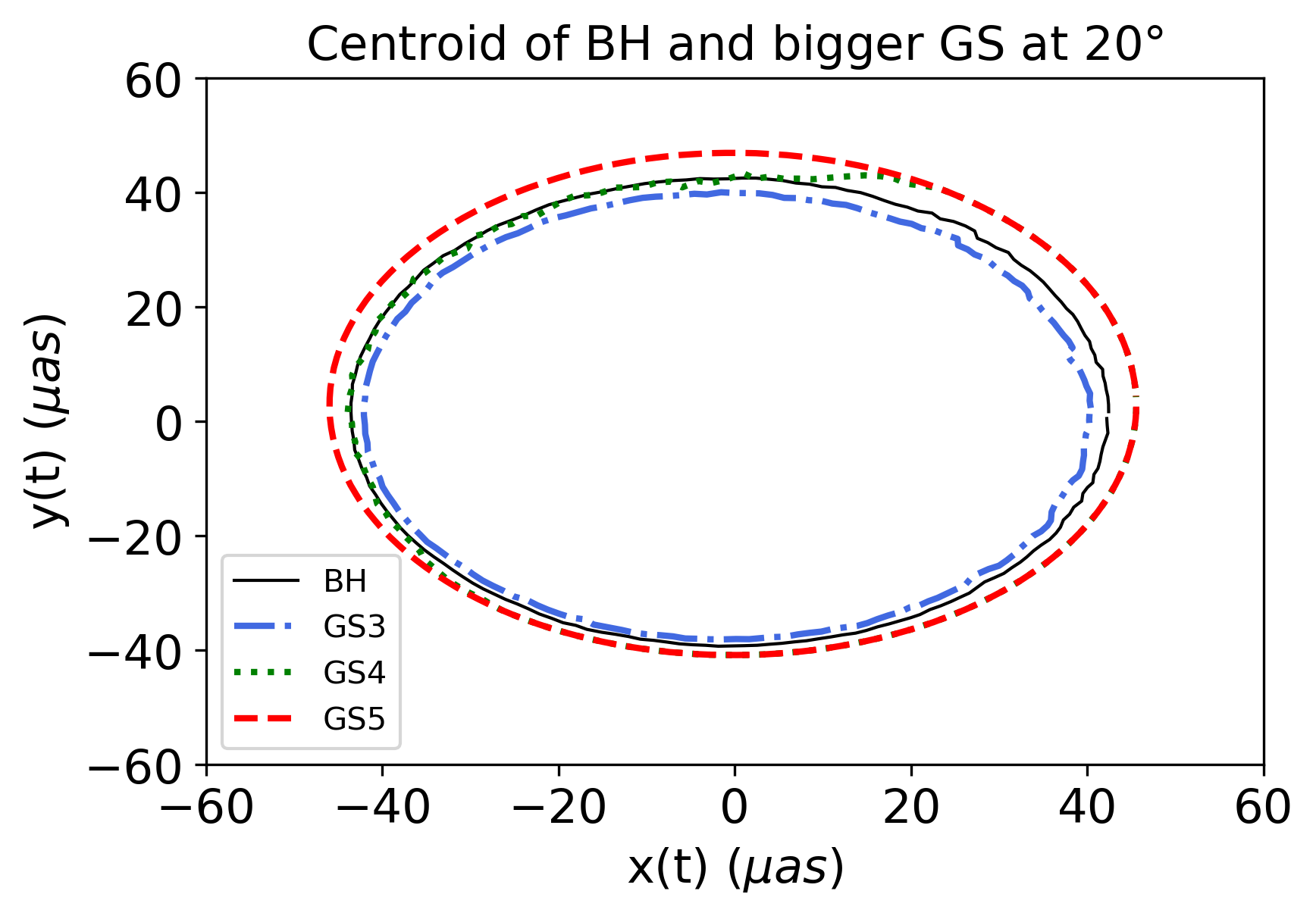} \qquad
    \includegraphics[scale=0.4]{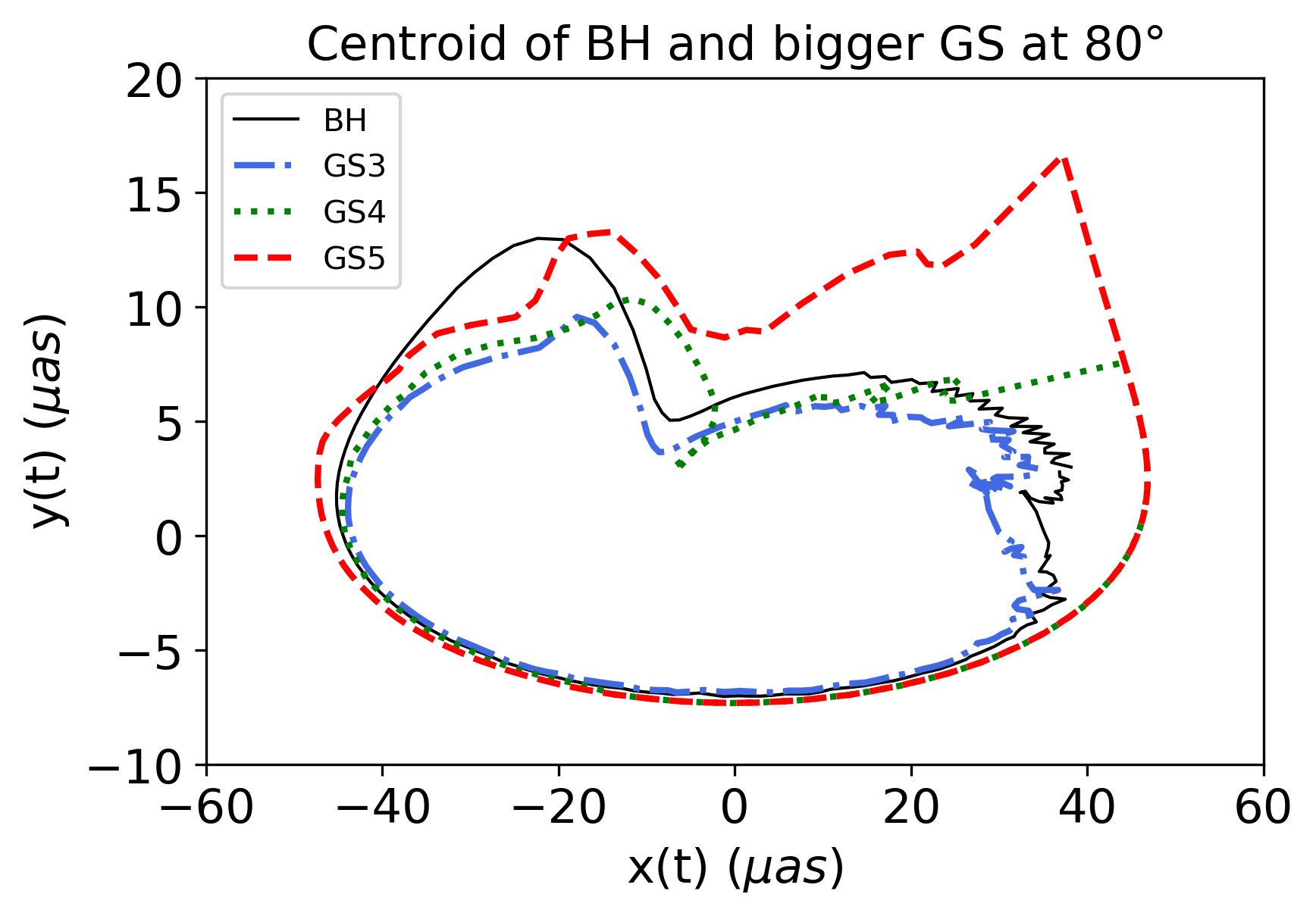}\\
    \caption{Temporal centroids of the orbital motion of hot-spots for the Schwarzschild BH and the models GS1 to GS5 for an observation inclination of $\theta=20^\circ$ (left column) and $\theta=80^\circ$ (right column), with $\alpha=1$ and $r_o=8M$. The non-ultra compact models GS4 and GS5 strongly deviate from their BH counterparts, while the ultra compact models GS1 to GS3 display comparable centroid tracks.}
    \label{fig:centroid}
\end{figure*}

\begin{figure*}
    \centering
    \includegraphics[scale=0.35]{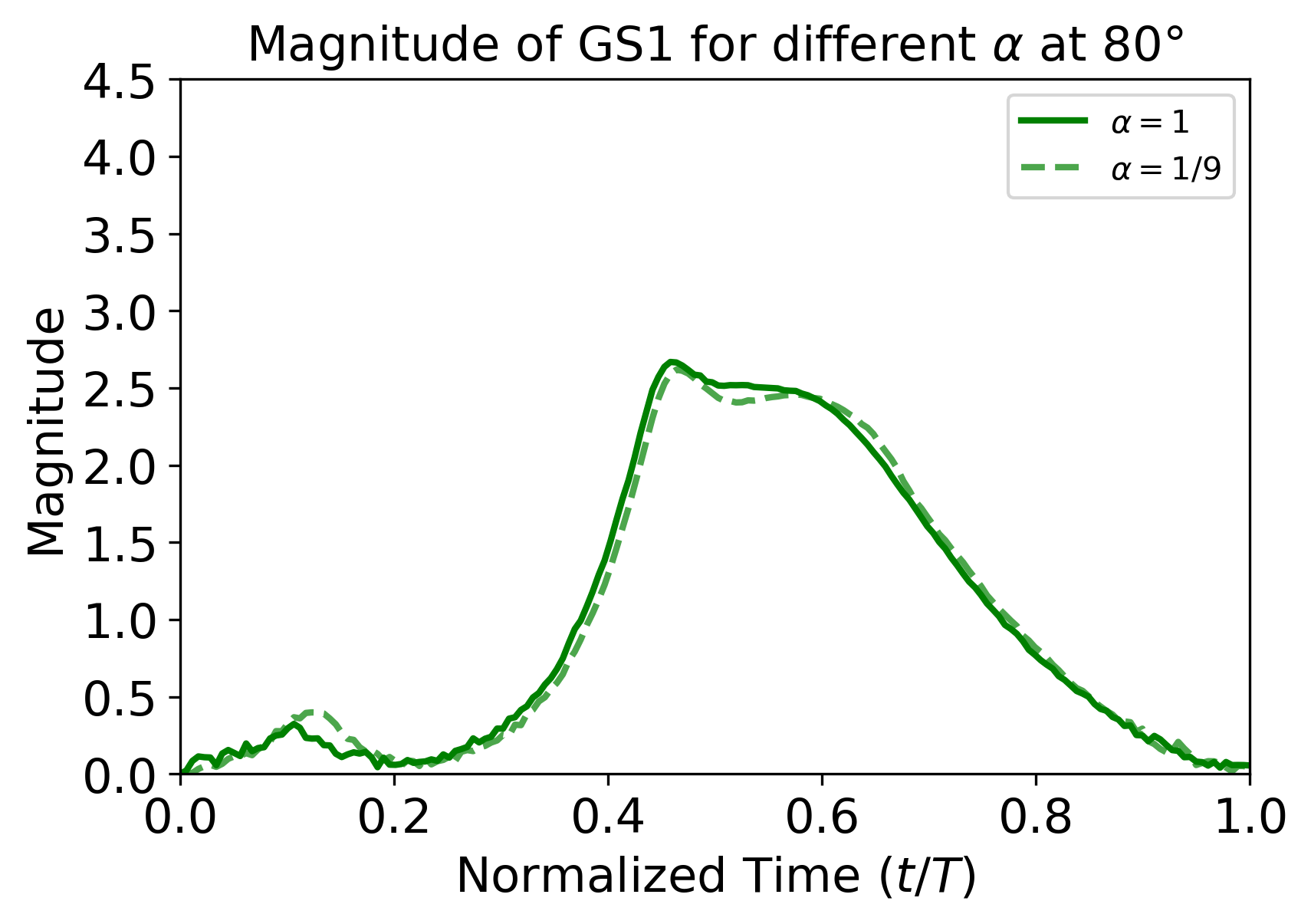} \qquad
    \includegraphics[scale=0.35]{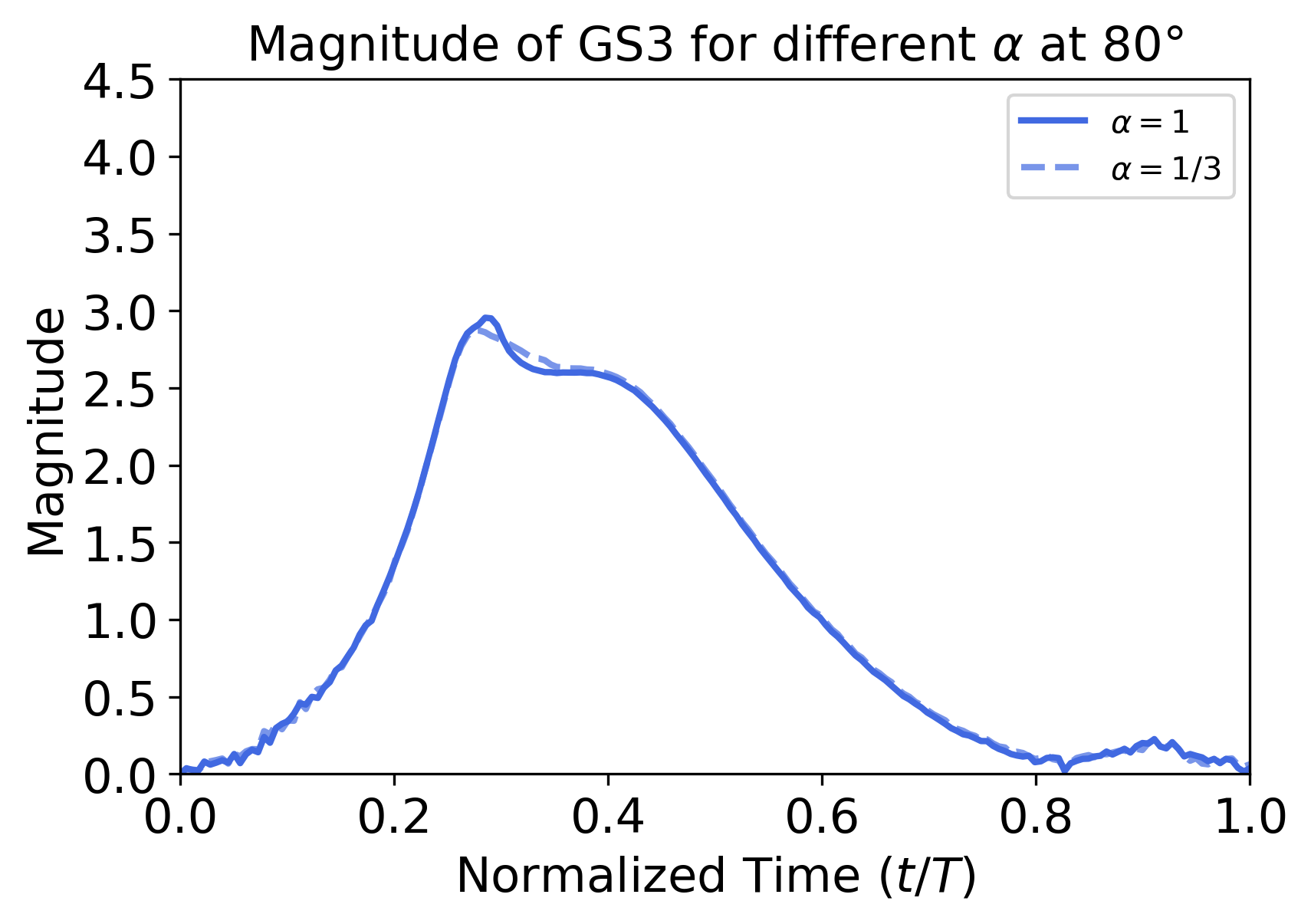} \qquad
    \includegraphics[scale=0.35]{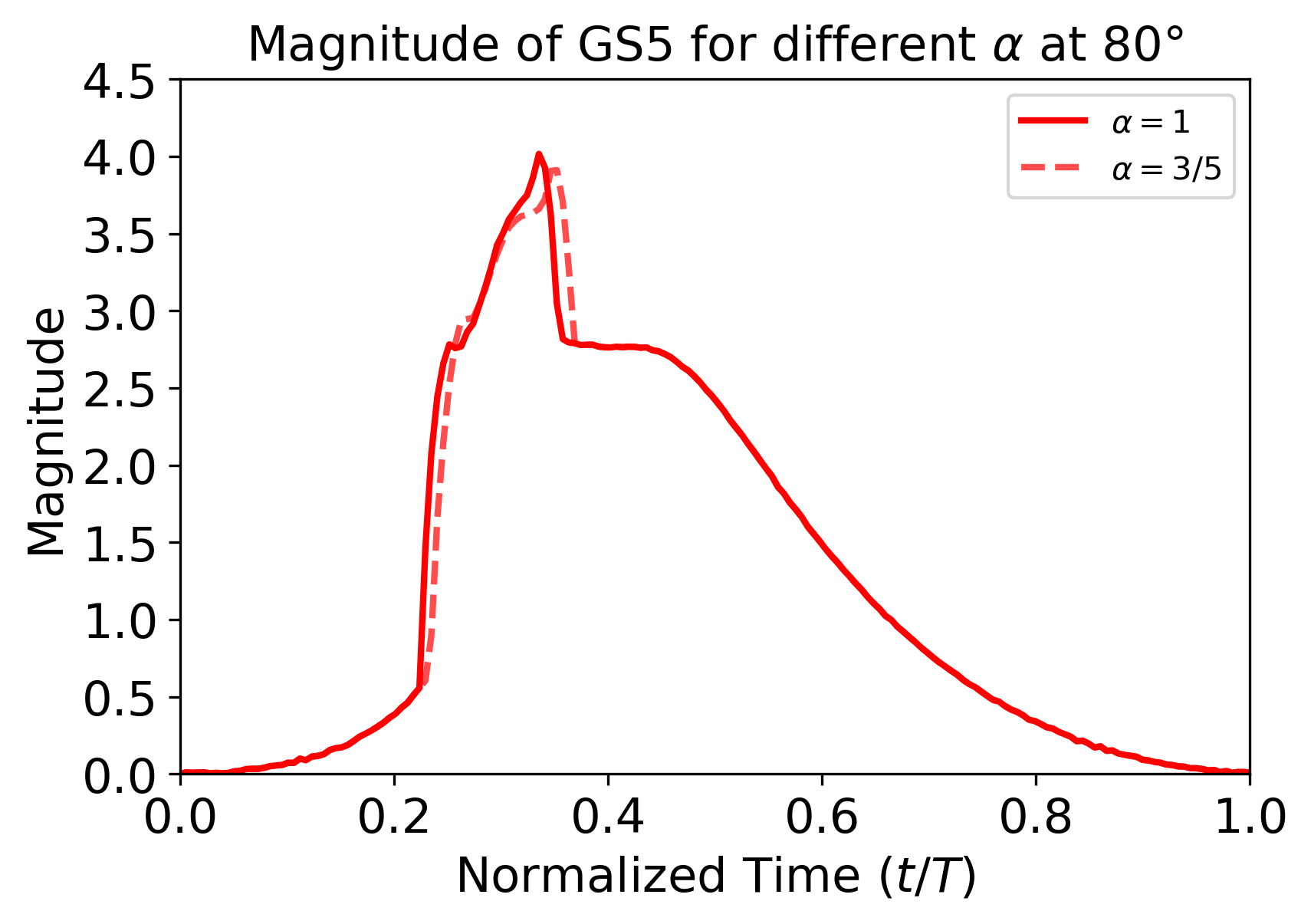}\\
    \includegraphics[scale=0.35]{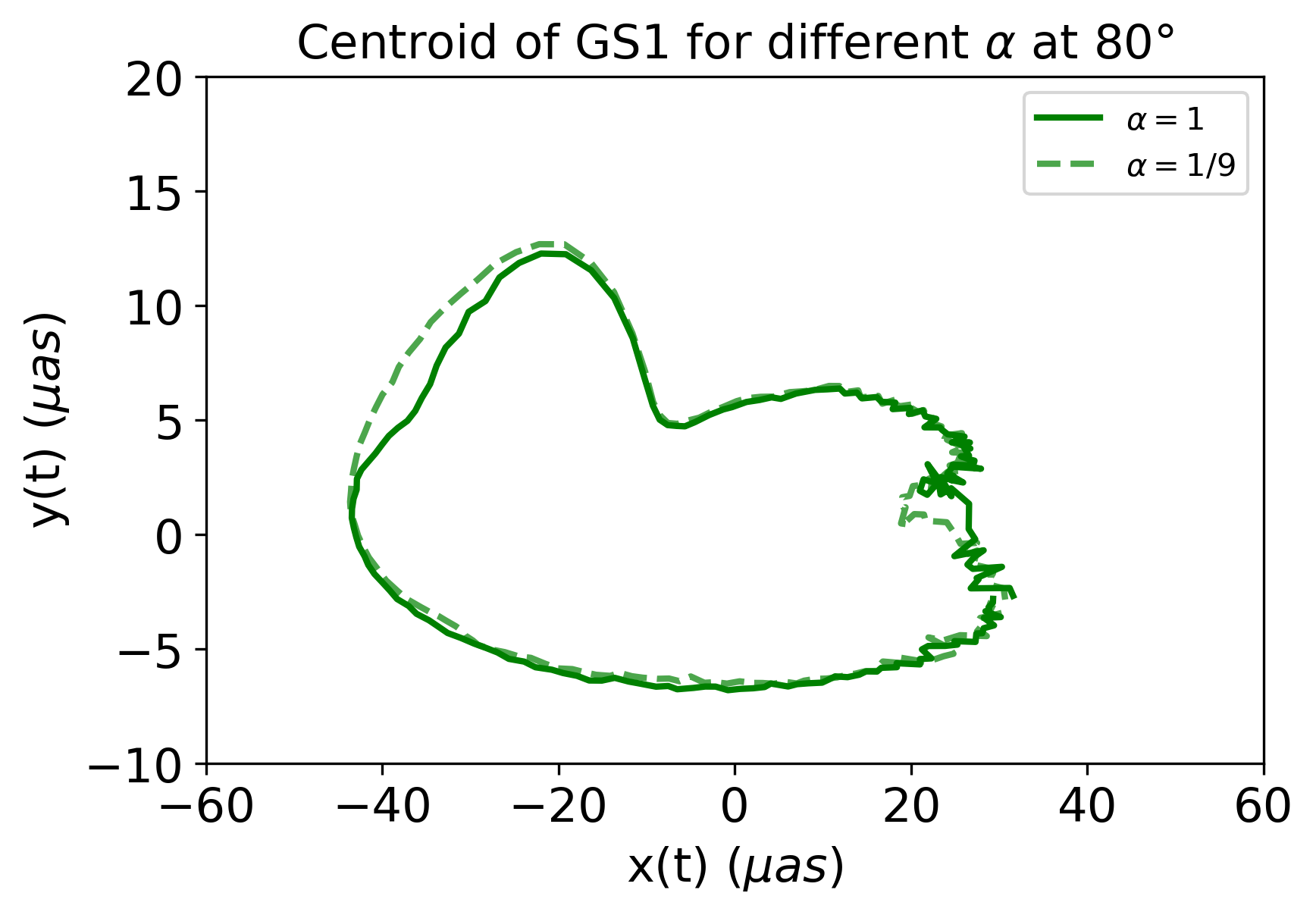} \qquad
    \includegraphics[scale=0.35]{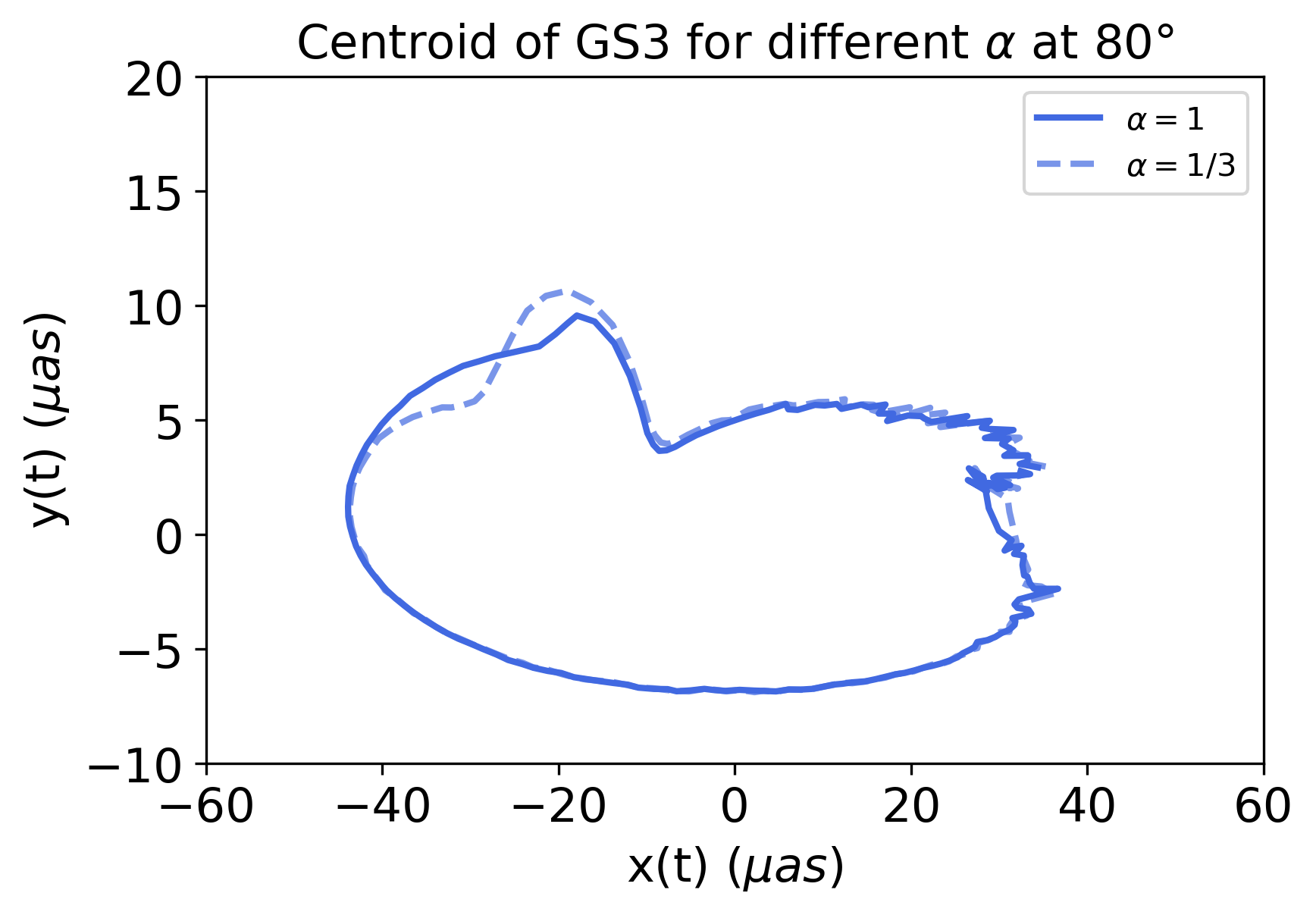} \qquad
    \includegraphics[scale=0.35]{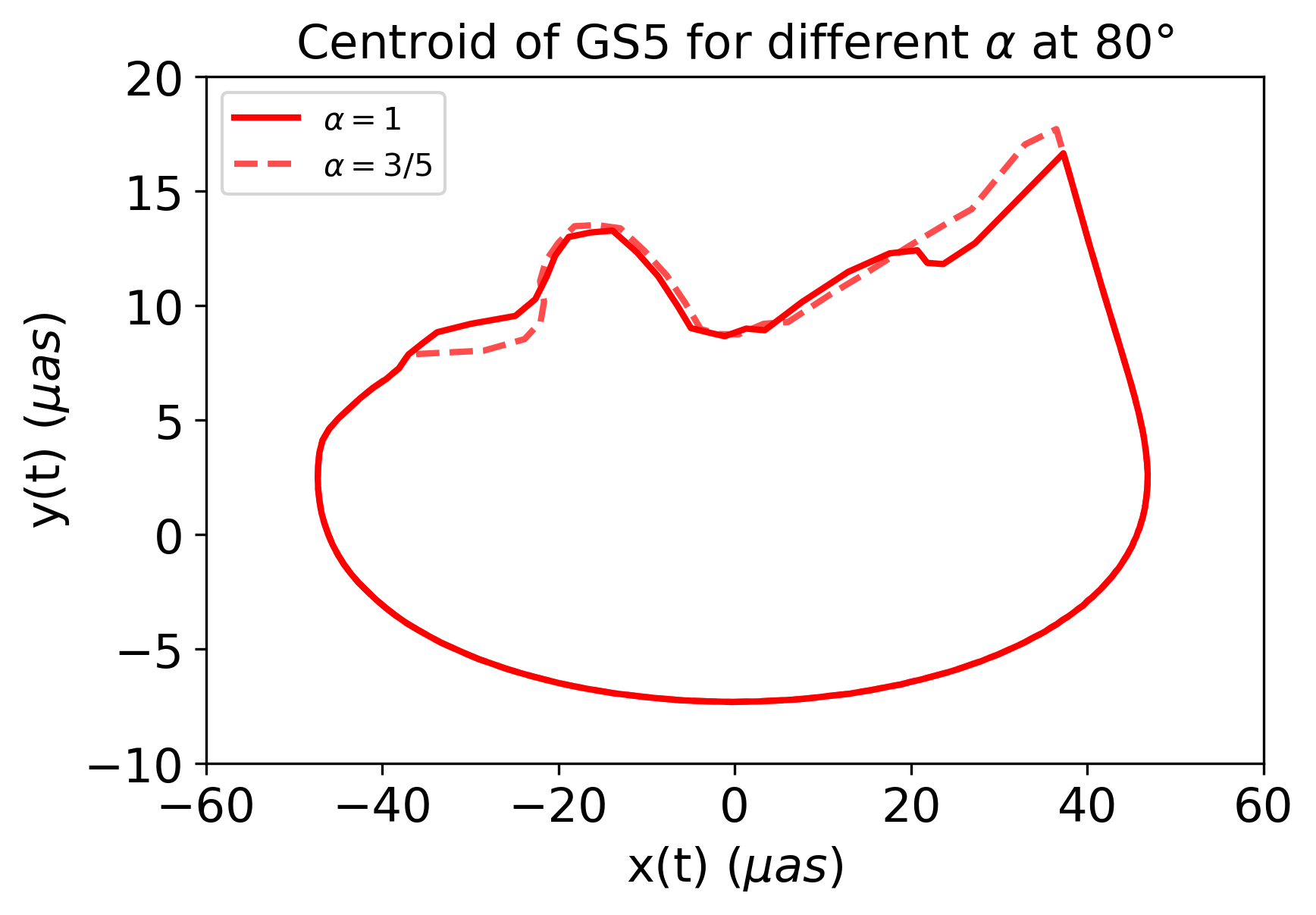}\\
    \caption{Temporal magnitudes (top row) and temporal centroids (bottom row) of the orbital motion of hot-spots for the models GS1 (left column), GS3 (middle column) and GS5 (right column) for $\alpha=1$ and $\alpha=\alpha_{\rm min}$, with $\theta=80^\circ$ and $r_o=8M$. The parameter $\alpha$ is shown to affect negligibly the astrometric observables.}
    \label{fig:astroalpha}
\end{figure*}

\begin{figure*}
    \centering
    \includegraphics[scale=0.35]{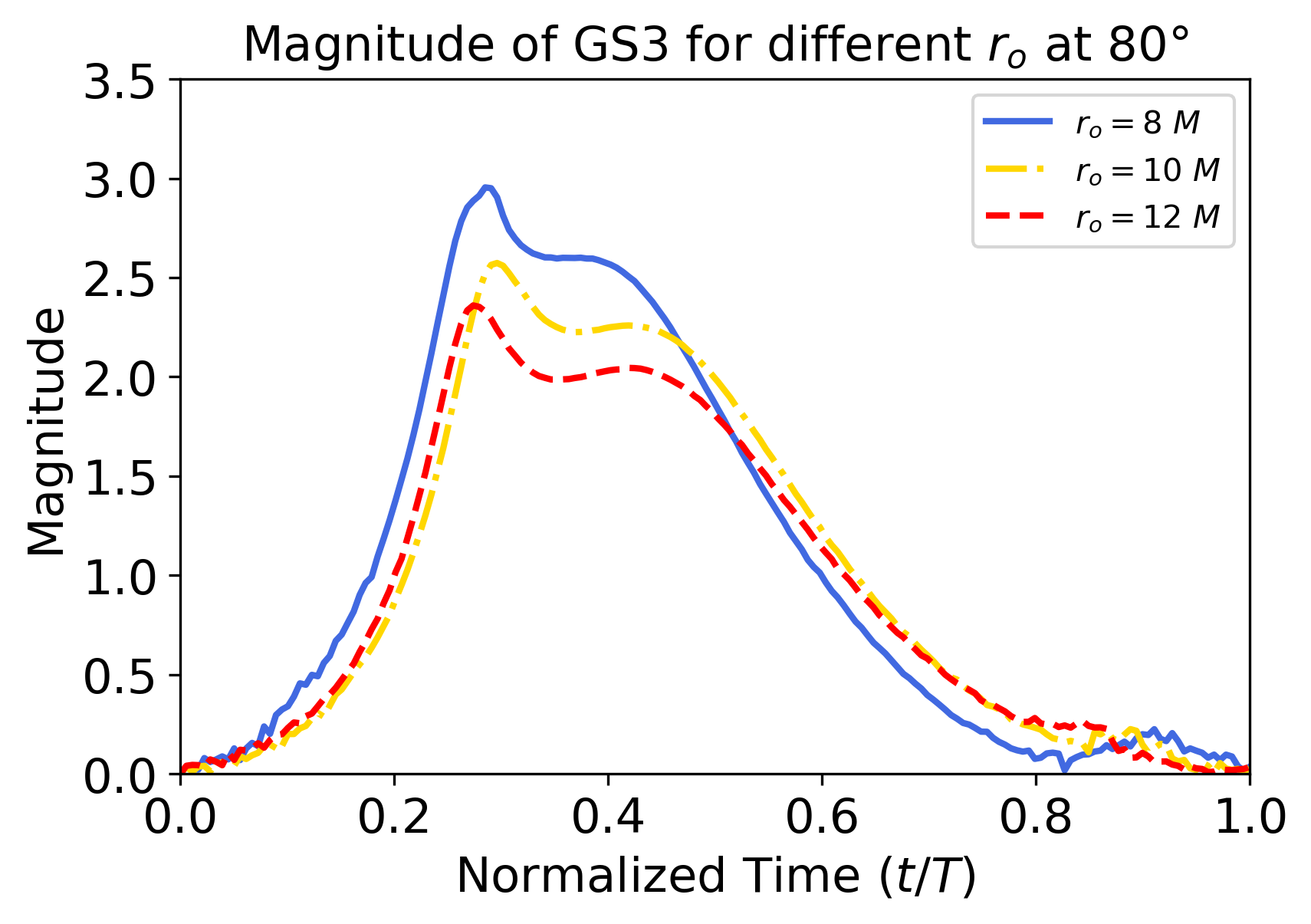} \qquad
    \includegraphics[scale=0.35]{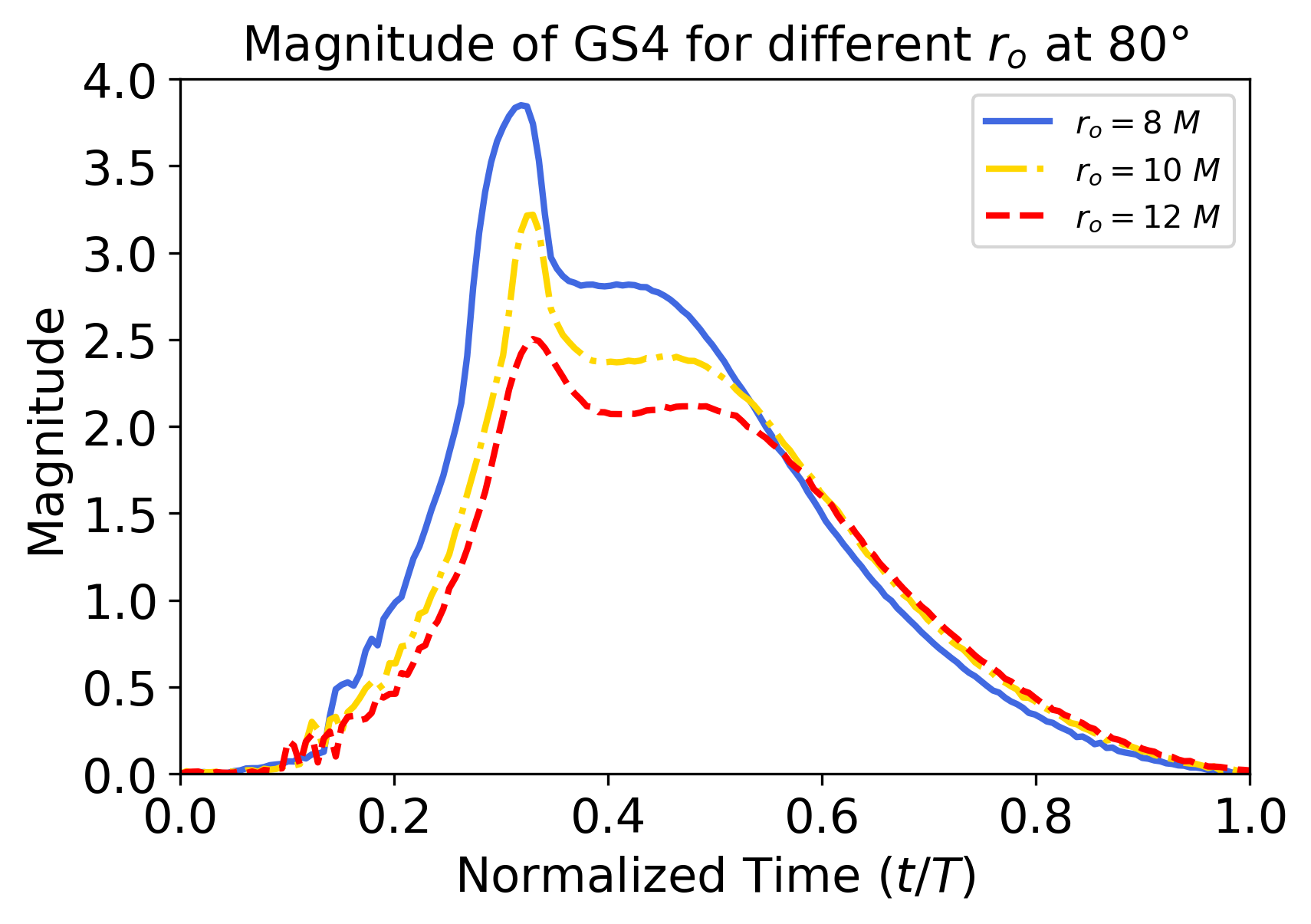} \qquad
    \includegraphics[scale=0.35]{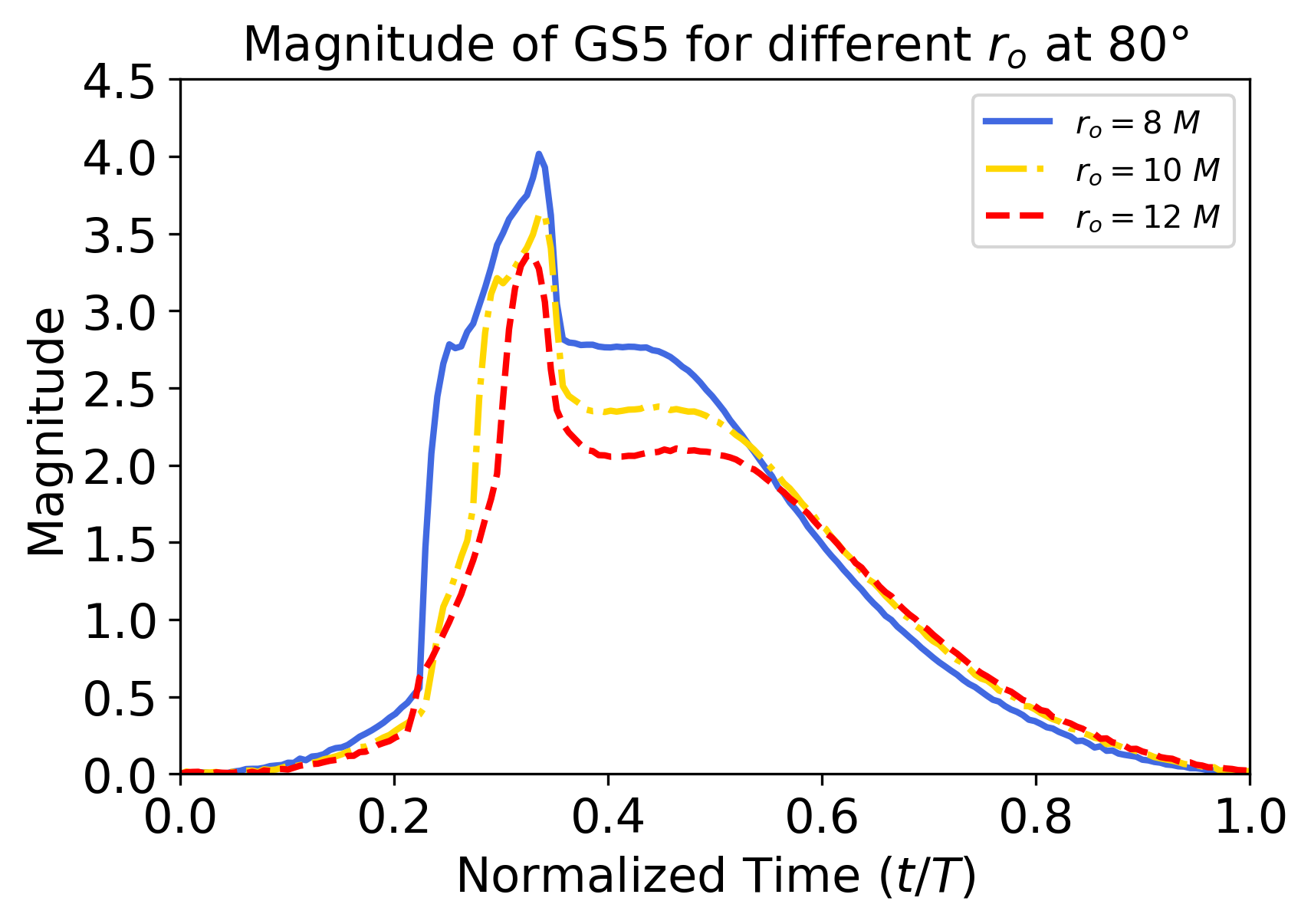}\\
    \includegraphics[scale=0.35]{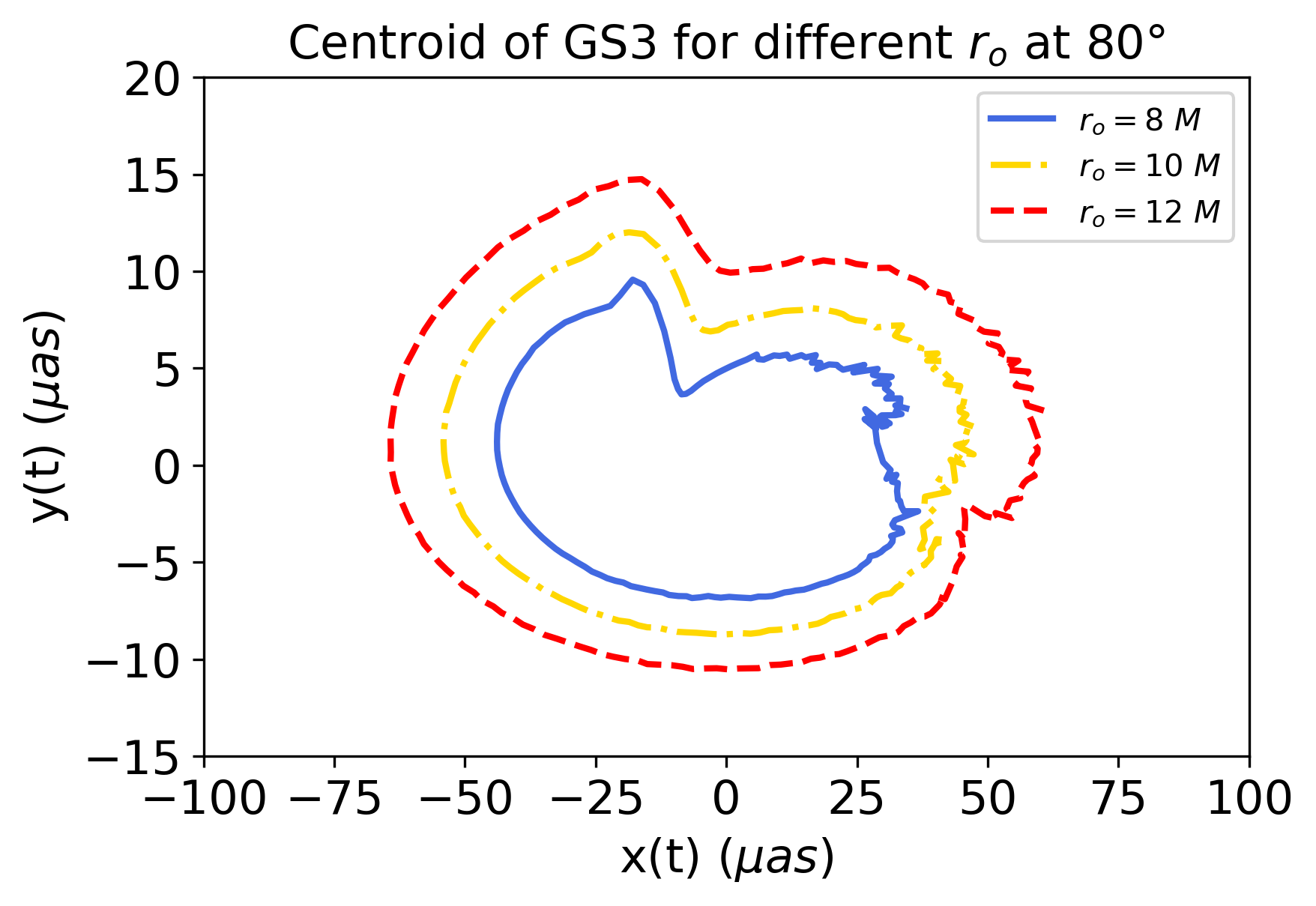} \qquad
    \includegraphics[scale=0.35]{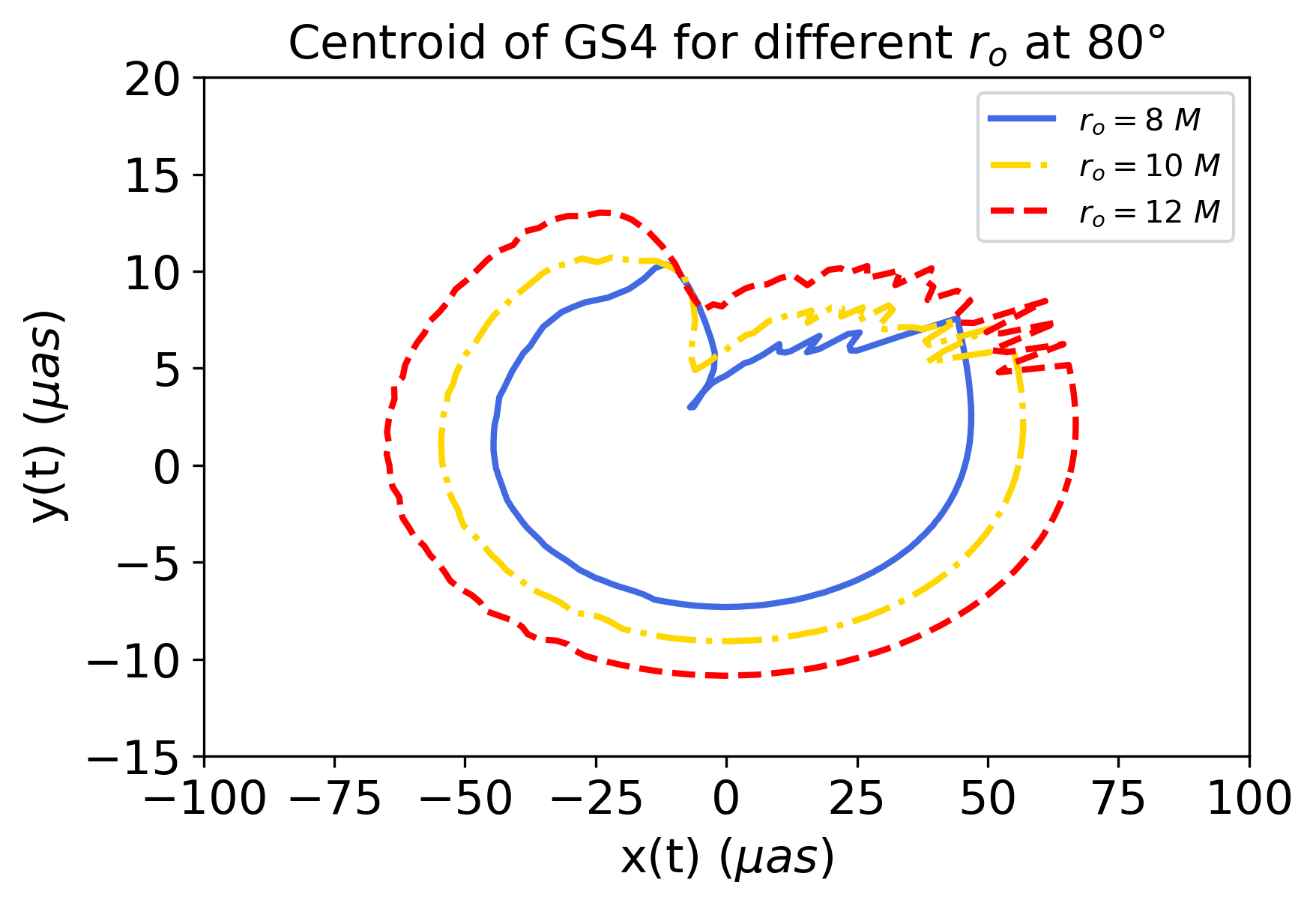} \qquad
    \includegraphics[scale=0.35]{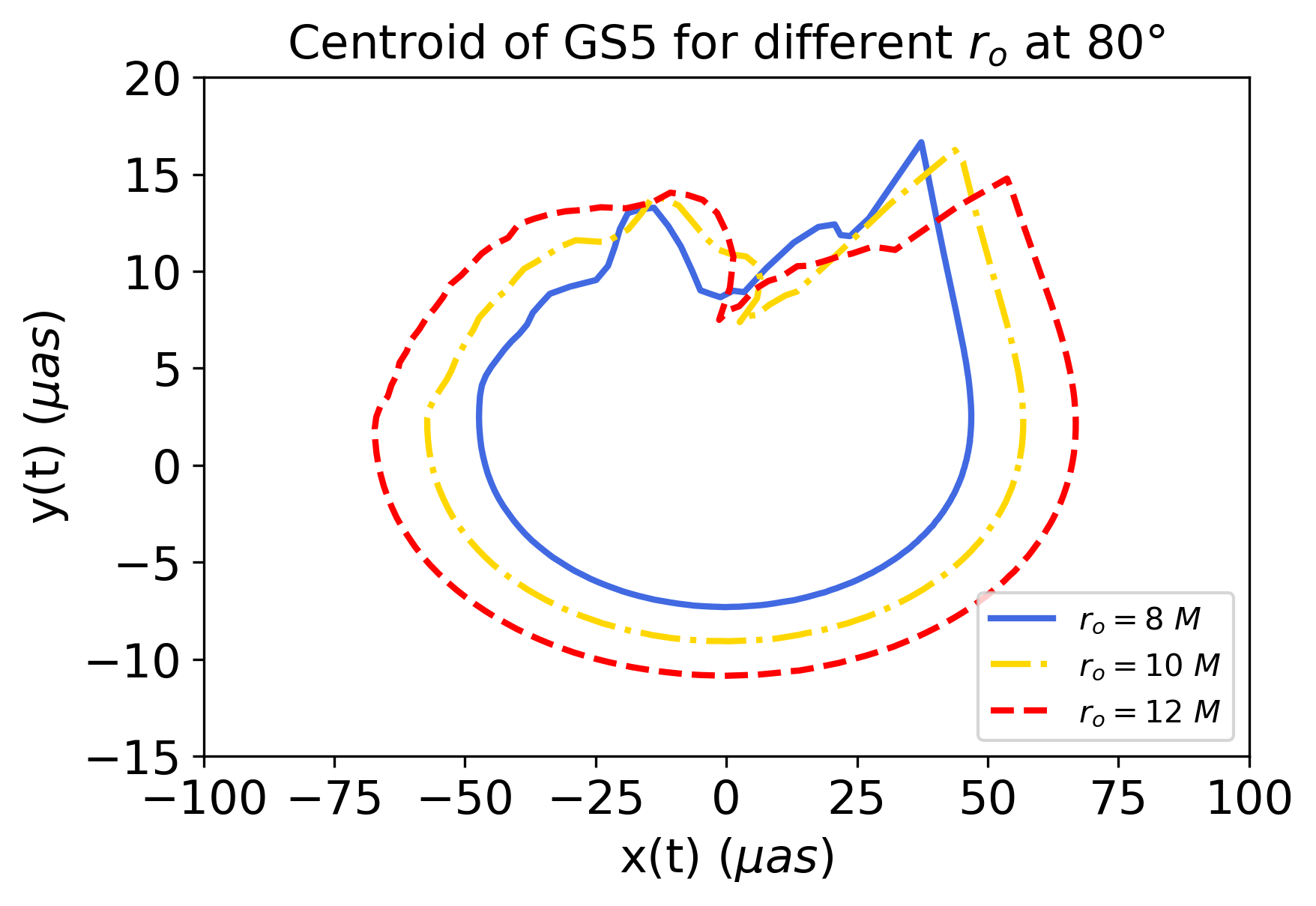}\\
    \caption{Temporal magnitudes (top row) and temporal centroids (bottom row) of the orbital motion of hot-spots for the models GS3 (left column), GS4 (middle column) and GS5 (right column) for orbital radii of $r_o=\{8M, 10M, 12M\}$, with $\theta=80^\circ$ and $r_o=8M$. The models GS1 and GS2 were omitted since their astrometric observables are qualitatively similar to those of model GS3. The astrometric observables for the non-ultra compact models GS4 and GS5 are shown to more significantly depend on the orbital radius of the hot-spot.}
    \label{fig:astroradius}
\end{figure*}

The astrometric observables defined above are analyzed for five different gravastar radii $R=\{2.25M, 2.5M, 3M, 4M, 5M\}$, labeled as GS1, GS2, GS3, GS4 and GS5 respectively, for two different values of $\alpha$, namely $\alpha=\{1, \alpha_{\rm min}\}$, with $\alpha_{\rm min}=1-2\bar M$, for two different observation inclination angles $\theta=\{20^\circ, 80^\circ\}$, and for three different orbital radii $r_o=\{8M, 10M, 12M\}$. The results are organized as follows: the effects of the observation inclination $\theta$, value of $\alpha$, and orbital radius $r_o$ in the time-integrated fluxes are shown in Figs. \ref{fig:fluxincline}, \ref{fig:fluxalpha}, and \ref{fig:fluxradius}, respectively; the effects of the observation inclination $\theta$ in the magnitude and centroid are shown in Figs. \ref{fig:magnitude} and \ref{fig:centroid}, respectively; the effects of $\alpha$ in the magnitude and centroid are given in Fig. \ref{fig:astroalpha}; and the effects of the orbital radius $r_o$ in the magnitude and centroid are given in Fig. \ref{fig:astroradius}. 



\subsection{Time-integrated flux}

For ultra-compact configurations, i.e., with $R\leq 3M$, one observes that the integrated fluxes are qualitatively different from the black-hole scenario in the sense that one additional secondary track is present inside the LR contribution for the entirety of the orbit. As one approaches the limit of non-ultra compacticity, one observes the rise of an additional plunge-through contribution in the center of the image. In the transition between ultra and non-ultra compact, the two secondary tracks merge into a single closed secondary track that appears solely for a limited portion of the orbit, while the plunge-through contribution increases in size and intensity. For the configurations with $R>3M$, one observes that an increase in the gravastar radius induces a growth in the plunge-through contribution, eventually leading to the formation of a closed plunge-through track with two simultaneous images, and also to a decrease in the width of the closed secondary track, bringing the two secondary images close together. It is also noteworthy the fact that, if the compacticity is low enough, the secondary track might not be visible in low observation inclinations due to the weaker effects of light deflection.

The results summarized in the previous paragraph are also qualitatively different from the ones obtained in the framework of other alternative ECO models e.g. fluid and bosonic stars \cite{Tamm:2023wvn,Rosa:2023qcv}. Indeed, those models feature two additional secondary contributions in the integrated flux in the ultra-compact regime, thus implying a more pronounced deviation from the Schwarzschild black-hole results than the gravastar model. Furthermore, the results for low-compacticity are also fundamentally different from the ones observed in fluid and bosonic star models, for which a single additional closed secondary track is present in low-compactness configurations. These discrepancies between the observed properties of gravastars in comparison with other ECO models are caused by the fact that the interior of a gravastar spacetime is a repulsive de-Sitter core, fundamentally different to the massive fluid and bosonic star interiors, and thus whichever light contributions reach the observer after passing through the interior of the gravastar should carry an alternative imprint in comparison with the other ECO models.

Regarding the effects of the orbital radius $r_o$ one observes that, for ultra-compact configurations, a change in the orbital radius leads to an increase in the width of the primary and outer secondary tracks, but it does not qualitatively change the results, i.e., the number and type of tracks visible in the image remain unchanged. The same is not true for low-compactness configurations with $R>3M$, for which one observes that an increase in the orbital radius not only affects the primary and exterior secondary tracks (the latter now merged to the interior one), but also induces non-neglectable changes in the plunge-through contribution. Indeed, an increase in the orbital radius narrows the plunge-through contribution, eventually leading to its disappearance if the orbital radius is large enough. Such an effect is explained by the fact that an increase in the orbital radius while maintaining the same light-deflection effect (given that the model remains unchanged) leads to a decrease in the photons that reach the observer for that particular observation inclination. Finally, one verifies that the effects of modifying the parameter $\alpha$ are subdominant both in the ultra-compact and low-compactness regimes, as the integrated fluxes remain virtually unchanged both quantitatively and qualitatively under a change in $\alpha$, consistently with the predictions of Sec. \ref{sec:geodesics}.

\subsection{Magnitude}

For low inclinations, one observes that the magnitude of the observation corresponds to a single peak of intensity caused by the doppler shifting of the images present in the observer's screen. For the configurations GS1, GS2, GS3 and GS5, the number of images remains constant throughout the entire orbit, and thus no additional features are present in the magnitude profile. However, for the GS4 model, one verifies that there is a finite time interval (corresponding to a limited portion of the orbit) for which the magnitude increases above the other models, before returning back to the original profile. This temporary increase corresponds to the time interval for which the secondary images are present, as expected from the analysis of the integrated flux for this model. 

The situation changes for higher inclinations. Indeed, one observes that additional features besides the central doppler peak are present in the magnitude. For ultra-compact configurations, an additional peak in the magnitude is visible before the doppler peak. This peak is caused by the increase in the intensity of the secondary images for high inclinations, caused by the beaming of these images, as is visible in the integrated fluxes for the same models. For low compactness configurations, one observes an even larger peak of magnitude caused by the appearance of both the secondary and plunge-through contributions, which are absent during most of the orbital period. For the GS5 model, one can even notice the usual three-peak structure caused by the appearance, splitting, and merging of the two images composing the interior closed plunge-through track, similarly to what is observed for low compacticity fluid and bosonic stars \cite{Tamm:2023wvn,Rosa:2023qcv}, although in the latter cases these are caused by the closed secondary tracks, and not the plunge-through track. 

As for a variation in the orbital radius $r_o$ one verifies that, for ultra-compact configurations, the main effect induced by an increase in $r_o$ is a decrease in the height of the magnitude peak while preserving its qualitative shape. This is so because, as previously mentioned for the integrated fluxes, in the high-compactness regime the orbital radius does not affect the number and type of images present, and thus the only noticeable modification is a decrease in the doppler shifting due to a smaller orbital velocity. In the low-compactness regime however, the decrease in the plunge-through contributions with an increase in the orbital radius leads to a decrease in the height and width of the additional peak in the magnitude, and one can even observe that the three-peak structure induced by the closed plunge-through track in the GS5 model becomes a single peak when $r_o$ is large enough to prevent the splitting of the plunge-through images. Finally, one again verifies that the modifications to the magnitude induced by changes in the parameter $\alpha$ are subdominant in comparison to the remaining free parameters.

\subsection{Centroid}

For low inclinations, one observes that the centroid trajectory follows a slightly distorted ellipse. For ultra-compact configurations, the size of this ellipse is smaller than the black-hole counterpart due to the existence of additional secondary tracks in the interior of the LR, whereas for configurations with $R>3M$ the absence of secondary tracks implies that the ellipse is larger than the black-hole counterpart. Interestingly, during the portion of the orbit for which a secondary track is present in the GS4 model, see the corresponding integrated fluxes, one observes a transition from an exterior trajectory (similar to the outer ellipse of the GS5 case) to a more interior trajectory that follows closely the one in the black-hole case. This happens since, during this portion of the orbit, both the GS4 and the black-hole feature the same number of secondary tracks.

The centroid trajectories change abruptly for larger inclinations. Indeed, the previously mentioned effects of the increase in the intensity of the secondary images due to beaming in the ultra-compact situations, as well as the appearance and disappearance of the secondary and plunge-through contributions for low-compactness configurations, induce a shifting of the centroid trajectory towards the center. This shifting is only temporary, lasting for as long as the beaming of the secondary images is strong for ultra-compact configurations, or for as long as the additional contributions are present in the low compactness configurations. One also observes that the shifting occurs at different instants depending on the compacticity of the model, implying that these changes in the intensity of the secondary and additional contributions do not happen simultaneously for every model. It is noteworthy the fact that the shifting when plunge-through contributions are present is qualitatively different from the cases where only the secondary tracks are responsible for the effect.

Regarding the effects of the orbital radius $r_o$, again the effect is minor for ultra-compact configurations since the number and type of tracks present in the observation remains unaffected by changes in this parameter. The same is not true for low-compactness configurations, as we have observed that changes in $r_o$ induce changes in the qualitative aspect of the observed tracks. Indeed, given that the portion of the orbit for which the plunge-through contributions are present in low-compacticity configurations decreases with an increase in $r_o$, one observes that the shifting in the centroid becomes smoother with an increase in $r_o$. Finally, the effects of the parameter $\alpha$ are again proven to be subdominant in the astrometric observations.

%%%%%%%%%%%%%%%%%%%%%%%%%%%%%%%%%%%%%%%%%%%%%%%%%%%%%%%%%%%%%%%%%%%%%%%%%%%%
\section{Conclusions}\label{sec:concl}
%%%%%%%%%%%%%%%%%%%%%%%%%%%%%%%%%%%%%%%%%%%%%%%%%%%%%%%%%%%%%%%%%%%%%%%%%%%%

In this work we have studied the observational properties of thin-shell gravastar models under two different astrophysical systems, namely the orbits of isotropically emitting sources (hot spots) and optically-thin accretion disks. Two parameters of the model were analyzed, namely the radius of the gravastar and the parameter $\alpha$ which controls the proportion of the total mass that is distributed volumetrically and at the thin-shell. Our results show that, for certain combinations of parameters, the observational properties of thin-shell gravastar configurations closely resemble the ones for black-hole spacetimes, thus functioning as models for black-hole mimickers.

Two models for optically-thin accretion disks were considered: the ISCO model, for which one assumes that the extension of the accretion disk is limited to the region where circular timelike orbits are stable, and the Center model, that assumes a peak of emission from the center of the gravastar, in an attempt to simulate the emission of a potential accumulation of matter gathered by accretion. Similarly to other horizonless ECO models, the absence of an event horizon allows for additional contributions in the observed intensity profiles to arise in the observer's screen. However, due to the lack of a sufficiently strong gravitational redshift effect even in the most compact configurations, smooth gravastar models do not give rise to a shadow-like feature when emission from the center of the configuration is assumed. Indeed, our results indicate that, to reproduce such a feature, not only the model has to be extremely compact, but also the majority of the mass of the gravastar must be distributed on the thin-shell. Such a distribution of mass induces a decrease in the central value of $g_{tt}$, which is consequently responsible for the gravitational redshift effect that produces the shadow. Nevertheless, it is also important to mention that due to the absence of the event horizon, the effective size of the shadow is smaller for the gravastar models in comparison with the black-hole case.

Regarding the astrometrical observables from the orbits of relativistic hot-spots, we verified that the time-integrated fluxes, magnitudes, and centroids of the observations in ultra-compact gravastar spacetimes closely resemble the ones in the Schwarzschild black-hole case. Indeed, these observational features were proven to be even more similar to those of the Schwarzschild black-hole than the ones previously obtained for fluid and bosonic stars, since gravastars feature a single additional secondary track whereas the other ECO models mentioned feature two additional such contributions. For the low-compacticity cases, the compatibility of the observable properties with the black-hole scenario is not guaranteed and depends strongly on the orbital radius and observation inclination, leading to additional qualitatively different contributions in every observable analyzed. Interestingly, the proportion of mass allocated at the thin-shell, which was proven essential in the accretion disk analysis to reproduce the observed shadow, is a subdominant parameter in the astrometrical analysis and does not induce any significant changes in the observables. This is somewhat expectable given that we have assumed that the hot-spot moves along a circular geodesic of constant radius, and thus the effects of gravitational redshift are constant through the entire orbit.

Summarizing, our results indicate that, in order for the gravastar model to provide observational properties compatible with those expected from the black-hole spacetimes, two requirements are necessary: the parameter $\alpha$ must be close to its minimum value, corresponding to a large allocation of mass at the surface and, consequently, inducing a strong gravitational redshift effect that produces a shadow in the observed intensity profile; and the radius $R$ of the gravastar must be smaller than $3M$, in order to guarantee that the spacetime features a LR and, consequently, the astrometrical observables and geodesic structure are similar to those of a black-hole spacetime. However, the major drawback of such a model remains the absence of an acceptable natural formation mechanism.

The main aim of our analysis is to extend our awareness of how different alternatives to the black-hole hypothesis influence their observational properties and provide a set of predictions that could potentially be falsifiable in the future. The slight qualitative differences predicted in the observables for the gravastar model in comparison not only with the black-hole model but also with other ECO models like relativistic fluid and bosonic stars successfully accomplishes that goal. Although the current observations might not be precise enough to discriminate between different models, it is our hope that the next generation of interferometric experiments in gravitational physics, namely the ngEHT and the GRAVITY+ instrument, could be a major step towards the acquiring of precise data that could eventually lead to the confirmation or exclusion of some particular models and shed light on the nature of compact supermassive objects.

%%%%%%%%%%%%%%%%%%%%%%%%%%%%%%%%%%%%%%%%%%%%%%%%%%%%%%%%%%%%%%%%%%%%%%%%%%%%
\begin{acknowledgments}
The Mathematica code used to produce the results of Sec.~\ref{sec:disks} was developed in collaboration with Diego Rubiera-Garcia and Gonzalo Olmo. JLR acknowledges the European Regional Development Fund and the programme Mobilitas Pluss for financial support through Project No.~MOBJD647, project No.~2021/43/P/ST2/02141 co-funded by the Polish National Science Centre and the European Union Framework Programme for Research and Innovation
Horizon 2020 under the Marie Sklodowska-Curie grant agreement No. 94533, Fundação para a Ciência e Tecnologia through project number PTDC/FIS-AST/7002/2020, and Ministerio de Ciencia, Innovación y Universidades (Spain), through grant No. PID2022-138607NB-I00. CFBM acknowledge
Fundação Amazônia de Amparo a Estudos e Pesquisas
(FAPESPA), Conselho Nacional de Desenvolvimento
Científico e Tecnológico (CNPq) and Coordenação de
Aperfeiçoamento de Pessoal de Nível Superior (CAPES)
– Finance Code 001, from Brazil, for partial financial.
support.
DSJC and FSNL acknowledge funding through the research grants UIDB/04434/2020, 
UIDP/04434/2020 and PTDC/FIS-AST/0054/2021.
FSNL acknowledges support from the Funda\c{c}\~{a}o para a Ci\^{e}ncia e a Tecnologia (FCT) Scientific Employment Stimulus contract with reference CEECINST/00032/2018.  
\end{acknowledgments}
%%%%%%%%%%%%%%%%%%%%%%%%%%%%%%%%%%%%%%%%%%%%%%%%%%%%%%%%%%%%%%%%%%%%%%%%%%%%

%%%%%%%%%%%%%%%%%%%%%%%%%%%%%%%%%%%%%%%%%%%%%%%%%%%%%%%%%%%%%%%%%%%%%%%%%%%%

\end{document}